\documentclass{aa}
\usepackage[pdftex]{graphicx}
\usepackage[flushleft]{threeparttable}
\usepackage{amsmath}
\usepackage{hyperref}

\def\la{\lower.5ex\hbox{$\; \buildrel < \over \sim \;$}}
\def\ga{\lower.5ex\hbox{$\; \buildrel > \over \sim \;$}}

\begin{document}      

   \title{Thick turbulent gas disks with magnetocentrifugal winds in active galactic nuclei}
   \subtitle{Model infrared emission and optical polarization}

   \author{B.~Vollmer\inst{1}, M.~Schartmann\inst{2,3,4}, L.~Burtscher\inst{2,5}, F.~Marin\inst{1}, S.~H\"{o}nig\inst{6}, R.~Davies\inst{2}, R.~Goosmann\inst{1}}

   \offprints{B.~Vollmer, e-mail: Bernd.Vollmer@astro.unistra.fr}

   \institute{Observatoire astronomique de Strasbourg, Universit\'e de Strasbourg, CNRS, UMR 7550, 
              11 rue de l'Universit\'e, F-67000 Strasbourg, France \and
	      Max-Planck-Institut f\"{u}r extraterrestrische Physik, Postfach 1312, Gie{\ss}enbachstr., D-85741, Garching, Germany \and
              University Observatory Munich, Scheinerstra{\ss}e 1, D-81679 M\"{u}nchen, Germany \and
              Centre for Astrophysics and Supercomputing, Swinburne University of Technology, P.O. Box 218, Hawthorn, Victoria 3122, Australia \and
              Sterrewacht Leiden, Universiteit Leiden, Niels-Bohr-Weg 2, 2300 CA Leiden, The Netherlands \and
              Department of Physics and Astronomy, University of Southampton, Southampton SO17 1BJ, UK
              }

   \date{Received / Accepted}

   \authorrunning{Vollmer et al.}
   \titlerunning{Thick gas disks with magnetocentrifugal winds in AGN}

\abstract{
Infrared high-resolution imaging and interferometry have shown that the dust distribution is frequently elongated along the polar
direction of an AGN. In addition, interferometric mm line observations revealed a bipolar outflow in a direction nearly perpendicular to the nuclear disk.  
To explain these findings, we developed a model scenario for the inner $\sim 30$~pc of an AGN.
The structure of the gas within this region is entirely determined by the gas inflow from larger scales.
We assume a rotating thick gas disk between about one and ten parsec. External gas accretion adds mass and injects energy via gas compression into this gas disk 
and drives turbulence. We extended the description of a massive turbulent thick gas disk developed by Vollmer \& Davies (2013) by adding a magnetocentrifugal wind.
Our disks are assumed to be strongly magnetized via equipartition between the turbulent gas pressure and the energy density of the magnetic field.
In a second step, we built three dimensional density cubes based on the analytical model, illuminated
them with a central source, and made radiative transfer calculations.
In a third step, we calculated MIR visibility amplitudes and compared them to available interferometric observations.
We show that magnetocentrifugal winds starting from a thin and thick gas disk are viable in active galaxy centers.
The magnetic field associated with this thick gas disk plays a major role in driving a magnetocentrifugal wind
at a distance of $\sim 1$~pc from the central black hole. Once the wind is launched, it is responsible for the transport of angular momentum and
the gas disk can become thin. A magnetocentrifugal wind is also expected above the thin magnetized gas disk.
The structure and outflow rate of this wind is determined by the properties of the thick gas disk. 
The outflow scenario can account for the elongated dust structures, outer edges of the thin maser disks, and molecular outflows observed in local AGN.
The models reproduce the observed terminal wind velocities, the scatter of the MIR/intrinsic X-ray correlation, and point source fractions.
An application of the model to the Circinus~Galaxy and NGC~1068 shows that the IR SED, available MIR interferometric observations, and optical
polarization can be reproduced in a satisfactory way, provided that (i) a puff-up at the inner edge of the thin disk is present and (ii) 
a local screen with an optical depth of $\tau_{\rm V} \sim 20$ in form of a local gas filament and/or a warp of the thick disk hide a significant fraction 
of both nuclei. Our thick disk, wind, thin disk model is thus a promising scenario for local Seyfert galaxies.
}

\maketitle

\section{Introduction \label{sec:intro}}

The standard paradigm of type~1 and type~2 active galactic nuclei (AGN) postulates that obscuration by circum-nuclear dust 
in a torus geometry is responsible for the observed dichotomy (see Netzer 2015 for a recent review).
In type~1 sources the torus is seen face-on, whereas in type~2 sources it is seen edge-on.
The torus/unification model has been successful in explaining a number of observations including the 
detection of polarized broad lines (e.g., Ramos Almeida et al. 2016), the collimation of ionization cones (e.g., Fischer et al. 2013), 
its correspondence with the fraction of obscured sources 
(e.g., Maiolino \& Rieke 1995), and the overall spectral energy distribution from the near- to far-infrared (e.g., Netzer et al. 2016).
However, it is not clear at which distance the obscuring material is sitting and which physical configuration it has. 
Whereas Elitzur (2006) prefers a slow wind as the torus that is located very close to the central black hole near the broad 
line region, Vollmer et al. (2008) advocate a thick accretion disk at a distance of several parsec. 
Additional obscuration by galactic structure at kpc scales cannot be excluded either (e.g., Matt 2000, Prieto et al. 2014).

VLT SINFONI H$_2$ (Hicks et al. 2009) and interferometric CO/HCN/HCO$^+$ observations (Sani et al. 2012; Lin et al. 2016; Garcia-Burillo et al. 2016; 
Gallimore et al. 21016) showed that there
are massive rotating thick molecular gas disks sitting at distances of $10$-$50$~pc from the central black hole.
These gas disks contain dust which obscures the central engine if seen edge-on.

On the other hand, VLBI radio continuum observations of nearby AGN led to the discovery of thin molecular maser disks
at distance below $\sim 1$~pc (Greenhill et al. 1995, 1996, 2003). To insure velocity coherence, the velocity dispersion of the
disk must be low, i.e. the disk has to be thin. The massive thick molecular gas disk thus apparently
becomes thin at distances around $\sim 1$~pc from the central black hole (Greenhill 1998; Fig.~8 of Greenhill et al. 2003).

A challenge to the unification theory is that type~1 and type~2 AGNs essentially follow the same mid-IR/intrinsic X-ray relation from low
to high luminosities (e.g., Asmus et al. 2015). Much of the uncertainty about the geometry and dynamics of the torus comes from the fact that the
circum-nuclear dust in AGNs is usually unresolved in single-dish high-resolution images -- a deficiency that
infrared interferometry has partly solved in the last decade (see, e.g. Burtscher et al. 2013, Burtscher et al. 2016). 
Mid-infrared (MIR) interferometric observations of
nearby AGN reveal the geometry of warm ($\sim 300$~K) dust at scales of a few tenth to a few parsec.
The best studied cases are the Circinus galaxy (Tristram et al. 2014) and NGC~1068 (Lopez-Gonzaga et al. 2014). 
Moreover, detailed interferometric MIR observations of NGC~3783 (H\"{o}nig et al. 2013) and NGC~424 (H\"{o}nig et al. 2012) are available.
These observations revealed that the bulk of the MIR emission comes from extended elongated structures 
along the polar axis of the AGN, i.e. in the direction of the ionization cone. In the case of the Circinus galaxy and NGC~1068
thin elongated structures with the same geometry as the maser disks are observed in addition to the polar extended emission.
Most recently, Lopez-Gonzaga et al. (2016) found that 5 of 7 MIR structures in local AGNs observed with MIR interferometry are 
significantly elongated, all in polar direction. Polar dust emission with orientation consistent with that found by interferometry 
was also observed in high-resolution MIR imaging  (Asmus et al. 2016).  
The fact that the polar emission, which is less prone to geometrical extinction, dominates the total MIR emission is consistent with 
the tight mid-IR/X-ray relation (H\"{o}nig \& Kishimoto 2017).
 
The most appealing physical configuration which can explain the molecular line, maser, and MIR observations is a
structure containing three components: (i) an outer thick gas disk which is observed in molecular lines (HCN, HCO$^+$),
(ii) an inner thin disk which is the source of maser emission, and (iii) a polar wind which is responsible for the
bulk of the MIR emission. This wind might be even molecular as advocated by Gallimore et al. (2016) for NGC~1068 
who interpreted the velocities of maser clouds that did not follow the overall rotation pattern as a disk outflow.

Our view of an AGN is from outside in. A certain amount of gas is located at scales of $\sim 1$ to a few $10$~pc into
which a certain amount of energy is injected by external accretion. This energy injection leads to turbulence which
makes the gas disk thick. Since the injection timescale is smaller than the turbulent dissipation timescale, the gas that rotated within
$\sim 10$~pc is adiabatically compressed. The enhanced turbulence leads to overpressured gas clouds that cannot
collapse, i.e. star formation is suppressed (Vollmer \& Davies 2013). The gas mass and turbulent velocity dispersion set
the mass accretion rate of the thick gas disk (typically $\sim 1$~M$_{\odot}$yr$^{-1}$), which is determined by the external mass supply. 
At a certain radius, the poloidal magnetic fields associated with the
thick gas disk are bent outward and a magnetocentrifugal wind is launched (Blandford \& Payne 1982).
The wind takes angular momentum away from the disk, permitting it to become thin. Mass flux conservation leads to
an about two times smaller mass accretion rate of the thin disk compared to the thick gas disk, i.e. about half of the
thick disk mass accretion rate is expelled by the wind. Further in, a broad line
region (BLR) wind is expected (e.g., Gaskell 2009). The final accretion rate onto the central black hole is thus at least four
times smaller than the mass accretion rate of the thick gas disk. This final accretion rate $\dot{M}_{\rm final}$ sets the AGN luminosity.

In this article we elaborate a simple analytical model which takes into account these three components and links them physically.
We note that we are mainly interested in the thick gas disk and the transition between the thick and thin gas disks
involving a magnetocentrifugal wind. The detailed geometry of the inner thin gas disk is not subject of this article.
The role of radiation pressure, which is not an explicit part of our model, is depicted in Sect.~\ref{sec:radpress}.
The models of the thick and thin gas disks are described in Sect.~\ref{sec:thickdisk} and Sect.~\ref{sec:thindisk}, the wind model in Sect.~\ref{sec:wind}.
The link between the components is explained in Sect.~\ref{sec:windlink} and an expression for the critical
radius where the wind sets in is given in Sect.~\ref{sec:rw}. The model parameters are given in Sect.~\ref{sec:modelparam}.
The conditions under which these winds are viable are explored in Sect.~\ref{sec:viable}.
Their terminal wind speeds are presented in Sect.~\ref{sec:speed}.
Axisymmetric (Sect.~\ref{sec:3dmodels}) and non-axisymmetric (Sect.~\ref{sec:helical}) 
3D density distributions are computed and a full
radiative transfer model is applied to the model cubes (Sect.~\ref{sec:fullrt}). 
The model IR luminosities, central extinctions, and spectral energy distributions are compared to observations in Sect.~\ref{sec:central},
\ref{sec:sed}, and \ref{sec:luminosities}. To compare our models with MIR
interferometric observations, we compute the expected visibilities from the model MIR images (Sect.~\ref{sec:visibilities}).
The models are then applied to the Circinus galaxy and NGC~1068 (Sect.~\ref{sec:application}). The influence of our model geometry on
the optical polarization is investigated in Sect.~\ref{sec:polarization}.
Finally, we give our conclusions in Sect.~\ref{sec:conclusions}.

\section{The model \label{sec:model}}

\begin{figure}
  \centering
  \resizebox{\hsize}{!}{\includegraphics{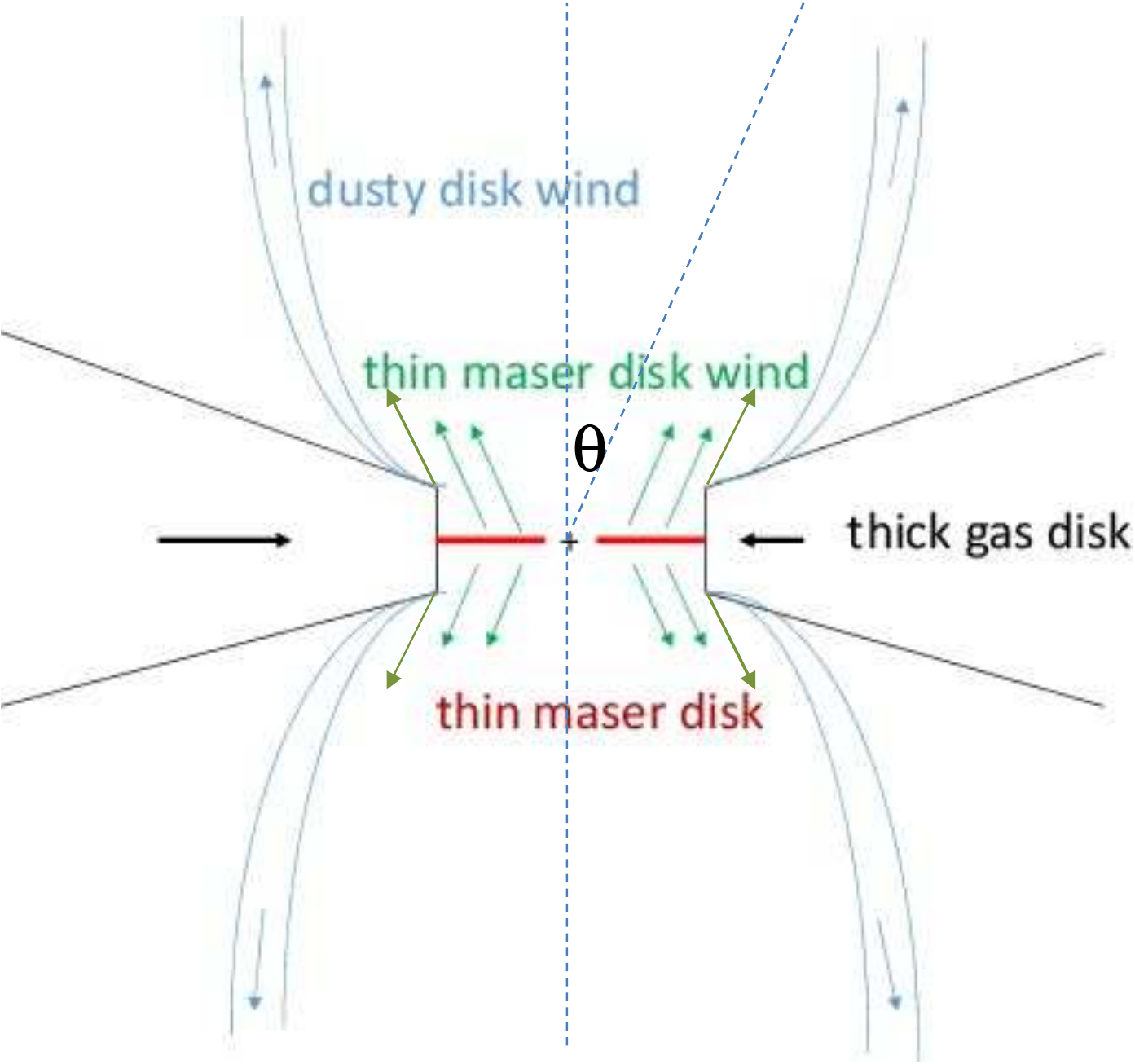}}
  \caption{Schematic ingredients of the model. The dusty disk wind corresponds to the magnetocentrifugal wind.
    The wind emanates from the thin and thick disks with a half-opening angle $\theta$. 
  \label{fig:modelsketch}}
\end{figure}

Our analytical model consists of three different structures: (i) a thick turbulent clumpy gas disk, (ii)
a magnetocentrifugal wind, and (iii) a thin gas disk (Fig.~\ref{fig:modelsketch}). In addition, it is expected that a wind
also emanates from the thin gas disk.
The thick gas disk is fed by externally infalling gas. The external mass accretion and energy injection rates are so high that
the disk has to increase its viscosity to be able to cope with the gas inflow. By increasing its viscosity, the disk becomes thick.
The poloidal magnetic field is dragged with the radial flow and, eventually bends at an angle of $\sim 30^{\circ}$ at the radius,
where the magnetocentrifugal wind sets in (Blandford \& Payne 1982). Since the wind takes over the angular momentum transfer, the gas
disk can become thin at smaller radii. We show in Sect.~\ref{sec:thindisk} that a magnetocentrifugal wind arises naturally from
a magnetized thin accretion disk around a massive black hole.

For the radial distribution of the magnetocentrifugal wind, we assume that the wind starts at the inner edge of the thick disk 
and continues over the thin maser disk. 
Radiation pressure pushes the part of the wind which is located well above the thin disk to larger radii, increasing
the wind angle with respect to the disk vertical. Potentially this can lead to a more radial/equatorial wind (see, e.g., Fig.~7 of Chan \& Krolik 2017). 
We can only speculate that the subsequent radial bending of vertical magnetic field lines
at the inner edge of the thick disk leads to a wind angle that is sufficient for the launching of a magnetocentrifugal wind at this position ($\ga 30^{\circ}$).
Such a bending of the magnetic field lines is plausible, because at the point where the wind is launched above the thick disk the
radiation pressure is in approximate equilibrium with the energy density of the magnetic field (see Sect.~\ref{sec:modelparam}).
A detailed analysis of this issue is beyond the scope of this article. The existence of a magnetocentrifugal wind is consistent
with the finding of Das et al. (2006), that in the NLR wind of NGC~1068 the outflow velocity cannot be simply accounted for by radiative forces driving the gas clouds.

The wind outflow rate is estimated at the inner edge of the thick gas disk.
We thus assume that it is not much different across the thin maser disk.
The physical parameters of the thick disk are determined by the external mass accretion rate, the gas surface density, 
and the turbulent velocity dispersion. The magnetocentrifugal wind sets in at a radius $r_{\rm wind}$.
At $r < r_{\rm wind}$ the disk becomes thin, because the wind extracts the angular momentum from the disk 
making mass accretion possible. The critical radius $r_{\rm wind}$ is set by mass flux conservation:
\begin{equation}
\dot{M}_{\rm thick\ disk}-\dot{M}_{\rm wind}-\dot{M}_{\rm thin\ disk}=0\ ,
\end{equation}
where $\dot{M}_{\rm thick\ disk}$, $\dot{M}_{\rm wind}$, and $\dot{M}_{\rm thin\ disk}$ are the mass accretion rates of
the thick accretion disk, the magnetocentrifugal wind, and the thin accretion disk.

\subsection{Radiation pressure \label{sec:radpress}}

Another cause for the onset of a wind is radiation pressure. 
For an optically thick medium the outward force exerted by radiation is $F=L/c$, where $L$ is the luminosity and $c$ the light speed.
If the near-infrared optical depth of the gas $\tau_{\rm NIR}$ is higher than unity, the force becomes $F=\tau_{\rm NIR} L/c$ (e.g., Roth et al. 2012).
We assume that the magnetocentrifugal wind has a hollow cone structure with an optical depth $\tau_{\rm V}$ of a few. The NIR optical depth is thus
smaller than unity. The magnetohydrodynamic (MHD) equation of motion of the gas in the presence of radiation pressure reads:
\begin{equation}
\rho \frac{{\rm d}\vec{v}}{{\rm d}t}=-\rho \nabla \Phi - \nabla p + \frac{1}{4\pi}(\nabla \times \vec{B}) \times \vec{B} + \frac{L \rho \kappa}{4 \pi R^2 c} \,
\label{eq:mhd}
\end{equation}
where $\vec{B}$ is the magnetic field, $v$ the gas velocity, $\rho$ the gas density, $\Phi$ the gravitational potential, $p$ the gas pressure, 
and $\kappa$ the dust absorption coefficient.
We assume that in the wind region the large-scale magnetic field and thus magnetic tension dominates: 
\begin{equation}
\frac{1}{4\pi}(\nabla \times \vec{B}) \times \vec{B} \sim \frac{1}{4\pi}(\vec{B} \cdot \nabla) \vec{B}\ .
\end{equation}
As stated by Roth et al. (2012), modeling the force from radiation pressure, and predicting
by what factor it exceeds $L/c$, becomes a difficult problem to tackle analytically in the absence of spherical symmetry.

The semianalytic model developed by Everett (2005) includes magnetic acceleration and radiative acceleration of a continuous self-similar wind launched from an 
accretion disk. In this model the central continuum radiation first encounters a purely magnetocentrifugally accelerated wind, which is referred to as a 
"shield''. The shield was introduced as a separate component in order to cleanly differentiate the effect of shielding from radiative acceleration; 
radiative driving of the shield was therefore not considered. Beyond that shield is an optically thin, radiatively and magnetically accelerated wind;
the radiation coming from an underlying thin gas disk. Everett (2005) considered radiative acceleration by bound-free (``continuum driving'') and 
bound-bound (``line driving''). They found that shielding by a magnetocentrifugal wind can increase the efficiency 
of a radiatively driven wind. For luminosities smaller than a tenth of the Eddington luminosity, magnetic driving dominates the mass outflow rate.
Keating et al. (2012) added the continuum opacity of interstellar medium (ISM) dust grains to the model of  Everett (2005) and
produced IR SEDs for a wide range of parameter space. They found that models with high column densities, Eddington ratios, and black hole masses
were able to adequately approximate the general shape and amount of power expected in the IR as observed in a composite of optically luminous 
Sloan Digital Sky Survey quasars. 

Roth et al. (2012) used 3D Monte Carlo radiative transfer calculations to determine the radiation force on dusty gas residing within approximately 30~parsecs 
from an accreting supermassive black hole. Static smooth and clumpy thick gas disk distributions were considered.
In the absence of a coupling between the radiative transfer calculation and a hydrodynamic
solver in a time-dependent calculation, they could not determine the dynamics of the gas. 
Roth et al. (2012) found that these dust-driven winds can carry momentum fluxes of $1$-$5$ times $L/c$ and can correspond to mass-loss 
rates of $10$-$100$~M$_{\odot}$yr$^{-1}$ for a $10^8$~M$_{\odot}$ black hole radiating at or near its Eddington limit.

Wada (2012) used a three-dimensional, multi-phase hydrodynamic model including radiative feedback
from the central source, i.e., radiation pressure on the dusty gas and the X-ray heating of cold, warm, and hot ionized gas to 
study the dynamics of a thick gas and dust disk located in the inner $30$~pc around the central black hole.
Only  the  radial  component  of the central radiation flux was considered for radiative heating and pressure. 
Wada (2012) showed that a geometrically and optically thick  torus with a biconical outflow ($v_{\rm outflow} \sim 100$~km\,s$^{-1}$) can  be  naturally  
formed in the central region extending tens of parsecs around a low-luminosity AGN.  

Chan \& Krolik (2016, 2017) performed three-dimensional, time-dependent radiative magnetohydrodynamics simulations of AGN tori featuring quality 
radiation transfer and simultaneous evolution of gas and radiation. The simulations solved the magnetohydrodynamics equations simultaneously 
with the infrared and ultraviolet radiative transfer equations. 
Their thick gas torus achieved a quasi-steady state lasting for more than an orbit at the inner edge, and potentially for much longer. 
The associated central wind is propelled by IR and UV radiation. 
Despite the gas torus being magnetized, the outflow is not a magnetocentrifugal wind because meandering loops of magnetic fields in the outflow 
are too weak to exert much force.

The physical model of Dorodnitsyn et al. (2016) describes the time-evolution of a three-dimensional distribution of gas and dust in the gravitational
field of a supermassive black hole, adopting radiation hydrodynamics in axial symmetry (2.5D calculations on a uniform cylindrical grid). 
Radiation input from X-ray and UV illumination was taken
into account. Dorodnitsyn et al. (2016) showed that in the absence of strong viscosity the conversion of external UV and X-ray
into IR radiation becomes important at Eddington ratios in excess of $0.01$. Gas located closer to the black hole escapes in the form of a fast thermally 
driven wind with a characteristic velocity of $100$-$1000$~km\,s$^{-1}$. An IR-driven wind exists farther away from the black hole.
For times in excess of a few $10^4$~yr, the wind outflow rates are $M_{\rm wind} \la 0.1$~M$_{\odot}$yr$^{-1}$.

In the following we argue that for our massive, highly turbulent gas disks with strong magnetic fields (under the assumption of
energy equipartition between the gas pressure and the energy density of the magnetic field) the magnetocentrifugal outflow rate
exceeds that induced by radiation pressure.

The near-infrared optical depth of the wind is much smaller than unity. Thus, the UV luminosity dominates the radiation pressure
in a thin layer of column density $N \sim 5 \times 10^{20}$~cm$^{-2}$. Radiation pressure will then radially push the gas and magnetic fields 
in the wind, until the point of equilibrium between the magnetic tension and radiation pressure. This equilibrium sets the wind opening angle,
which is defined as twice the angle between the inner edge of the wind and the polar axis.
Since we assume energy equipartition between the turbulent kinetic energy density and that of the magnetic field, the strength of the polar magnetic
field above the thin disk is significantly smaller than that located above the thick gas disk.
Radiation pressure exceeds the magnetic pressure in the region above the thin disk, making the wind more radial/equatorial there.
The wind is expected to bend upwards at the inner edge of the thick gas disk creating a hollow cone.
The observed elongated polar structures in local AGN (Tristram et al. 2014; H\"{o}nig et al. 2012, 2013; Asmus et al. 2016) are in favor
of a scenario where magnetic tension dominates already at relatively small wind opening angles. 
Indeed, the comparison of the magnetic pressure in the wind $p_{\rm B}$ and the radiation pressure $p_{\rm rad}$ (assuming an optically thick
medium) at the critical radius $r_{\rm wind}$ (Table~\ref{tab:output}) in NGC~1068 and Circinus shows that the radiation pressure is comparable to the
magnetic pressure at the critical radius where the wind sets in.
Moreover, we argue in Sect.~\ref{sec:irinterferometry} that the observed MIR visibilities are consistent with such a homogeneous wind of column densities 
$N_{\rm wind} \sim 5 \times 10^{21}$~cm$^{-2}$ ($\tau_{\rm V} \sim 3$, Fig.~\ref{fig:tau_inc_SURVEY1bnew}). 

Within the thick disk, the NIR optical depth is high and IR radiation pressure has to be taken into account.
The condition for a disk in which radiation pressure dominates is given by Chan \& Krolik (2016; Eq.~15 and 29):
\begin{equation}
(\frac{L_{\rm UV}}{4 \pi R^2 c}) \times (\frac{2 C_{\rm UV}}{1-C_{\rm IR}}) \sim \rho v_{\rm rot}^2\ , 
\end{equation}
where $C_{\rm UV}$ and $C_{\rm IR}$ are the UV and IR covering fractions, $\rho$ the midplane gas density, and $v_{\rm rot}$ the rotation velocity of the thick gas disk.
With the assumed wind and disk opening angles (Fig.~\ref{fig:modelcutcircinus}) we set $C_{\rm UV}=1-\cos(70^{\circ})$ and $C_{\rm IR}=1-\cos(30^{\circ})$.
Furthermore, the gas density is given by $\rho=v_{\rm rot}^2/(R^2 \pi G Q)$, where $G$ is the gravitational constant and $Q$ the Toomre parameter.
Inserting the disk properties for Circinus and NGC~1068 (Table~\ref{tab:input}), yields UV luminosities of $L_{\rm UV}=5 \times 10^{44}$~erg\,s$^{-1}$
for Circinus and $L_{\rm UV}=6 \times 10^{45}$~erg\,s$^{-1}$ for NGC~1068. These luminosities are about a factor of $20$ higher than their actual
luminosities (Table~\ref{tab:input}). We thus conclude that within our massive thick gas disk turbulent gas pressure exceeds by far radiation pressure.

As a further test, we calculated the expected mass outflow rates and terminal velocities for radiation pressure-driven winds (Eq.~34 and 35 of Chan \& Krolik 2015).
We found $\dot{M}=0.06$~M$_{\odot}$yr$^{-1}$ and $v_{\infty}=1870$~km\,s$^{-1}$ for Circinus and $\dot{M}=0.34$~M$_{\odot}$yr$^{-1}$ and $v_{\infty}=3500$~km\,s$^{-1}$
for NGC~1068. These mass outflow rates are about two times lower, the terminal wind velocities more than three times higher than our values (Table~\ref{tab:output}).
The terminal wind speeds of the radiation-pressure-driven winds are significantly higher than those observed in local AGN by M{\"u}ller-Sanchez et al. (2011).
The energy density of our model disk is thus dominated by kinematics (turbulence), that of the wind by the magnetic field (Table~\ref{tab:output}).

We thus conclude that IR radiation pressure does not play a major role in our thick disks, because they are massive and strongly magnetized.
In the following we will thus ignore radiation pressure, keeping in mind that it will certainly shape the wind above the thin gas disk and
probably even the inner rim of the wind above the thick disk, being responsible for the
wind opening angle (Fig.~\ref{fig:modelsketch}). In our model, we assume a parabolic hollow wind cone with a half-opening angle of 
$\sim 25^{\circ}$ at a height of $\sim 4$~pc. 
Magnetocentrifugal winds can be recognized by their relatively low terminal wind speeds (Fig.~\ref{fig:mdiskvwind}) and high
rotation velocities.

\subsection{The thick turbulent clumpy gas disk \label{sec:thickdisk}}

Gas disks around central galactic black holes contain clumps of high volume densities (e.g., Krolik \& Begelman 1988, 
G\"{u}sten et al. 1987). The formation of regions of overdense gas is caused by thermal instabilities and, if present, selfgravity (e.g., Wada et al. 2002).
\footnote{Another possibility consists of supernova-driven turbulence (e.g., Wada et al. 2009).}
In turbulent galactic disks, gas clumps are of transient nature with lifetimes of about a turbulent crossing time (e.g., Dobbs \& Pringle 2013).
The governing gas physics of such disks are highly time-dependent and intrinsically stochastic. 
Over a long-enough timescale, turbulent motion of clumps is expected to redistribute angular momentum 
in the gas disk like an effective viscosity would do. This allows accretion of gas towards the center and makes it possible to treat 
the disk as an accretion disk (e.g., Pringle 1981). This gaseous turbulent accretion disk rotates in a given gravitational
potential $\Phi$ with an angular velocity $\Omega=\sqrt{R^{-1}\frac{{\rm d}\Phi}{{\rm d}R}}$, where $R$ is the disk radius. 

Vollmer \& Davies (2013) developed an analytical model for turbulent clumpy gas disks where the energy to drive turbulence is supplied
by external infall or the gain of potential energy by radial gas accretion within the disk.
The gas disk is assumed to be stationary ($\partial \Sigma/\partial t=0$) and the external mass accretion rate to be close to the 
mass accretion rate within the disk (the external mass accretion rate feeds the disk at its outer edge). 
The external and disk mass accretion rates averaged over the viscous timescale are assumed to be constant.
Within the model, the disk is characterized by the disk mass accretion rate $\dot{M}$ and the Toomre $Q$ parameter which is
used as a measure of the gas content of the disk for a given gravitational potential.
Vollmer \& Davies (2013)  suggested that the velocity dispersion of the torus gas is increased through adiabatic compression by the 
infalling gas. The gas clouds are not assumed to be selfgravitating. The disk velocity dispersion is
fixed by the mass accretion rate and the gas surface density via the Toomre parameter $Q$. 
Turbulence is assumed to be supersonic, creating shocks in the weakly ionized dense molecular gas. For not too high shock velocities 
($<50$~km\,s$^{-1}$) these shocks will be continuous (C-type). The cloud size is determined by the size of a C-shock at a given 
velocity dispersion. Typical cloud sizes are $\sim 0.02$~pc at the inner edge of the thick disk and $\sim 0.1$~pc at a radius of $5$~pc (Vollmer \& Davies 2008).

In such a turbulent clumpy gas disk the area filling factor is
\begin{equation}
\Phi_{\rm A}=  \Phi_{\rm V} H/r_{\rm cl} = 11.6 \sqrt{\frac{v_{\rm A,0}}{Q v_{\rm turb}}} \ ,
\label{eq:phiaturb}
\end{equation}
where $r_{\rm cl}$ is the cloud radius, $v_{\rm A,0}=1$~km\,s$^{-1}$ the Alfv\'en velocity and $v_{\rm turb}$ the turbulent velocity dispersion of the disk.
The Toomre parameter is given by
\begin{equation}
\label{eq:q}
Q=\frac{v_{\rm turb}}{v_{\rm rot}}\frac{M_{\rm dyn}}{M_{\rm gas}}\ ,
\end{equation}
where $v_{\rm rot}$ is the rotation velocity, $M_{\rm dyn}$ the dynamical mass, and $M_{\rm gas}$ the disk gas mass.

The disk mass accretion rate is given by 
\begin{equation}
\dot{M}_{\rm thick\ disk}=2\,\pi \nu \Sigma = 2\,\pi \Phi_{\rm A} v_{\rm turb} H^2 \rho\ ,
\end{equation}
where $\nu=\Phi_{\rm A} v_{\rm turb} H$ is the gas viscosity, $\Sigma=\rho H$ the gas surface density, $\rho$ the gas density, and $H$ the disk thickness.
In a disk of constant $Q$ in hydrostatic equilibrium 
\begin{equation}
\label{eq:QQ}
\rho=\Omega^2/(\pi G Q)\ , 
\end{equation}
where $\Omega$ is the angular velocity and $G$ the gravitation constant
(e.g., Vollmer \& Beckert 2002). This leads to
\begin{equation}
\label{eq:mdotdisk}
\dot{M}_{\rm thick\ disk}=2\,\Phi_{\rm A} \frac{v_{\rm turb}^3}{G Q}\ .
\end{equation} 

We parametrize the model with $M_{\rm dyn}/M_{\rm gas}$ and the turbulent velocity of the disk $v_{\rm turb}$.
This leads to the Toomre $Q$ parameter (Eq.~\ref{eq:q}), the area filling factor (Eq.~\ref{eq:phiaturb}), and the
disk mass accretion rate (Eq.~\ref{eq:mdotdisk}).

\subsection{Launching a wind from a thin disk \label{sec:thindisk}}

Wardle \& K\"onigl (1993) investigated the vertical structure of magnetized thin accretion disks that power centrifugally
driven winds. The magnetic field is coupled to the weakly ionized disk material by ion-neutral and electron-neutral collisions.
The resulting strong ambipolar diffusion allows a steady state field configuration to be maintained against radial inflow and azimuthal 
shearing. They showed that the presence of a magneto-centrifugal wind implies that the thin disk has to be confined by magnetic
stresses rather than by the tidal field. These authors derived criteria for viable thin-disk-wind models based on the 
ratio of the dynamical timescale to the neutral-ion coupling time $\eta=\eta_{\rm in} x_{\rm i} \rho \Omega^{-1} $,
where $\eta_{\rm in}=3.7 \times 10^{13}$~cm$^{3}$s$^{-1}$g$^{-1}$ (Draine et al. 1983) is the collision coefficient and
$x_{i}$ the ionization fraction, the ratio of the Alfv\'en speed to the turbulent velocity or sound speed $a=v_{\rm A}/c$, and the Mach number associated with the inward
radial drift of the neutral gas at the midplane $\epsilon=v_{\rm r}/c$:
(i) $\eta > 1$ insures a pure ambipolar diffusion regime and (ii) $(2 \eta)^{-\frac{1}{2}} \la a \la 2 \la \epsilon \eta$ insures (1) that
the disk rotates sub-Keplerian, (2) that the disk is confined by magnetic stresses, (3) the validity of the wind launching criterion, (4) a wind starting point that
lies well above the disk scale height.

The ionization fraction is given by 
\begin{equation}
x_{i}= \gamma \big(\frac{\zeta_{\rm CR}}{n_{\rm H}}\big)^{\frac{1}{2}}\ ,
\end{equation}
where $\gamma=600$~cm$^{-\frac{3}{2}}$s$^{\frac{1}{2}}$, $\zeta_{\rm CR}=2.5 \times 10^{-15}$~s$^{-1}$ (Vollmer \& Davies 2013),
and $n_{\rm H}=\rho/(2.3 \times m_{\rm p})$.
With Eq.~\ref{eq:QQ} and $Q=1$ we obtain
\begin{equation}
\eta=\gamma \eta_{\rm in} \sqrt{\frac{\zeta_{\rm CR} 2.3 m_{\rm p}}{\pi G Q}} \simeq 4 \ .
\end{equation}
With an Alfv\'en speed of $v_{\rm A}=1$~km\,s$^{-1}$ (Vollmer \& Davies 2013), a turbulent/sound speed of $c=1.5$~km\,s$^{-1}$, and a radial inflow velocity
$v_{\rm r}=1.5$~km\,s$^{-1}$, we obtain $a=0.7$ and $\epsilon=1$. This set of parameters is close to that of the typical solutions of Wardle \& K\"onigl (1993) and 
fulfills all criteria cited above.

Since in the model of Wardle \& K\"onigl (1993) the transition from a sub-Keplerian quasi-hydrostatic disk to a centrifugally driven outflow occurs
naturally, we conclude that radiation pressure is a priori not needed to launch the wind from the thin maser disk. On the other hand,
we expect that radiation pressure accelerates the centrifugally launched gas to higher velocities and larger radii until the point
where the pressure of the azimuthal magnetic field equals the radiation pressure. 

\subsection{The magnetocentrifugal wind \label{sec:wind}}

Since we want to describe the magnetocentrifugal wind with ideal MHD, we have to assess the role of ambipolar diffusion
in the thick gas disk and the wind.
According to McKee et al. (2010) the Reynolds number for ambipolar diffusion is 
\begin{equation}
R_{\rm AD}=4 \pi \eta_{\rm in} \rho_{\rm i} \rho_{\rm n} l v B^{-2}\ ,
\end{equation}
where $\eta_{\rm in}=3.7 \times 10^{13}$~cm$^3$g$^{-1}$s$^{-1}$ (McKee et al. 2010) is the ion--neutral coupling coefficient,
$\rho_{\rm i}$ and $\rho_{\rm n}$ are the ion and neutral densities, respectively, $l$ and $v$ the characteristic length scale and velocity,
and $B$ the magnetic field strength. Note that the parameter $\beta$ which describes the coupling between the gas and the
magnetic field (e.g., Pudritz \& Norman 1983) is the inverse of the Reynolds number for ambipolar diffusion $\beta=t_{\rm ni}/t_{\rm flow}=R_{\rm AD}^{-1}$,
where $t_{\rm ni}$ is the neutral-ion collision timescale and $t_{\rm flow}=l_{\rm flow}/v_{\rm flow}$ the timescale of the flow.
Assuming energy equipartition $B^2/(8\,\pi)=1/2 \rho v^2$, i.e. the wind speed equals the Alfv\'enic velocity, the Reynolds number is
\begin{equation}
R_{\rm AD}=\eta_{\rm in} \rho_{\rm i} l v^{-1}\ .
\end{equation}
We assume a degree of ionization 
\begin{equation}
x_{\rm i}=\frac{n_{\rm i}}{n_{\rm n}}=\gamma \big(\frac{\zeta_{\rm CR}}{n_{\rm H}}\big)^{\frac{1}{2}}\ .
\end{equation}

For the thick gas disk we assume a mean density of $n_{\rm n}=\Omega^2/(\pi G Q)=10^6$~cm$^{-3}$, a characteristic length scale equal to the disk height 
$l \sim H \sim 0.5$~pc, and a characteristic velocity equal to the velocity dispersion $v=40$~km\,s$^{-1}$ (see Table~\ref{tab:input}).
This yields an ionization fraction $x_{\rm i}=3 \times 10^{-8}$, an ion density of $\rho_{\rm i}=30 \times 3 \times 10^{-8} n_{\rm n}$ (the factor 
$30$ is due to the heavy ion approximation), and a Reynolds number for ambipolar diffusion $R_{\rm AD}=20 \gg 1$.
Therefore, ambipolar diffusion does not play an important role\footnote{$R_{\rm AD} = \infty$ corresponds to ideal MHD.}  in the thick disk
and the approximation of ideal MHD is justified. 

For typical wind densities of $n_{\rm wind}=10^5$~cm$^{-3}$, wind velocities of $v_{\rm wind}=300$~km\,s$^{-1}$ (see Sect.~\ref{sec:speed}), 
flow lengthscale of $l_{\rm wind}=1$~pc, and $R_{\rm AD}=20$ or $\beta=0.05$, we obtain an ionization fraction $x_{\rm i}=n_{\rm i}/n_{\rm n}=10^{-6}$.
The ionization rate caused by cosmic rays is $x_{\rm i,CR}=10^{-7}$.
The ten times higher ionization rate, which is required for the application of ideal MHD, can be easily sustained by the X-ray emission of the central engine
which directly illuminates the wind (X-ray dominated region XDR; Meijerink \& Spaans 2005). 

The equations of stationary, axisymmetric, ideal MHD are the conservation of mass, the equation of motion,
the induction equation for the evolution of the magnetic field, and the solenoidal condition on the magnetic field:
\begin{equation}
\label{eq:mhdeq}
\begin{gathered}
\nabla \cdot (\rho \vec{v})=0 \\
\rho \vec{v} \cdot \nabla \vec{v}=-\nabla p - \rho \nabla \Phi + \frac{1}{4\pi}(\nabla \times \vec{B}) \times \vec{B}\\
\nabla \times (\vec{v} \times \vec{B})=0\\
\nabla \cdot \vec{B} = 0\ ,
\end{gathered}
\end{equation}
where $\rho$ is the gas density, $\vec{v}$ the gas velocity, $p$ the gas pressure, $\Phi$ the gravitational potential,
and $\vec{B}$ the magnetic field.
The angular momentum equation for an axisymmetric flow is described by the azimuthal component of the equation of motion.
For simplicity we ignore stresses that would arise from turbulence and neglect the pressure and gravitational potential (see, e.g., K\"{o}nigl \& Pudritz 2000).
The solution is thus only valid for the freely flowing part of the wind far away from the gas disk ($z/H \gg 1$).
\footnote{Within this approximation the winds above the thin and thick disk have to be regarded separately. The wind structure  in the transition region
is more complex and its study is beyond the scope of this article.
}
With the separation of poloidal and toroidal field components $\vec{B}=\vec{B}_{\rm p}+B_{\rm t}\vec{\hat e}_{\rm t}$ and
$\vec{v}=\vec{v}_{\rm p}+v_{\rm t}\vec{\hat e}_{\rm t}$ we obtain
\begin{equation}
\label{eq:momentumeq}
\rho \vec{v}_{\rm p} \cdot \nabla(r v_{\rm t})=\frac{1}{4\pi} \vec{B}_{\rm p} \cdot \nabla (rB_{\rm t})\ .
\end{equation}

The induction equation links the velocity field and the magnetic field. Because of axisymmetry, the poloidal velocity 
vector is parallel to the poloidal component of the magnetic field (K\"{o}nigl \& Pudritz 2000), which implies
\begin{equation}
\label{eq:Bk}
\rho \vec{v}_{\rm p} = k\,\vec{B}_{\rm p}\ ,
\end{equation}
with the mass load per unit time and unit magnetic field flux, which is preserved along streamlines from the rotator
\begin{equation}
\label{eq:k}
k=\frac{\rho v_{\rm p}}{B_{\rm p}}=\frac{{\rm d} \dot{M}_{\rm wind}}{{\rm d} \Phi}\ ,
\end{equation}
where ${\rm d}\dot{M}_{\rm wind}=\rho v_{\rm p} {\rm d}A$ is the mass loss rate of the wind and 
${\rm d}\Phi = B_{\rm p}{\rm d}A$ is the magnetic flux.
The mass load is determined by the physics of the underlying rotator, i.e. the accretion disk.

The induction equation also determines the field of the flow (K\"{o}nigl \& Pudritz 2000)
\begin{equation}
B_{\rm t}=\frac{\rho\,r}{k}(\omega-\omega_0)\ ,
\end{equation}
where $\omega=v_{\rm t}/r$ is the angular velocity, and $\omega_0$ is the angular velocity at the disk midplane.

The application of Eq.~\ref{eq:Bk} to the momentum equation with $k=$const yields a
constant angular momentum per unit mass along a streamline
\begin{equation}
\label{eq:lrv}
l=r\,v_{\rm t} - \frac{r B_{\rm t}}{4\pi k}\ .
\end{equation}
This means that the specific angular momentum of a magnetized flow is carried by both the rotating gas and
the twisted field. The value of $l$ can be found by 
\begin{equation}
\label{eq:rv}
r\,v_{\rm t} = \frac{lm^2-r^2\omega_0}{m^2-1}\ ,
\end{equation}
where $m=v_{\rm p}/v_{\rm A}$ is the Alfv\'en Mach number and $v_{\rm A}=B_{\rm p}/\sqrt{4\pi \rho}$.

Once the wind speed equals the Alfv\'en speed at a point called ``Alfv\'en point'', magnetic field lines that are 
carried and stretched by the wind open up, and all the mass at this point is considered lost from the disk. Another way 
to look at this process is to think of the magnetic field lines as rods that are attached to the rotating disk at one end, 
whereas the other ends of the open field lines are radially stretched beyond the Alfv\'en point. As a result, each field 
line applies a torque on the disk and spins it down. This torque is proportional to the momentum of the wind at the 
Alfv\'en point, to the disk rotation rate, and to the distance of the Alfv\'en point (the lever arm that applies the torque). 
The imaginary surface that represents all Alfv\'en points is called ``Alfv\'en surface'' and the integral of the mass flux 
through this surface is the mass loss rate of the disk to the wind. The Alfv\'en surface is defined by $r=r_{\rm A}$ 
on the outflow field lines where $m=1$ (Eq.~\ref{eq:rv}). The flow along any field line corotates with the accretion disk until this surface is reached.

From the regularity condition at the Alfv\'en critical point where the denominator of Eq.~\ref{eq:rv}
vanishes, it follows
\begin{equation}
\label{eq:l}
l=\omega_0 r_{\rm A}\ .
\end{equation}
The index $0$ denotes quantities which are evaluated in the disk plane.
The terminal speed of the flow is approximately
\begin{equation}
\label{eq:vinfty1}
v_{\infty} \simeq \sqrt{2} \omega_0 r_{\rm A}\ .
\end{equation}
Michel (1969) found that the terminal speed of a cold MHD wind is of the order of 
\begin{equation}
\label{eq:vinfty2}
v_{\infty} \sim \big( \frac{\omega^2 \Phi^2}{\dot{M}_{\rm wind}}\big)^{\frac{1}{3}}\ ,
\end{equation}
with the conservation of the magnetic flux $\Phi=B_{\rm p} r_{\rm A}^2=B_{0{\rm p}} r_0^2$ (Pudritz \& Norman 1986).

Combining Eqs.~\ref{eq:vinfty1} and \ref{eq:vinfty2} leads to 
\begin{equation}
\label{eq:ra}
r_{\rm A} \sim \frac{v_{\infty}}{\sqrt{2} \omega_0}=\frac{1}{\sqrt{2}} \big( \frac{\Phi^2}{\omega_0 \dot{M}_{\rm wind}} \big)^{\frac{1}{3}}=\big( \frac{B_{\rm p}^2 r_0^4}{\omega_0 \dot{M}_{\rm wind}} \big)^{\frac{1}{3}} \ .
\end{equation}

The mass outflow rate in a high density regime ($\beta \ll 1$) is given by
\begin{equation}
\label{eq:mdotwind}
\dot{M}_{\rm wind}=\int_{\rm A}\rho \vec{v_{\rm p}} \cdot {\rm d}\vec{A} \sim 4\pi (\rho v_{\rm p})_{r_{\rm A}}r_{\rm A}^2 \Omega
\end{equation}
(Pudritz \& Norman 1983), where $A$ is the Alfv\'enic surface and $4 \pi \Omega$ the solid angle that it subtends.
For a cone with a half-opening angle $\theta$, $\Omega=(1-\cos(\theta))$.
We estimate the gas density at the Alfv\'enic surface through conservation of mass flux within the thick gas disk and the wind:
\begin{equation}
\rho v_{\rm r} = 2 \rho_{\rm A} v_{\rm p}\ ,
\end{equation}
where $v_{\rm r}$ is the radial velocity of the disk gas and $\rho_{\rm A}$ the gas density at the Alfv\'enic surface.
The radial velocity of the thick disk gas is given by the gas viscosity $\nu=v_{\rm r} R=\Phi_{\rm A} v_{\rm turb} H$ and thus $v_{\rm r}=v_{\rm turb}^2/v_{\rm rot}$.
With $v_{\rm p} \sim v_{\rm rot}$ this leads to
\begin{equation}
\label{eq:rhoa}
\rho_{\rm A}=\frac{1}{2} \big(\frac{v_{\rm turb}}{v_{\rm rot}}\big)^2 \rho\ .
\end{equation}

Inserting Eq.~\ref{eq:ra} and Eq.~\ref{eq:rhoa} into Eq.~\ref{eq:mdotwind} yields
\begin{equation}
\label{eq:mdotw}
\dot{M}_{\rm wind}= \big( \xi \frac{1}{2} (\frac{v_{\rm turb}}{v_{\rm rot}})^2 \rho v_{\rm rot} B_{\rm p}^{\frac{4}{3}}r^{\frac{8}{3}}\omega_0^{-\frac{2}{3}} \big)^{\frac{3}{5}}\ ,
\end{equation}
where $\xi=4 \pi \Omega$.

We assume that the poloidal regular magnetic field is about $1/3$ of the total magnetic field. This is consistent
with the fraction of the regular large-scale to the total magnetic field in spiral galaxies (e.g., Beck 2016).
Energy equipartition between the gas energy density and the total magnetic field yields 
\begin{equation}
\label{eq:bpp}
B_{\rm p}=\frac{1}{3} \sqrt{4 \pi \rho v_{\rm turb}^2}\ ,
\end{equation}
where $v_{\rm turb}$ is the turbulent gas velocity dispersion of the accretion disk.
The density of the accretion disk in hydrostatic equilibrium and with a constant Toomre $Q$ parameter is given by
$\rho=\omega_0^2/(\pi G Q)$ (e.g., Vollmer \& Beckert 2002), where $G$ is the gravitation constant.

Inserting Eq.~\ref{eq:bpp} into Eq.~\ref{eq:mdotw} leads to our final expression for the wind mass loss rate:
\begin{equation}
\label{eq:mdotw1}
\dot{M}_{\rm wind}=2^{\frac{1}{5}} 3^{-\frac{4}{5}} \pi^{-\frac{3}{5}} \xi^{\frac{3}{5}} Q^{-1} G^{-1} v_{\rm rot} v_{\rm turb}^2 \ .
\end{equation}

From Eq.~\ref{eq:mdotw1} it becomes clear that there is a degeneracy between the factor $\frac{1}{3}$ between the poloidal and the
total magnetic field (Eq.~\ref{eq:bpp}) and the solid angle subtended by the wind $\xi$: an increase of the solid angle 
together with an increase of the poloidal magnetic field fraction lead to the wind mass loss.

\subsection{Linking the wind to the accretion disk \label{sec:windlink}}

To determine the mass accretion rate of the thin disk, we assume that the wind outflow rates from the thin disk and the inner
edge of the thick disk are comparable. Since the solid angle subtended by the wind from the thick disk (see Sect.~\ref{sec:rw}) is about
$3$ times larger than the solid angle subtended by the wind from the thin disk with a half-opening angle of $\theta \sim 20^{\circ}$,
this implies that the mass flux of the outflow from the inner disk is about $3$ times smaller than that from the inner edge of the thick disk.

To calculate the torque exerted by the wind on the underlying accretion disk, we apply the momentum equation
(Eq.~\ref{eq:momentumeq}) to the accretion disk (see K\"{o}nigl \& Pudritz 2000):
\begin{equation}
\label{eq:momvr}
\frac{\rho v_{\rm r}}{r_0} \frac{\partial (r_0 v_{\rm rot})}{\partial r_0}=\frac{B_{\rm r}}{4 \pi r_0} \frac{\partial (r_0 B_{\rm t})}{\partial r_0} + \frac{B_z}{4 \pi} \frac{\partial B_{\rm t}}{\partial z}\ .
\end{equation}
The specific angular momentum is thus removed from the accretion flow by magnetic torques associated with the radial/vertical shear of 
the toroidal field. We assume that for typical field inclination the second term of Eq.~\ref{eq:momvr} dominates.
This implies that the magnetic field lines are inclined less than $\sim 60^{\circ}$ with respect to the disk normal.

Mass conservation in an accretion disk gives the relation between the disk mass accretion rate and the radial velocity
\begin{equation}
\dot{M}_{\rm thin\ disk}=-2 \pi \Sigma v_{\rm r} r_0\ ,
\end{equation}
where $\Sigma=\rho H$ is the gas surface density and $H$ the disk thickness.
With Eq.~\ref{eq:momvr} we obtain
\begin{equation}
\dot{M}_{\rm thin\ disk} \frac{{\rm d}(r_0v_{\rm rot})}{{\rm d}r_0}=-r_0^2B_{\rm t}B_{\rm z}\ .
\end{equation} 
The angular momentum can be carried away by Alfv\'en waves or, when the magnetic field lines are inclined more than $\sim 30^{\circ}$
with respect to the disk normal, by a centrifugally driven wind (Blandford \& Payne 1982). Thus, the range of inclination angles 
$\theta_{\rm B}$ between the magnetic field lines and the disk normal is approximately $30^{\circ} \leq \theta_{\rm B} \leq 60^{\circ}$.

Rewriting Eq.~\ref{eq:lrv} as $rB_{\rm t}= 4 \pi k (r v_{\rm rot} - l)$ and inserting the expressions for $k$ (Eq.~\ref{eq:k}) and $l$
(Eq.~\ref{eq:l}) yields
\begin{equation}
\label{eq:mdotthin}
\dot{M}_{\rm thin\ disk}= f_{\rm g} \dot{M}_{\rm wind} \big(\frac{r_{\rm A}}{r_0} \big)^2 \ ,
\end{equation}
where $f_{\rm g}$ is a geometric factor, which depends on the geometry of the poloidal field.
Following Pudritz \& Norman (1986) we assume $f_{\rm g}=\frac{1}{3}$ for a polar wind.
This means that if the viscous torques in the disk are relatively unimportant, the angular momentum
loss is provided by the magnetocentrifugal wind.

\subsection{Where the wind sets in \label{sec:rw}}

Within the presented scenario the external mass inflow $\dot{M}_{\rm ext}=\dot{M}_{\rm disk}$
and the Toomre parameter $Q$ determine the physical properties of the thick disk (Sect.~\ref{sec:thickdisk}).
With high $\dot{M}_{\rm disk}$ and $Q$, the turbulent disk can be relatively thick.
The disk is permeated by a magnetic field which has a large-scale regular and a small-scale turbulent
magnetic field. The regular field has a poloidal and a toroidal component.
At a given radius or distance $r_{\rm wind}$ to the central black hole the angle between the poloidal field lines and the
disk normal exceeds $30^{\circ}$ and a magnetocentrifugal wind is launched.
At $r < r_{\rm wind}$ the wind provides the transport of angular momentum (see Sect.~\ref{sec:wind}) and the disk becomes thin.
The mass accretion rate of the thin disk is given by Eq.~\ref{eq:mdotthin}.
For simplicity, we assume a sharp transition between the thick and the thin disk at $r=r_{\rm wind}$.
Furthermore, we assume that the wind outflow rate is the same above the thick and the thin disk and that the
Alfv\'{e}n radii of the thin and thick disk are the same at the transition radius $r_{\rm wind}$. With Eqs.~\ref{eq:ra}, \ref{eq:bpp}, and \ref{eq:mdotw1}
this implies that the gas pressures of the thin and thick disk at $r_{\rm wind}$ are the same:
$\rho_{\rm thick} v_{\rm turb,\ thick}^2=\rho_{\rm thin} v_{\rm turb,\ thin}^2$. With Eq.~\ref{eq:QQ} we obtain 
$v_{\rm turb,\ thick}/v_{\rm turb,\ thin}=\Sigma_{\rm thin}/\Sigma_{\rm thick}=\sqrt{Q_{\rm thick}/Q_{\rm thin}}$.

In the absence of a detailed knowledge of the configuration of the magnetic field,
we assume that mass conservation determines the radius $r_{\rm wind}$ where the wind sets in:
\begin{equation}
\label{eq:mm}
\dot{M}_{\rm thick\ disk}-\dot{M}_{\rm wind}-\dot{M}_{\rm thin\ disk}=0\ .
\end{equation}
A constant turbulent velocity of the thick gas disk is assumed, which is consistent with the
SINFONI H$_2$ observations of Hicks et al. (2009). The observed extent of the thin disk gives $M_{\rm dyn}/M_{\rm gas}$
and thus determines $Q$ (Eq.~\ref{eq:q}), the mass accretion rate of the thick disk $\dot{M}_{\rm thick\ disk}$ (Eq.~\ref{eq:mdotdisk}),
wind outflow rate $\dot{M}_{\rm wind}$ (Eq.~\ref{eq:mdotw1}), the accretion rate of the thin disk $\dot{M}_{\rm thin\ disk}$
(Eq.~\ref{eq:mdotthin}), and the radius $r_{\rm wind}$ (Eq.~\ref{eq:mm}) where the wind sets in.

The dynamical mass in the galactic center is given by
\begin{equation}
M_{\rm dyn}=M_{\rm BH}+M_* r^{\frac{5}{4}}\ ,
\end{equation}
where $M_{\rm BH}$ is the mass of the central black hole and $M_*$ defines the mass of the
central star cluster (Vollmer \& Duschl 2001).
This parametrization of the dynamical mass leads to an approximately constant rotation curve
$v_{\rm rot}=\sqrt{M_{\rm dyn} G/r}$ beyond the sphere of the influence of the black hole.

\subsection{Model parameters \label{sec:modelparam}}

We apply our model on the two best studied nearby AGN, the Circinus galaxy ($D=4.2$~Mpc) and NGC~1068 ($D=14.4$~Mpc).
The input parameters are presented in Table~\ref{tab:input}. The adopted bolometric luminosities are consistent with 
the values estimated by Moorwood et al. (1996) for Circinus and Pier et al. (1994) and H\"{o}nig et al. (2008) for NGC~1068.
Pudritz \& Norman (1983) set the
solid angle subtended by the wind to $4 \pi \Omega=4 \pi 0.1$, which corresponds to a half-opening angle of $\theta =26^{\circ}$.
In our scenario, this cone is not filled as in the  Pudritz \& Norman model, but hollow.
The solid angle subtended by a hollow cone that reproduces IR interferometric observations is about $4 \pi \Omega \sim 0.025$.
This corresponds to inner and outer half-opening angles of $\theta_{\rm in}=20^{\circ}$ and $\theta_{\rm out}=24^{\circ}$, comparable to
the narrow line region cones of Mrk~1066, NGC~4051, NGC~4151 (Fischer et al. 2013), and NGC~1068 (M\"{u}ller-Sanchez et al. 2011).

The choice of the turbulent velocities is motivated by Plateau de Bure Interferometer HCN and HCO$^+$ observations presented in Sani et al. (2012) and Lin et al. (2016).
The observed velocity dispersion of the dense gas (HCN, HCO$^+$) is about a factor of $1.5$ lower than that derived from
SINFONI H$_2$ observations presented in Davies et al. (2007) and Hicks et al. (2009). The black hole masses are taken from
Greenhill et al. (2003) and Lodato \& Bertin (2003). The outer radii of the thin maser disks are
$\sim 0.4$~pc for the Circinus galaxy (Greenhill et al. 2003) and $\sim 1.1$~pc for NGC~1068 (Greenhill \& Gwinn 1997).
Since the maser disks seem to be warped these radii are lower limits. In addition, the transition between
the thin and the thick disk might not be sharp as assumed by our simple model. We thus adopted $\sim 30$\,\% larger
radii for the transition between the thin and the thick disk (Table~\ref{tab:input}). This nicely reproduced the elongated compact
components of the MIR interferometric observations (Sect.~\ref{sec:irinterferometry}) and is comprised within the model uncertainties.
\begin{table*}
\begin{center}
\caption{Model input parameters.\label{tab:input}}
\begin{tabular}{lcccccccc}
\hline
 & $D$ & $L_{\rm bol}^{\rm a}$ & $M_{\rm BH}$ & $M_*^{\rm b}$ & $v_{\rm rot}^{\rm c}$ & $v_{\rm turb}^{\rm d}$ & $\Omega^{\rm e}$ & $r_{\rm wind}^{\rm f}$ \\
 & (Mpc) & (erg\,s$^{-1}$) & (M$_{\odot}$)  & (M$_{\odot}$pc$^{-\frac{5}{4}}$) & (km\,s$^{-1}$) & (km\,s$^{-1}$) & & (pc) \\
\hline
Circinus & $4.2$ & $3 \times 10^{43}$ & $1.6 \times 10^6$  & $1.0 \times 10^6$ & 100 & $30$ & $0.022$ & $0.53$ \\
NGC 1068 & $14.4$ & $3 \times 10^{44}$ & $8.6 \times 10^6$ & $3.0 \times 10^6$ & 170 & $50$ & $0.022$ & $1.5$ \\
\hline
\end{tabular}
\begin{tablenotes}
  \item $^{\rm a}$ Moorwood et al. (1996), Pier et al. (1994)
  \item $^{\rm b}$ leading to a flat rotation curve at $R=3$~pc for Circinus and $R=5$~pc for NGC~1068
  \item $^{\rm c}$ rotation velocity at $r=10$~pc
  \item $^{\rm d}$ assumed turbulent velocity dispersion of the thick disk
  \item $^{\rm e}$ assumed solid angle subtended by the wind
  \item $^{\rm f}$ outer radii of the thin maser disks (Greenhill et al. 2003, Greenhill \& Gwinn 1997)
    \end{tablenotes}
\end{center}
\end{table*}

The resulting parameters of the thick disk/wind/thin disk model are presented in Table~\ref{tab:output}.
\begin{table*}
\begin{center}
\caption{Model results for the thick disk, magnetocentrifugal wind, and the thin disk.\label{tab:output}}
\begin{tabular}{lcccccccccc}
\hline
 & $M_{\rm gas}/M_{\rm dyn}$ & $Q$ & $\Phi_{\rm A}$ & $B_{\rm p}^{\rm a}$ & $r_{\rm A}$ & $\dot{M}_{\rm thick\ disk}$ & $\dot{M}_{\rm wind}$ & $\dot{M}_{\rm thin\ disk}$ & $p_{\rm B}(r_{\rm wind})$ & $p_{\rm rad}(r_{\rm wind})$ \\
 & & & & (mG) & (pc) & (M$_{\odot}$yr$^{-1}$) & (M$_{\odot}$yr$^{-1}$) & (M$_{\odot}$yr$^{-1}$) & (erg\,cm$^{-3}$) & (erg\,cm$^{-3}$) \\
\hline
Circinus & $0.010$ & $22$ & $0.45$ & $14$ & $0.85$ & $0.26$ & $0.14$ & $0.12$ & $7.4 \times 10^{-6}$ & $5.1 \times 10^{-6}$ \\
NGC 1068 & $0.016$ & $15$ & $0.42$ & $15$ & $2.4$ & $1.57$ & $0.85$ & $0.73$ & $8.5 \times 10^{-6}$ & $6.2 \times 10^{-6}$ \\
\hline
\end{tabular}
\begin{tablenotes}
  \item $^{\rm a}$ large-scale polar magnetic field in the wind with $B_{\rm p}=1/3\,B_0$.
    \end{tablenotes}
\end{center}
\end{table*}
The Toomre $Q$ parameter of the two thick gas disks is $Q \sim 15$-$20$. This is higher than the values assumed
by Vollmer et al. (2008) which where based on gas masses derived from NIR observations of warm H$_2$ (Davies et al. 2007)
with an uncertain conversion factor. The lower gas masses are corroborated by HCN observations of the central $50$~pc in nearby
AGN (Sani et al. 2012). The area filling factor of gas clouds in the disks is close to one $\Phi_{\rm A} \sim 0.5$.
The wind outflow rates are comparable to the mass accretion rates of the thin disk.
The thick/thin disks and magnetocentrifugal winds of the Circinus galaxy and NGC~1068 have very different mass accretion and outflow 
rates, the mass accretion and outflow rates of NGC~1068 being about $6$ times those of the Circinus galaxy. 
The Alfv\'en radii of the two galaxies are $\sim 1.6$ times larger than the critical radii.
The situation is different to that in protostellar outflows where this ratio is $\sim 3$.
The magnetic field strength of $\sim 15$~mG at the Alfv\'en radius is well comparable to the magnetic field strength
of the gas and dust torus in NGC~1068 inferred from NIR polarimetric observations (Lopez-Rodriguez et al. 2015).

\subsection{Viable magnetocentrifugal disk winds \label{sec:viable}}

\begin{figure}
  \centering
  \resizebox{\hsize}{!}{\includegraphics{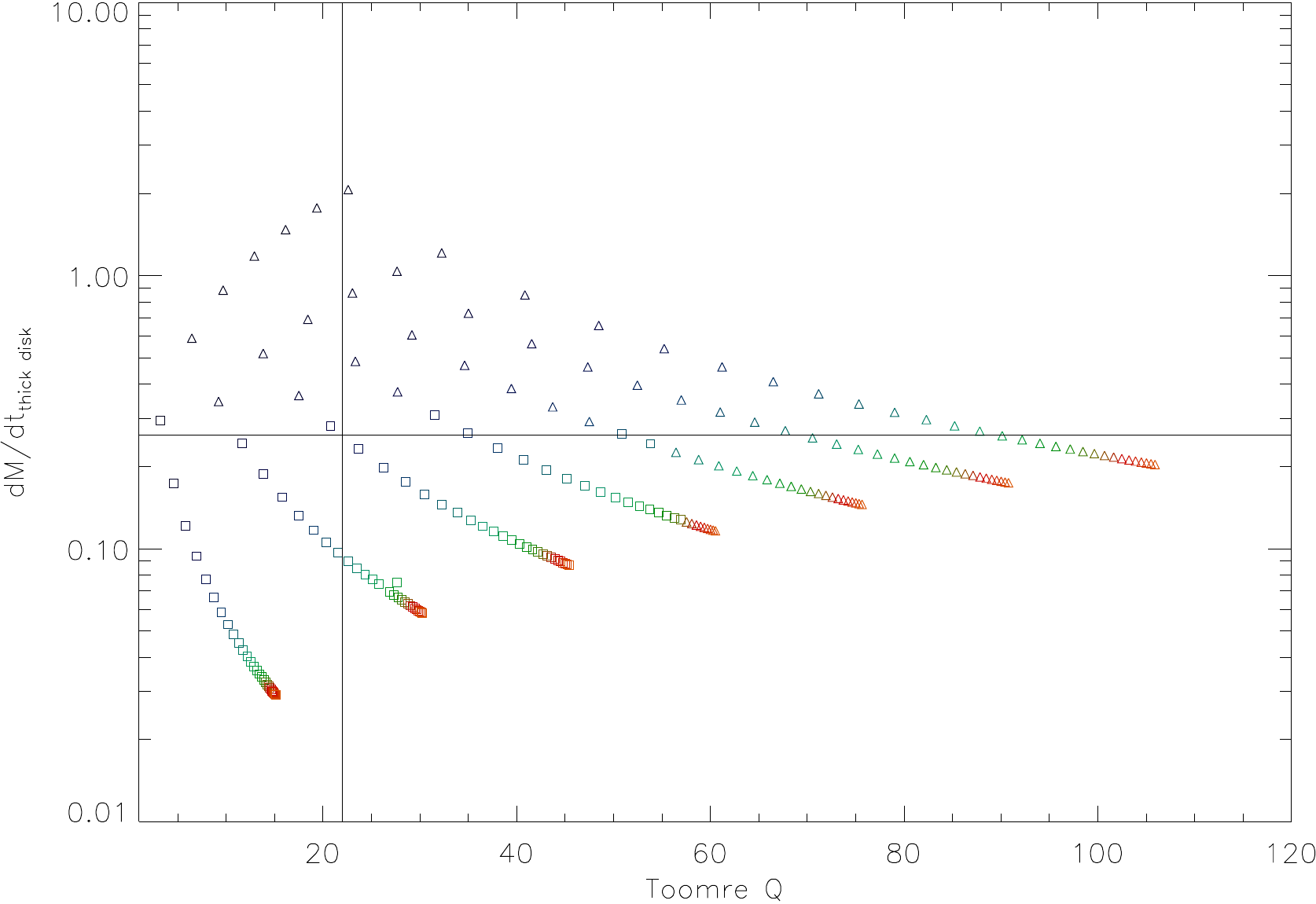}}
  \caption{Circinus galaxy: mass accretion rate of the thick gas disk (in $M_{\odot}$yr$^{-1}$) as a function of the Toomre $Q$ parameter.
    Each point corresponds to a given critical radius where the wind sets in and the accretion disk becomes thin.
    Each line corresponds to a velocity dispersion of $v_{\rm turb}=10, 20, 30, 40, 50, 60, 70$~km\,s$^{-1}$ (from left to right).
    The critical radii range from $0.2$~pc (blue) to $3$~pc (red) in steps of $0.1$~pc.
    Triangles: $p_{\rm rad}/p_{\rm B} < 0.5$; boxes: $p_{\rm rad}/p_{\rm B} \ge 0.5$.
    The intersection of the solid lines corresponds to the assumed critical radius $r_{\rm wind}=0.53$~pc and $v_{\rm turb}=30$~km\,s$^{-1}$.
  \label{fig:mdotqcircinus}}
\end{figure}

\begin{figure}
  \centering
  \resizebox{\hsize}{!}{\includegraphics{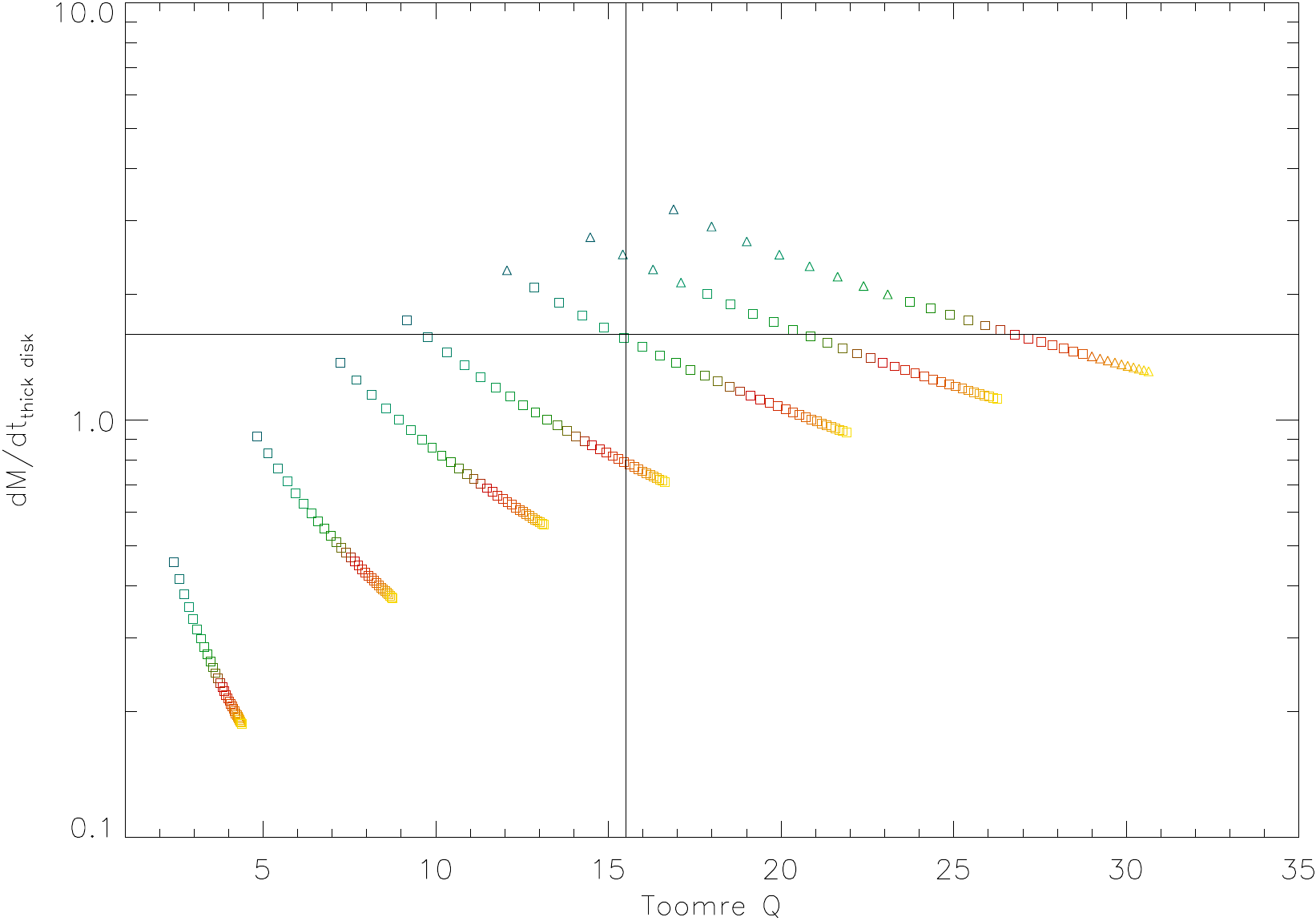}}
  \caption{NGC~1068: mass accretion rate of the thick gas disk (in $M_{\odot}$yr$^{-1}$) as a function of the Toomre $Q$ parameter.
    Each point corresponds to a given critical radius where the wind sets in and the accretion disk becomes thin.
    Each line corresponds to a velocity dispersion of $v_{\rm turb}=10, 20, 30, 40, 50, 60, 70$~km\,s$^{-1}$ (from left to right).
    The critical radii range from $1$~pc (green) to $4$~pc (yellow) in steps of $0.1$~pc.
    Triangles: $p_{\rm rad}/p_{\rm B} < 0.5$; boxes: $p_{\rm rad}/p_{\rm B} \ge 0.5$.
    The intersection of the solid lines corresponds to the assumed critical radius $r_{\rm wind}=1.5$~pc and $v_{\rm turb}=50$~km\,s$^{-1}$.
  \label{fig:mdotqn1068}}
\end{figure}

In the previous Section we calculated the Toomre $Q$ parameter, the mass accretion rates of the thick and thin disk,
and the wind outflow rate based on the observed outer radius of the thin maser disks.
We can now generalize and assume different transition radii $r_{\rm wind}$ between the thick and thin disks.
For a given turbulent velocity dispersion $v_{\rm turb}$, each choice of $r_{\rm wind}$ leads to a Toomre $Q$ parameter and a mass accretion rate of 
the thick disk $\dot{M}_{\rm thick\ disk}$. We varied the outer radius of the thin disk within reasonable ranges and the velocity dispersion from 
$10$~km\,s$^{-1}$ to $70$~km\,s$^{-1}$. The results of these calculations are presented in Fig.~\ref{fig:mdotqcircinus} and Fig.~\ref{fig:mdotqn1068}.
We observe a general trend that the thick disk accretion rate decreases with increasing $Q$.
For each $Q$ a range of $\dot{M}_{\rm thick\ disk}$ within about $1$~dex leads to viable disk-wind solutions.
If the radiation pressure is responsible for the angle of $\sim 30^{\circ}$ between the polar magnetic fields and the disk normal necessary to drive the wind, 
we expect that only solutions with $p_{\rm rad}/p_{\rm B} > 0.5$ are viable (boxes in Fig.~\ref{fig:mdotqcircinus} and Fig.~\ref{fig:mdotqn1068}).
This would greatly reduce the number of viable solutions for the Circinus model.
For $v_{\rm turb}=30$~km\,s$^{-1}$ and $Q > 50$ no disk-wind configuration is viable, i.e. this kind of accretion disks cannot have a wind.
The circumnuclear disk (CND) in the Galactic Center with $Q=100$-$200$ (Vollmer et al. 2004) is in this situation.

\subsection{The terminal wind speed \label{sec:speed}}

\begin{figure}
  \centering
  \resizebox{\hsize}{!}{\includegraphics{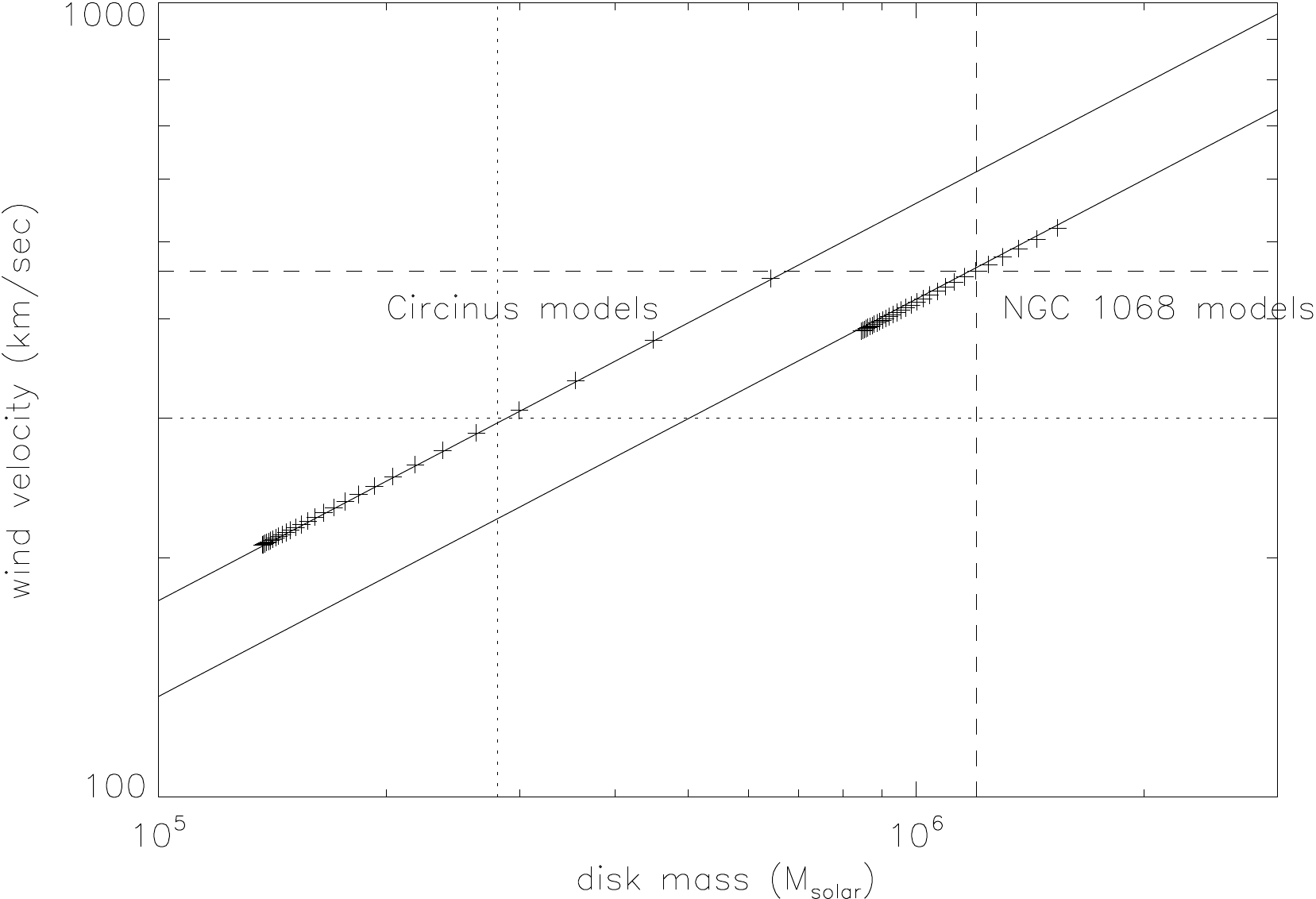}}
  \caption{Terminal wind speed (Eq.~\ref{eq:vinfty1}) as a function of the disk gas mass within $10$~pc for the models described in Sect.~\ref{sec:viable}.
    The dotted and dashed lines indicate the Circinus and NGC~1068 3D models.
  \label{fig:mdiskvwind}}
\end{figure}

The magnetocentrifugal wind has a terminal speed given by Eq.~\ref{eq:vinfty1}. We calculated the terminal wind speed 
for the models described in Sect.~\ref{sec:viable}. The results are presented in Fig.~\ref{fig:mdiskvwind}.
For a given gravitational potential (black hole mass and stellar mass distribution), the terminal wind speed increases with increasing disk
gas mass ($v_{\infty} \propto M_{\rm gas}^{\frac{1}{2}}$). On the other hand, we observe an offset between the relations for Circinus and
NGC~1068 which is approximately proportional to the total mass included within $10$~pc ($v_{\infty} \propto M_{\rm tot}^{-\frac{1}{4}}$). 
As expected, a deeper gravitational potential leads to a higher terminal wind speed. Our model terminal wind speeds are well
comparable with those given by M\"uller-Sanchez et al. (2011; Fig.~27). However, our model does not reproduce the extreme
terminal wind speed of $\sim 1000$~km\,s$^{-1}$ observed in NGC~1068. Since NGC~1068 has a high bolometric luminosity and an Eddington
ratio of $\sim 0.2$-$0.7$, we suggest that radiation pressure, which is not included in our model, might play an important role for the acceleration of
the gas and dust in the wind of NGC~1068.

\section{Axisymmetric 3D models \label{sec:3dmodels}}

\subsection{Density distribution \label{sec:ddist}}

With the analytical model described in Sect.~\ref{sec:model} we can construct a model of the 3D gas
distribution within a central mass distribution around a central black hole.
This 3D model has three ingredients:
\begin{itemize}
\item
a thick gas disk for $r > r_{\rm wind}$,
\item
a thin gas disk for $r < r_{\rm wind}$, and
\item
a magnetocentrifugal wind starting at $r=r_{\rm wind}$.
\end{itemize}
The transition between the thin and the thick disk is assumed to be sharp, i.e. there is an inner
vertical wall which is directly illuminated by the central AGN.

The structure of the thick accretion disk is given in Sect.~\ref{sec:thickdisk}. 
The disk height is determined by the hydrostatic equilibrium
\begin{equation}
\rho v_{\rm turb}^2=\rho G M_{\rm dyn} \frac{H^2}{(r^2+H^2)^{1.5}}\ .
\end{equation}
The vertical density distribution is assumed to be Gaussian $\rho(z)=\rho_0 \exp (-(z/H)^2)$.
For simplicity, we assume a smooth disk instead of a clumpy disk. Given the area filling factor of
$\Phi_{\rm A} \sim 0.5$ derived from Eq.~\ref{eq:phiaturb} for the Circinus and NGC~1068 models,
this approximation is acceptable. In a subsequent work we plan to extend the model to include a clumpy gas distribution.

A key ingredient of the model is the transition region between the thick and the thin disk which creates a directly illuminated inner 
wall of the thick gas and dust disk. Whereas the abrupt drop of the disk height might be exaggerated, we nevertheless expect
a rapid decrease of the disk height caused by the onset of the magnetocentrifugal wind.

The inner disk is assumed to have a velocity dispersion of $v_{\rm turb}=10$~km\,s$^{-1}$.
Its density is given by
\begin{equation}
\rho=\frac{\Omega^2}{\pi G Q} \frac{v_{\rm turb}^{\rm thick}}{(10~{\rm km\,s}^{-1})}\ .
\end{equation}
For the radiative transfer models we added a puff-up to the thin disk. 
As observed in young stellar objects (e.g., Monnier et al. 2006), the inner rim of the thin disk is puffed up and is much hotter than the rest of 
the disk because it is directly exposed to the AGN flux (Dullemond et al. 2001, Natta et al. 2001). The puff-up is located directly behind the dust 
sublimation radius. Within our model, the main reason for its existence is the need for an increased
NIR emission of the dust distribution to reproduce available NIR observations, since the dust temperature is elevated at these small distances. 
The puff-up thus naturally provides the necessary increase of the NIR emission. We do not intend to elaborate a
detailed model for a puff-up, which is beyond the scope of this article.
The region of increased NIR emission is then obscured by the thick gas and dust disk. We are mainly interested in the latter effect.
The puff-up is located at a radius of $r=0.75\,\sqrt{L_{\rm bol}/(8 \times 10^{44}\ {\rm erg\ s}^{-1})}$~pc, has a maximum height of
$h=0.225\,\sqrt{L_{\rm bol}/(8 \times 10^{44}\ {\rm erg\ s}^{-1})}$~pc, and has a width of one sixth of its radius.  
For the ``best fit'' NGC~1068 model the puff-up is located at a radius of $r=0.50\,\sqrt{L_{\rm bol}/(8 \times 10^{44}\ {\rm erg\ s}^{-1})}$, has a maximum height of
$h=0.150\,\sqrt{L_{\rm bol}/(8 \times 10^{44}\ {\rm erg\ s}^{-1})}$~pc, and has a width of one sixth of its radius. The vertical extent increases the solid angle of this structure
and therefore leads to a higher fraction of absorbed and re-radiated AGN emission at small distances from the central source.
The geometry of the puff-ups was chosen ad-hoc to reproduce the IR spectral energy distributions. They might be created by magnetic or radiation pressure.
Alternatively, the inner maser disk might be warped and/or tilted, which would have the same effect on the IR SED (Fig.~9 of Jud et al. 2017).

The wind is assumed to have a density distribution $\rho \propto (r/\sqrt{r^2+z^2})$ or $\rho \propto (r/\sqrt{r^2+z^2})^{2}$.
At the footpoint the wind has $1/50$ of the density of the disk. This heuristically determined description led
to MIR luminosities and visibility amplitudes which are consistent with observations.
The wind is located between
\begin{equation}
|H| < (\frac{r_0/{\rm 1~pc}}{H/r_0+0.15})^2~{\rm pc} \ {\rm and}\ |H| > (\frac{r_0/{\rm 1~pc}}{H/r_0-0.05})^2~{\rm pc}\ .
\end{equation}
This distribution has been designed adhoc and leads to a hollow wind cone, which is consistent with that
of the analytical model and comparable to the narrow line region cones of Mrk~1066, NGC~4051, and NGC~4151 (Fischer et al. 2013).

All model cubes have the dimension $501 \times 501 \times 501$~pixels.
The pixel size $\Delta$ is adapted to the bolometric luminosity of the central source in the following way:
\begin{equation}
\Delta=0.04 \sqrt{L_{\rm bol}/(8 \times 10^{44}~{\rm erg\,s}^{-1})}~{\rm pc}\ .
\end{equation}
Thus a gas disk with a high luminosity AGN is more extended than a gas disk with a central source
of low luminosity. This ensures that the inner disk radius, the sublimation radius, is resolved in our model cubes.

\begin{figure}
  \centering
  \resizebox{\hsize}{!}{\includegraphics{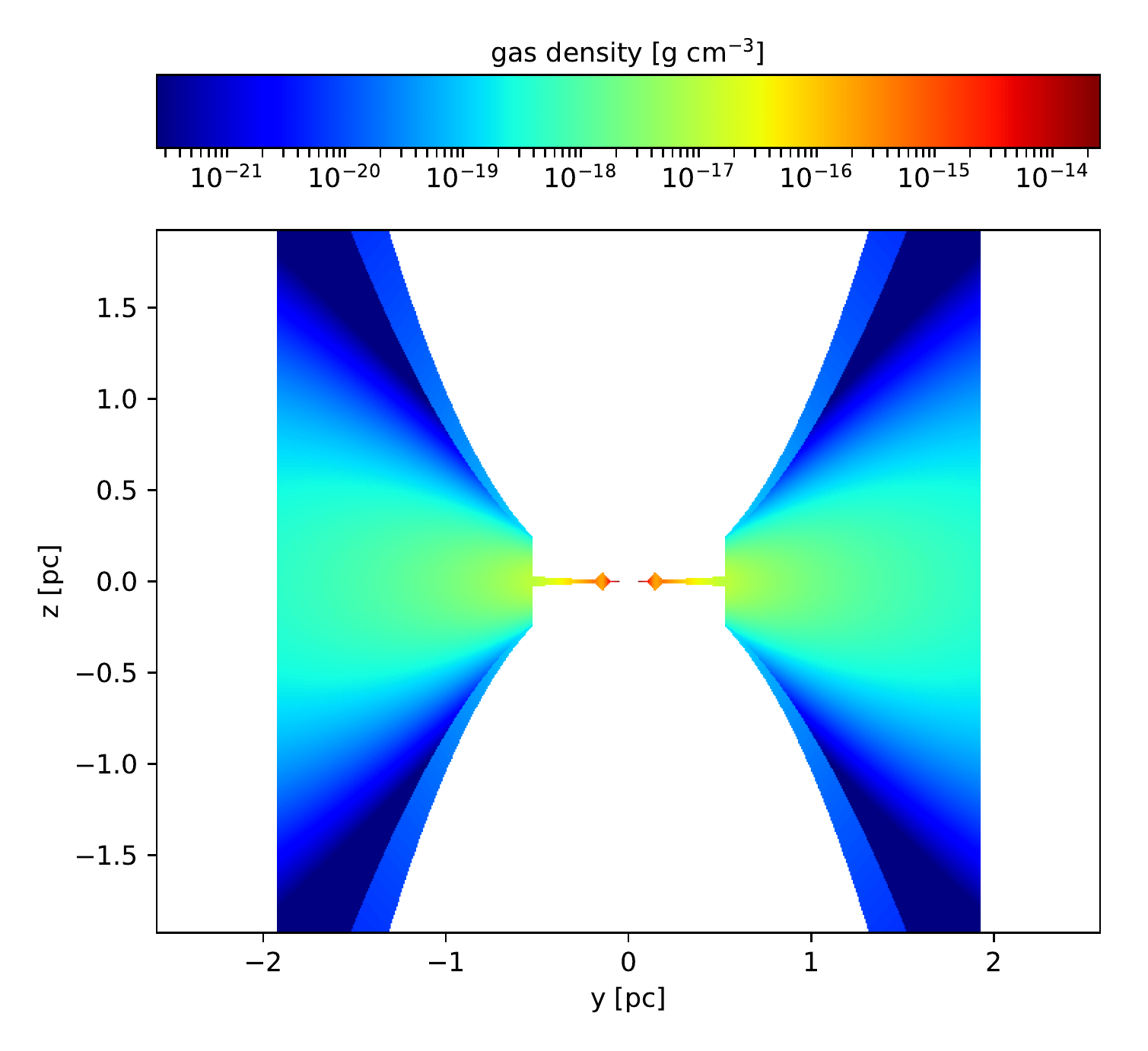}}
  \resizebox{\hsize}{!}{\includegraphics{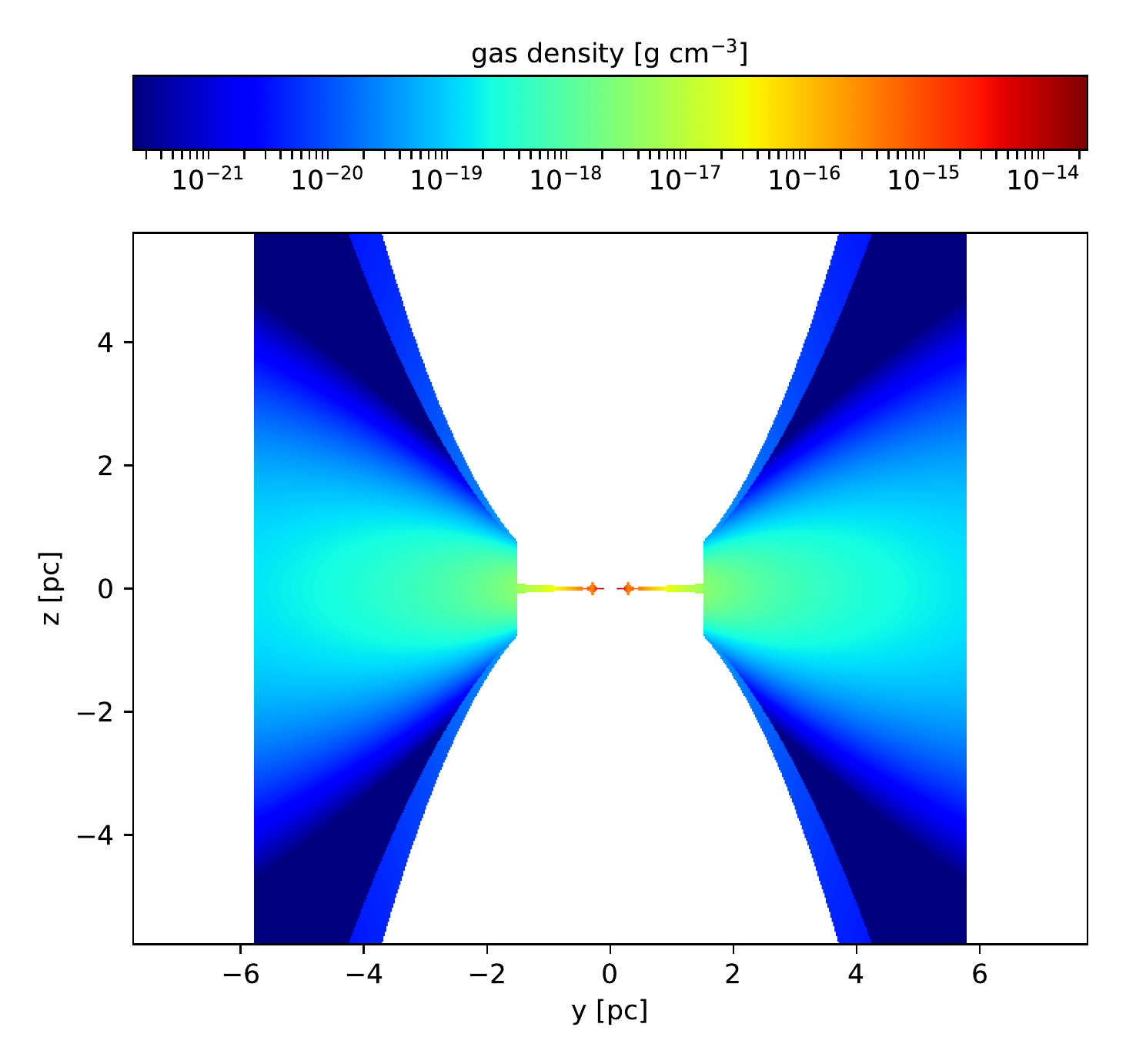}}
  \caption{Density cut through the model cube of the Circinus galaxy (upper panel) and NGC~1068 (lower panel). The scaling is logarithmic.
    The thick disk, thin disk, and the magnetocentrifugal wind are clearly visible. The thick gas ring near the
    inner edge of the thin disk was added ad hoc to enhance the NIR emission.
  \label{fig:modelcutcircinus}}
\end{figure}

Fig.~\ref{fig:modelcutcircinus} shows a cut through the density distribution of the Circinus model with 
$L_{\rm bol}=3 \times 10^{43}$~erg\,s$^{-1}$ and the NGC~1068 model with $L_{\rm bol}=3 \times 10^{44}$~erg\,s$^{-1}$.
The thick disk, thin disk, and the magnetocentrifugal wind are clearly visible.
The inner gap of the thin disk is due to the sublimation of dust (Barvainis 1987; Kishimoto et al. 2011).

We note that the models show sharp edges. Moreover, they are radially cut at $2$ and $6$~pc. These properties influence
the IR spectral energy distributions (H\"{o}nig \& Kishimoto 2010) and IR visibility amplitudes.

For comparison, we also set up a model of a thick gas disk without a wind (Fig.~\ref{fig:modelcuts_nowind}).

\subsection{Radiative transfer \label{sec:fullrt}}

From the dust density distribution discussed in Sect.~\ref{sec:ddist}, we calculate
spectral energy distributions as well as images in the near- and mid-infrared with the help of RADMC-3D (Dullemond 2012).
The latter is a modular and versatile three-dimensional radiative transfer code relying on the Monte Carlo method.
A constant gas-to-dust ratio of $150$ (e.g., Draine \& Lee 1984, Draine et al. 2007) was assumed.
The dust density model is binned onto a spherical, two-dimensional grid. It is illuminated by a central
energy source, which is point-like, isotropically emitting with a spectral energy distribution resembling the one
of quasars (see discussion in Schartmann et al. 2005), and normalized to the bolometric luminosities of NGC\,1068 and the Circinus galaxy.
The dust composition is according to a galactic dust model similar to the one employed in Schartmann et al. (2014). Five different grain sizes with
a size distribution as in Mathis et al. (1977) are used for each of the three different grain species: silicate and the two orientations
of graphite grains with optical properties adapted from Draine \& Lee (1984), Laor \& Draine (1993), Weingartner \& Draine (2001) 
and Draine (2003).
Following a thermal Monte Carlo simulation (Lucy 1999, Bjorkman \& Wood 2001), the resulting dust temperature distribution is used to simulate continuum
spectral energy distributions and images at near- and mid-infrared wavelengths.
As the models discussed in this work reach very high optical depths close to the midplane ($\tau_{\rm V} \sim 10^4$--$10^6$), 
we use the so-called {\it modified random walk} method (Fleck \& Canfield 1984, Robitaille 2010) 
to reduce computation times. In cells of very high optical depth photon packages might end up on a random walk with a very large number of absorption and 
re-emission or scattering events.
This is prevented by using the analytical solution to the diffusion equation within this cell. Min et al.~(2009) showed that this results in very good approximations 
of the radiation transfer in objects with optical depths as high as in our setup.

\section{Model results \label{sec:results}}

In this Section we use the central extinction to discriminate between type~1 and type~2 objects and
compare the model infrared luminosities, spectral energy distributions, and point source fractions to observations.
We illuminated the density distribution isotropically and with a $\cos(\theta)$ pattern. The inner puff-up is modelled by a Gaussian whose
maximum is placed at $2,3,4 \times r_{\rm sub}$, where the sublimation radius (Kishimoto et al. 2007, Eq.~10 of Burtscher et al. 2013) is given by
\begin{equation}
r_{\rm sub}=0.175 \times \sqrt{\frac{L_{\rm bol}}{8 \times 10^{44}\ {\rm erg\,s}^{-1}}}\ {\rm pc}\ .
\label{eq:rsub}
\end{equation} 
The FWHM of the Gaussian is half the radius of its maximum. The model with $4 \times r_{\rm sub}$ represents our basic model.

\subsection{The central extinction \label{sec:central}}

The extinction of the central pixel of the model image determines the optical classification between type~1 and type~2
objects. Schnorr-M\"uller et al. (2016) studied the broad-line region (BLR) of nine nearby Seyfert 1 galaxies.
They showed that type~1.5 objects have central extinctions $\leq 3$~mag, whereas type~1.8-1.9 objects have central extinctions 
between $4$ and $8$~mag. The extinction of type~2 objects thus exceeds $\sim 10$~mag (Burtscher et al. 2015 found $15$-$35$~mag).
Netzer (2015) argued that the ratio between type~1 (including types 1.8-1.9) and type~2
objects is about one (see also Mateos et al. 2017). This implies a ratio between the disk height and radius of $H/R \sim 0.6$ or
an angle between the disk height and the equatorial plane of $\sim 30^{\circ}$. Since type~1.8/1.9 objects are not necessarily obscured 
by the torus, but more likely by ``foreground'' (kpc-scale) dust lanes in the host galaxy (e.g., Prieto et al. 2014), this angle 
has to be regarded as an upper limit. We can check if our model results are consistent with these findings.

\begin{figure*}
  \centering
  \resizebox{\hsize}{!}{\includegraphics{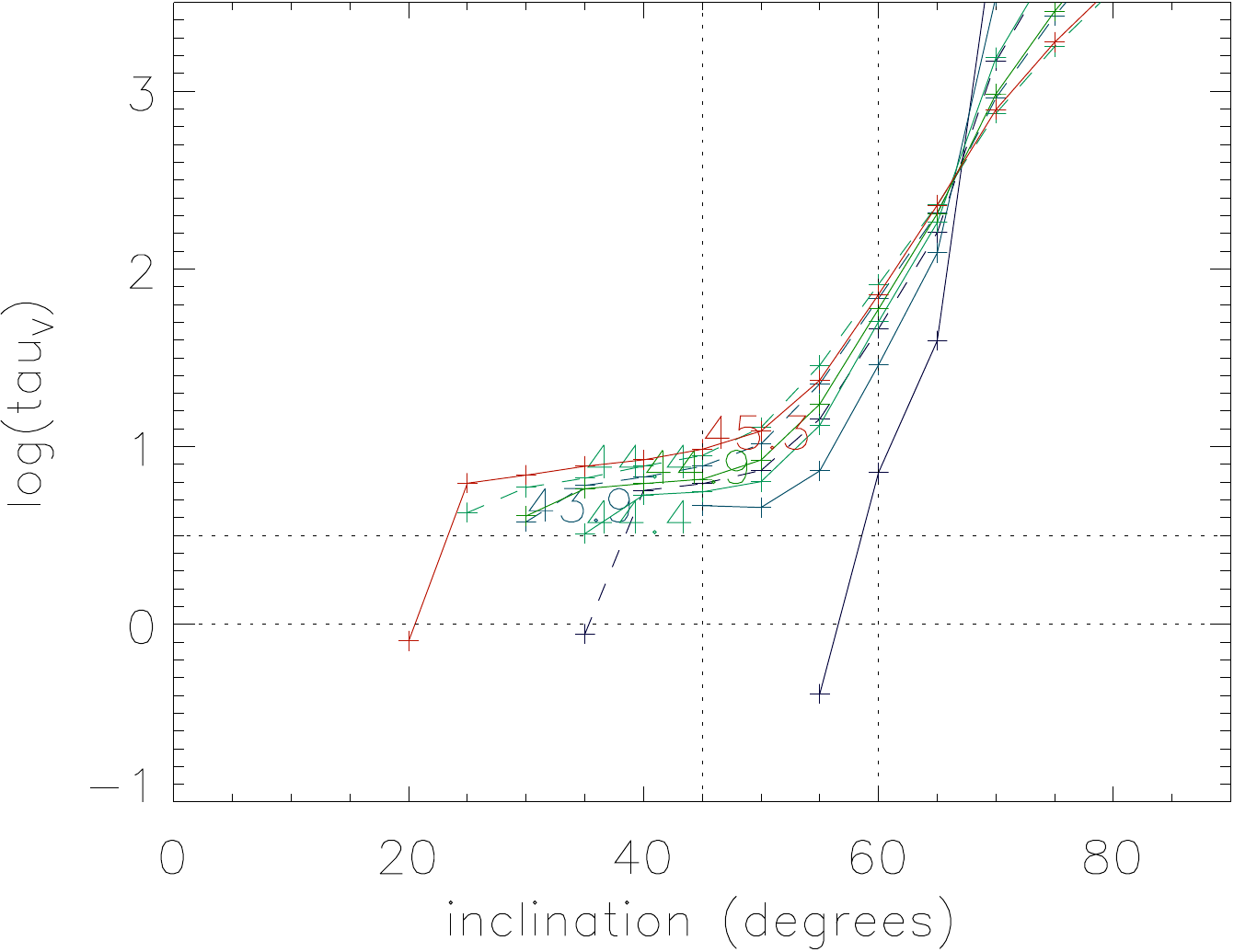}\includegraphics{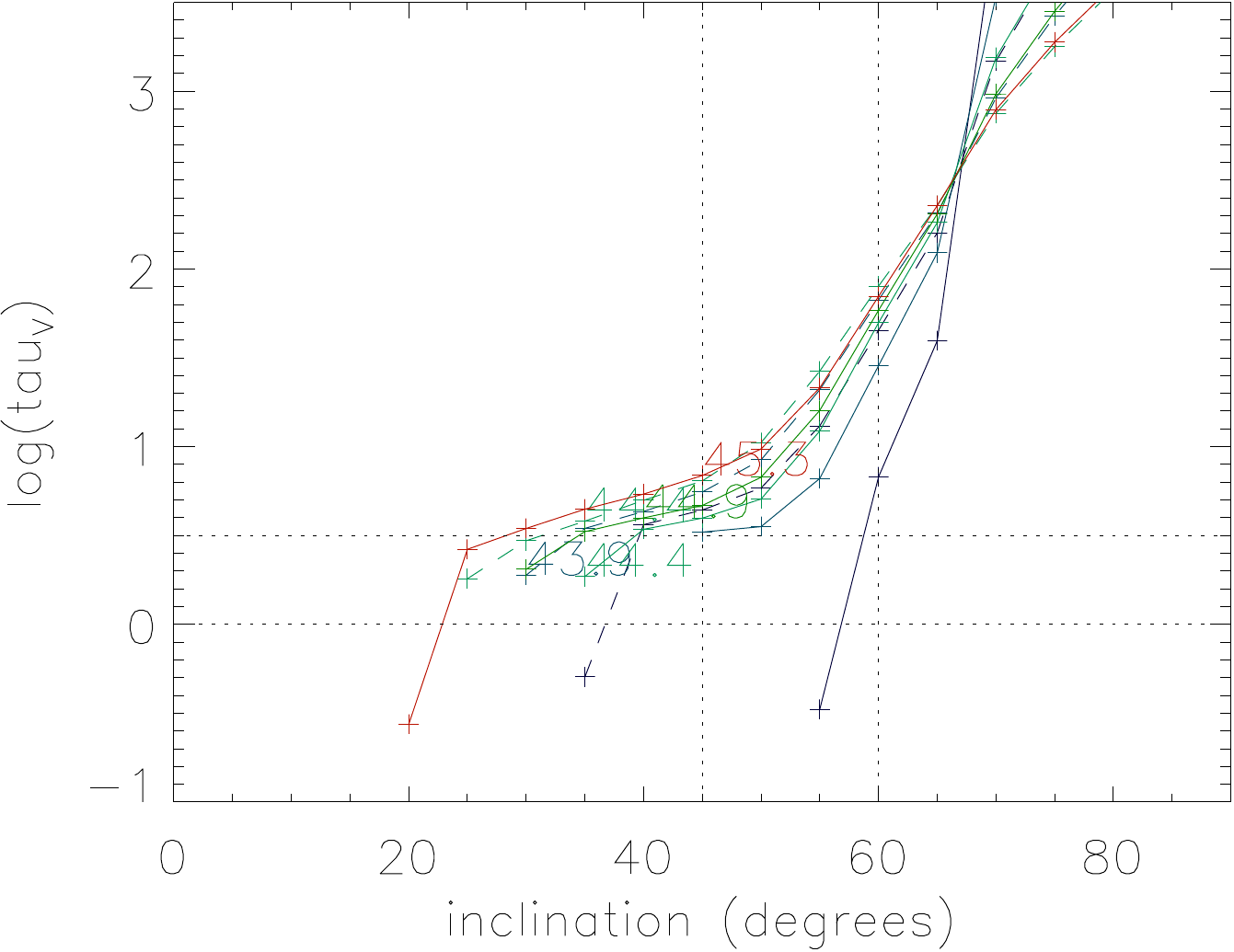}\includegraphics{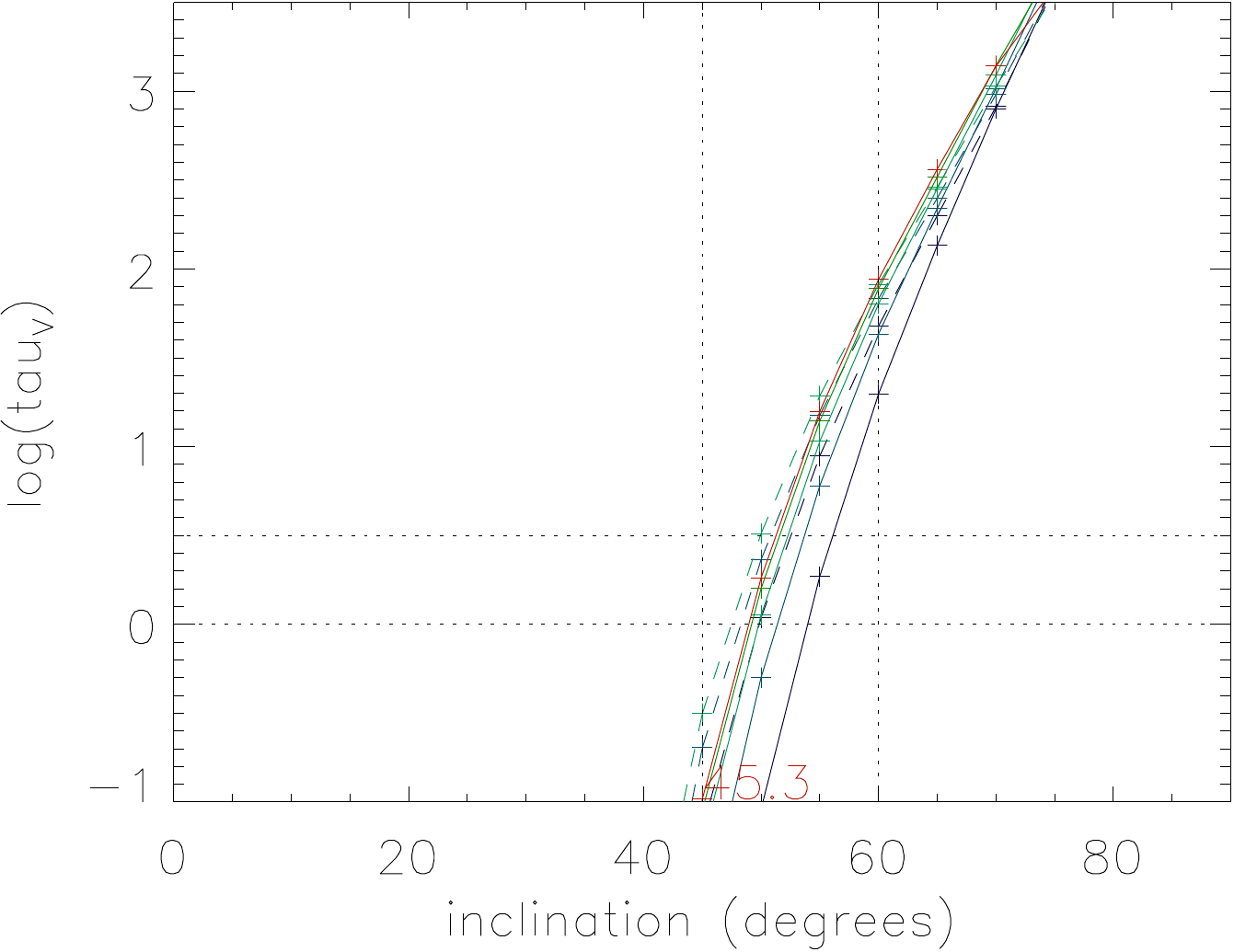}}
  \put(-470,115){\large $1/z$ wind}
  \put(-300,115){\large $1/z^2$ wind}
  \put(-130,115){\large no wind}
  \caption{Optical depth $\tau_{\rm V}$ as a function of the inclination angle of the gas disk. The colors correspond to
    different bolometric luminosities. Left panel: wind with $1/z$ density profile. Middle panel: wind with $1/z^2$ density profile.
    Right panel: no wind component. The numbers indicate the logarithm of the bolometric luminosities of the models in erg\,s$^{-1}$.
    The horizontal dotted lines correspond to $\tau_{\rm V}=1$ and $3$, i.e. the transition from type~1 to type~2 objects.
    The vertical dotted lines correspond to the range of inclination angles where the degree of polarized optical emission 
    changes from type~1 to type~2 objects (Marin 2014).
  \label{fig:tau_inc_SURVEY1bnew}}
\end{figure*}

The optical depth $\tau_{\rm V}$ as a function of
the inclination angle of the gas disk is presented in Fig.~\ref{fig:tau_inc_SURVEY1bnew}.
Without a wind component, the optical depth is unity for an inclination angle of $\sim 50^{\circ}$ for all models.
We find $\tau_{\rm V} \sim 50$ for $i=60^{\circ}$.
In the presence of a wind component the optical depth increases with respect to that of the model without wind
at $i=60^{\circ}$. Thus, the wind component provides the bulk of the central extinction at these inclinations.
We also observe a luminosity-dependence of the increase of the optical depth due to the wind, which can be
explained in the following way: 
since our model cubes always contain $501^3$ pixels and the pixel size varies with the square root of the 
bolometric luminosity, the extent of the cube also increases with $\sqrt{L_{\rm bol}}$. This leads to a longer
sightline through the wind, and thus a higher extinction. This extinction obviously depends on the 
geometry (opening angle and width) of the wind. 

The optical depth at $i < 40^{\circ}$ also depends on (i) the ratio between the wind density and the disk density at the footpoint of the wind,
(ii) the dust absorption coefficient,  and (iii) the gas-to-dust ratio of the wind. If, e.g., we decrease the ratio between the
wind and the disk density from $1/50$ to $1/150$, the central extinction decreases by a factor of $3$.
For the central extinction, we prefer these models, because they are consistent with a small fraction of type~1i objects (types $>1.5$)
among the type~1 objects.
These models show $1.6$ times smaller MIR luminosities and unchanged NIR luminosities. The IR emission distributions of the 
$1/z$ wind, $1/z^2$ wind, and a wind/disk ratio $1/150$ models are equivalent to the first order. In the following, we will use the
$1/z$ wind with a wind/disk ratio of $1/50$.

We conclude that our model is in broad agreement with existing observations.
The observed extinction of type~1.5 objects of $\leq 3$~mag and the higher extinction of type~1.8-1.9
objects (between $4$ and $8$~mag; Schnorr-M\"uller et al. 2016) might thus well be due to magnetocentrifugal winds.
We note that this is compatible with the finding of Stern \& Laor (2012) and Schnorr-M\"uller et al. (2016), that the structure obscuring the BLR
exists on scales smaller than the narrow line region. 

\subsection{Spectral energy distribution \label{sec:sed}}

\begin{figure}
  \centering
  \resizebox{\hsize}{!}{\includegraphics{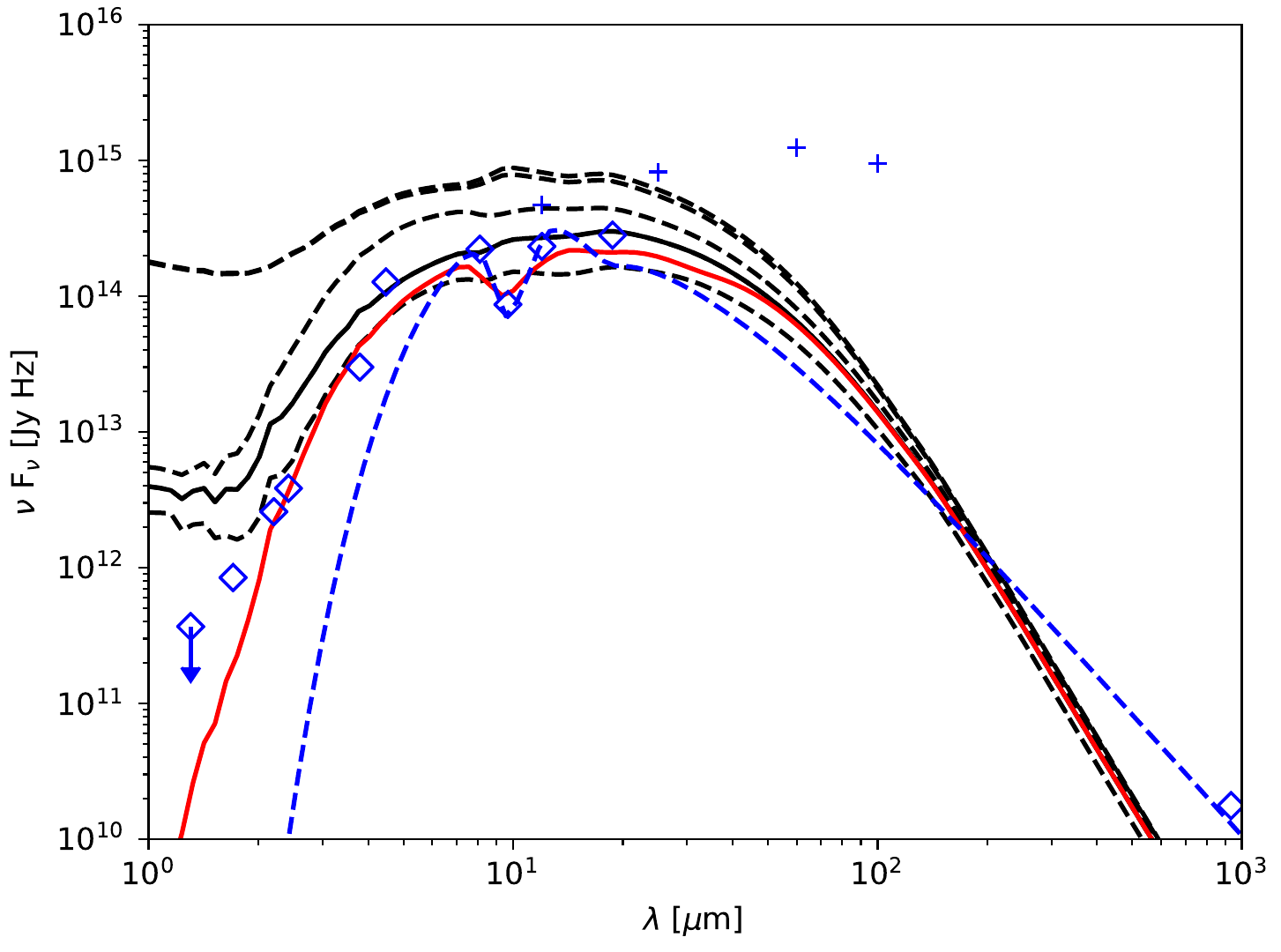}}
  \caption{Spectral energy distribution of the radiative transfer Circinus model for different
    inclination angles of the gas disk (solid red line: best-fit model with $i=65^{\circ}$ and with a homogeneous screen of $\tau_{\rm V}=20$; 
    solid black line: $i=65^{\circ}$ without a homogeneous screen;
    dashed black lines: $i=0,30,60,90^{\circ}$; blue dashed line: Tristram et al. (2014) model). Blue diamonds: observations from Prieto et al. (2010). 
    The typical uncertainties are of the order of 10\,\%.
    Blue pluses: low resolution IRAS photometry. These points can be regarded as upper limits. 
  \label{fig:circinus_comparison}}
\end{figure}

The spectral energy density distribution for the radiative transfer Circinus model 
(upper panel of Figs.~\ref{fig:modelcutcircinus}) is presented in 
Fig.~\ref{fig:circinus_comparison} for different inclination angles, together with the observations of Prieto et al. (2010).
In order to reproduce the observed silicate absorption at $12~\mu$m, we added a homogeneous screen of cold dust
and varied its optical depth.
It turned out that a screen with $\tau_{\rm V}=20$ reproduced the silicate feature for both AGN, Circinus and NGC~1068.
This optical depth is typical for galactic giant molecular clouds.
The observed MIR flux densities at wavelengths $\le 20$~$\mu$m are well reproduced by the model. The observed FIR flux densities were
obtained from large apertures and have thus to be considered as upper limits. 
The model NIR emission below $2~\mu$m is overestimated by an order of magnitude. It is not excluded
that this is due to a much lower intrinsic NIR luminosity. This would imply that the
central thick gas ring, as assumed by the present model (see Sect.~\ref{sec:3dmodels}), does not
exist in the Circinus galaxy. 
The observed silicate absorption feature at $12~\mu$m is not reproduced by the model without a cold dust screen, because directly
illuminated regions of the wind component, which are not extincted by foreground dust, dominate the emission at this wavelength.
The screen has to be located within the nuclear region ($<$ a few pc) since at slightly larger scales the polar region is well visible
in optical emission lines such as [O\ion{III}] (Fig.~5 of Wilson et al. 2000), which means that there cannot be much dust located in front of the outer parts of the polar wind. 
Alternatively, the screen might have a larger extent if it becomes clumpy/patchy at higher altitudes.

Wada et al. (2016) used three-dimensional radiation–hydrodynamic simulations to study the structure of a gas disk and an associated
outflow around a low-luminosity AGN. Their IR SED is well comparable to our results between $1$ and $10$~$\mu$m, but has higher 
flux densities between $10$ and $100$~$\mu$m. Motivated by high-quality VLT VISIR MIR imaging, 
Stalevski et al. (2017) proposed a phenomenological dust emission model for the AGN in the Circinus galaxy 
consisting of a compact dusty disc and a large-scale dusty cone shell, illuminated by a tilted accretion disc with an anisotropic emission pattern.
Our model geometry is closest to their hyperbolic geometry. For a realistic comparison with observations, Stalevski et al. (2017) needed
a foreground screen with an optical depth of $\tau_{\rm V}=34$, which is consistent with the optical depth of our model screen
($\tau_{\rm V}=20$). The resulting IR SED is well comparable to our results.

\begin{figure}
  \centering
  \resizebox{\hsize}{!}{\includegraphics{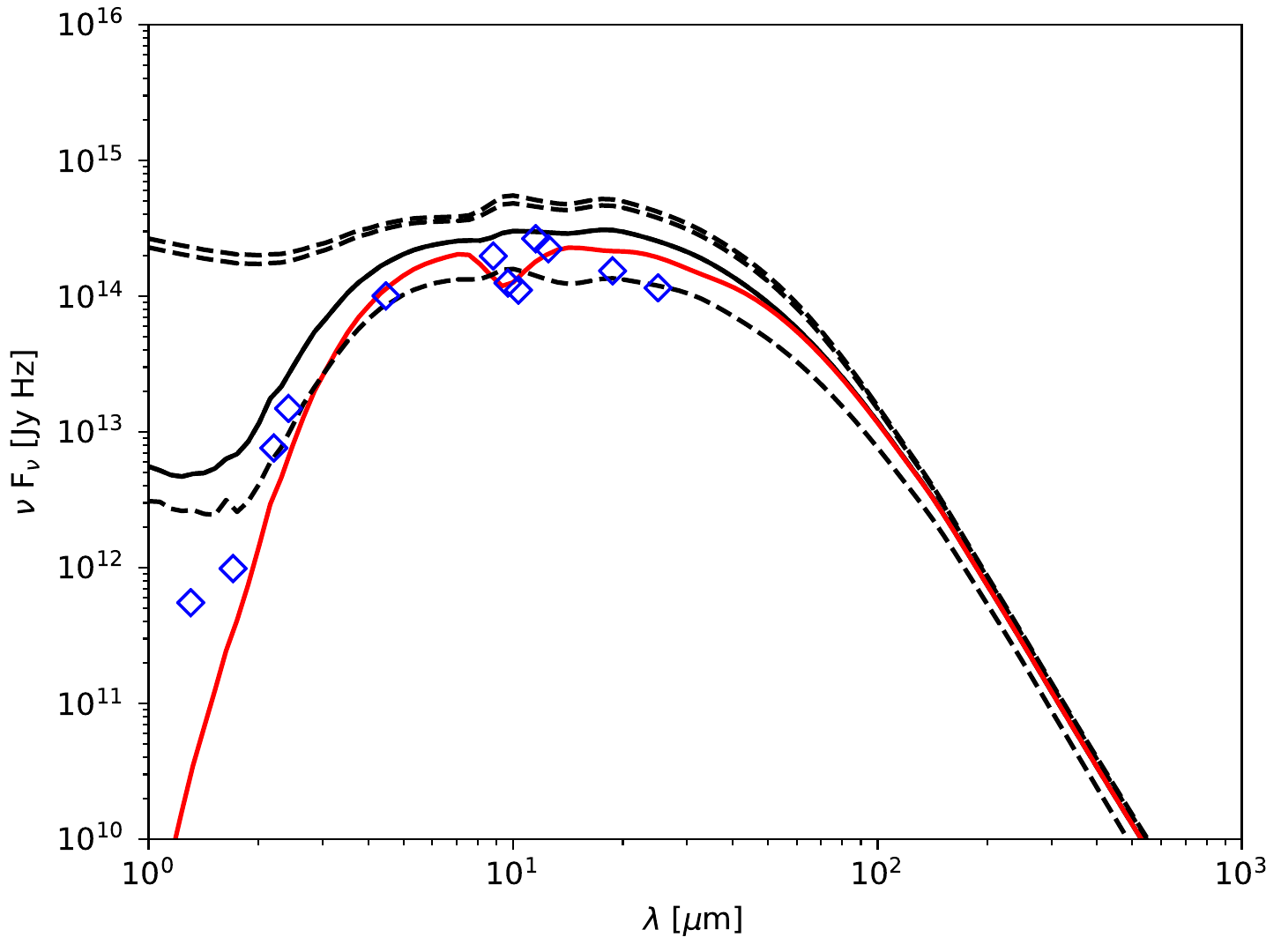}}
  \caption{Spectral energy distribution of the radiative transfer NGC~1068 model for different
    inclination angles of the gas disk (red solid line: best-fit model with $i=60^{\circ}$ and with a homogeneous screen of $\tau_{\rm V}=20$;
    black solid line: $i=60^{\circ}$ without a homogeneous screen; dashed black lines: $i=0,30,90^{\circ}$).
    Blue diamonds: observations from Prieto et al. (2010). The typical uncertainties are of the order of 10\,\%.
  \label{fig:ngc1068_comparison}}
\end{figure}

The spectral energy density distribution for the radiative transfer NGC~1068 model 
(lower panel of Figs.~\ref{fig:modelcutcircinus}) is presented in Fig.~\ref{fig:ngc1068_comparison} for different inclination angles.
The MIR flux densities between $5$~$\mu$m and $15$~$\mu$m are well reproduced by the model. 
The model $\sim 20$~$\mu$m flux densities are about $50$\,\% higher than the observed flux densities.
The NIR flux density of the model with an additional screen of $\tau_{\rm V}=20$ is more than a factor $10$ lower than the 
observed flux densities. This can be explained by additional stellar continuum emission.

We conclude that our model reproduces the observed IR spectral energy distribution in a satisfactory way.
The absence of the silicate absorption feature is due to the absence of sufficient dust absorption of the emission from
directly illuminated surfaces of the wind component. A cold dust screen with $\tau_{\rm V} \sim 20$ can provide
the necessary extinction to create the observed silicate absorption features.

\subsection{Infrared luminosities \label{sec:luminosities}}

Burtscher et al. (2015) combined two approaches to isolate the AGN luminosity at near-IR wavelengths and relate the near-IR pure AGN luminosity to
other tracers of the AGN. They showed that a significant offset exists between type~1 and type~2 sources in the sense that type~1
sources are about 10 times brighter in the NIR than in the MIR. 
We think that the models of the two AGN bracket the range of observed local AGN population in terms of black hole mass, rotation velocity, and bolometric luminosity. 
Therefore, we assume a fixed gas distribution for the low and high mass accretion case and
illuminate them with different bolometric luminosities. In the following we compare the resulting model MIR and NIR luminosities to observations.

\subsubsection{Isotropic illumination \label{sec:isotrop}}

\begin{figure*}
  \centering
  \resizebox{\hsize}{!}{\includegraphics{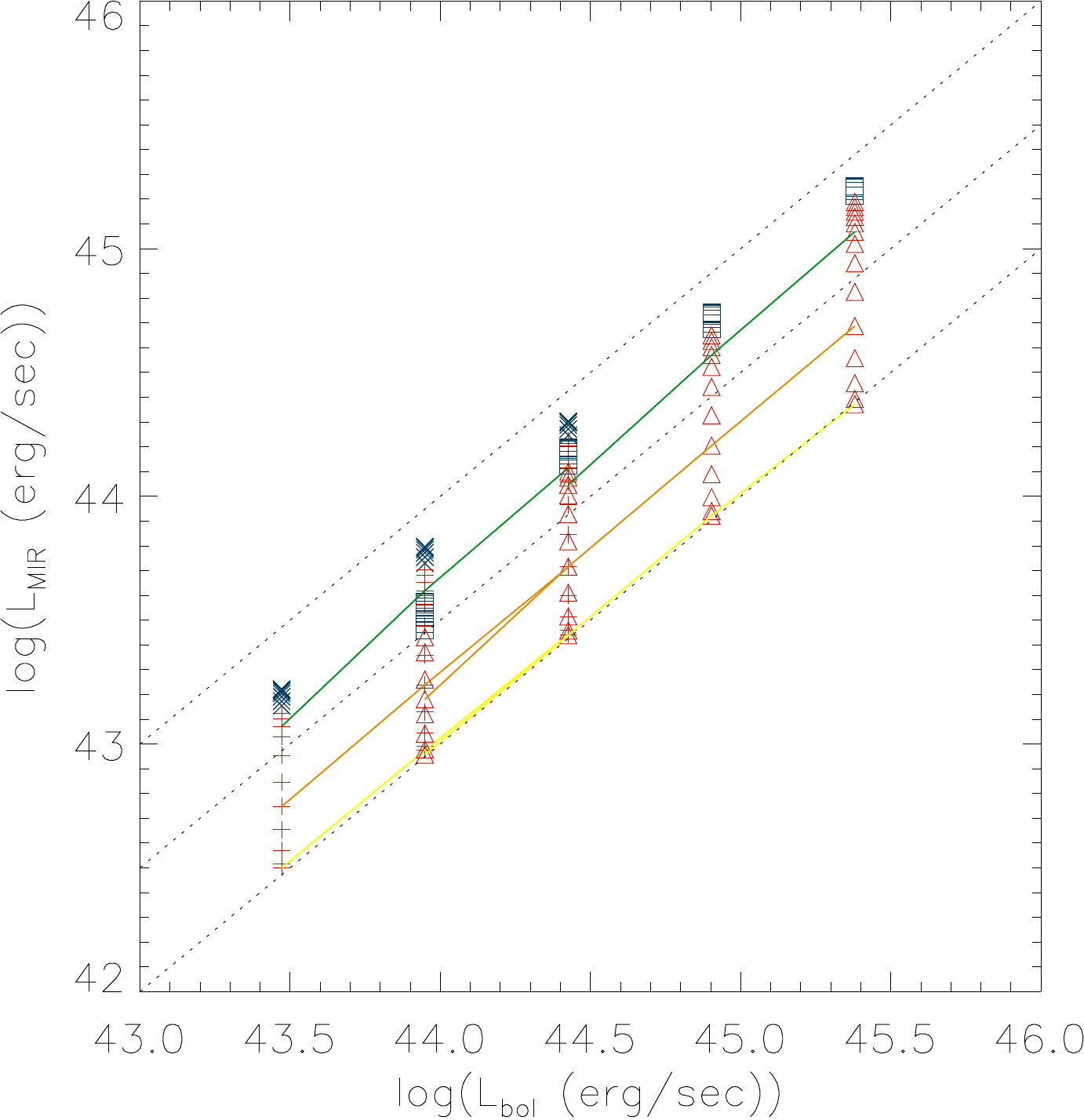}\includegraphics{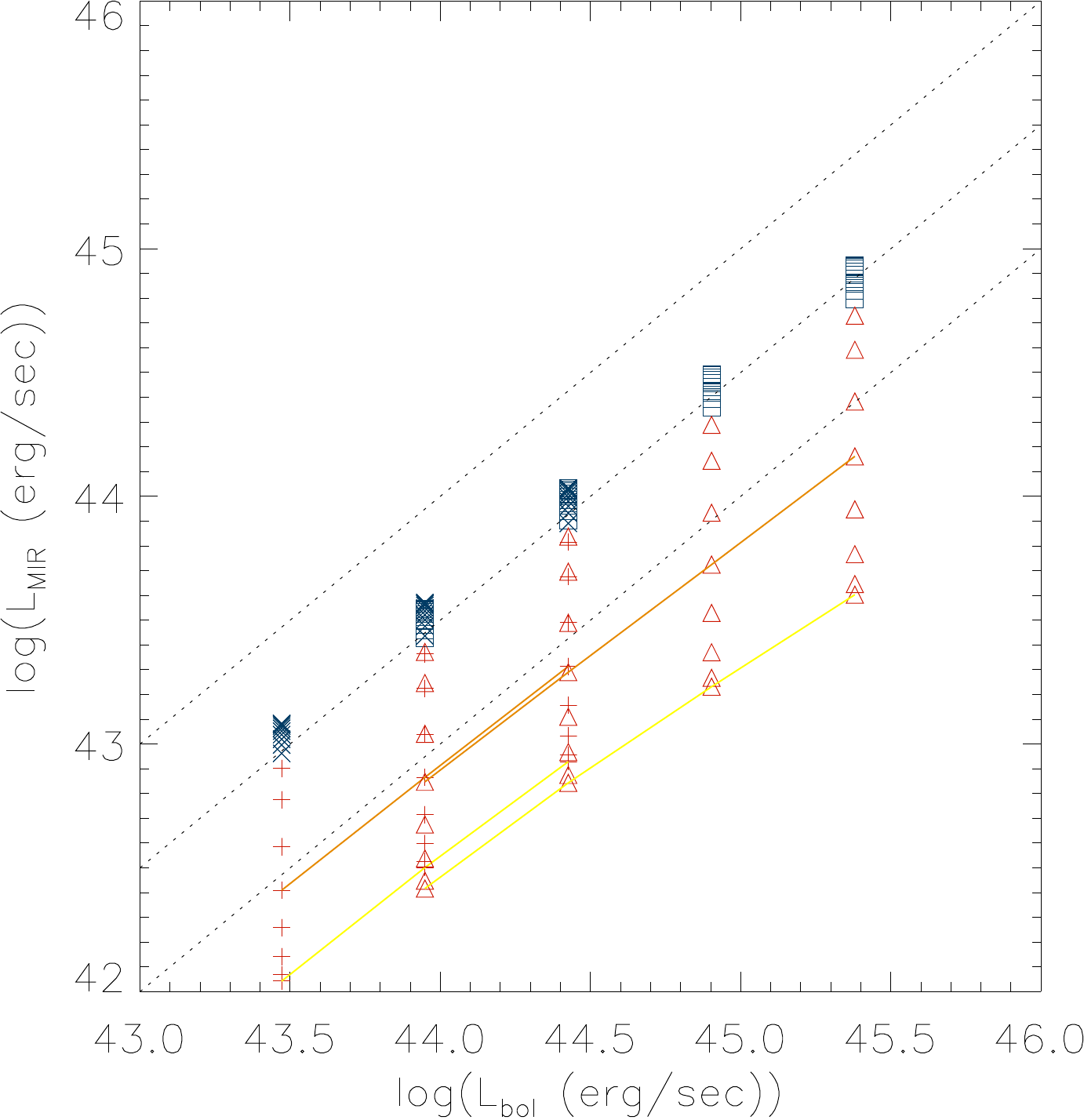}}
  \put(-450,240){\Large $1/z$ wind}
  \put(-200,240){\Large no wind}
  \caption{Model MIR luminosities as a function of the bolometric luminosity. Blue: type~1, red: type~2. Circinus model: 
    crosses/pluses; NGC~1068 model: squares/triangles. The inclination angles of type~2 objects are indicated: $i=50^{\circ}$ (green line),
    $i=70^{\circ}$ (orange line), $i=90^{\circ}$ (yellow line).
    Left panel: wind with $1/z$ density profile. Right panel: no wind component. The dotted lines correspond to $L_{\rm MIR}=\xi \, L_{\rm bol}$
    with $\xi=1,1/3,1/10$.
  \label{fig:plots_SURVEY1bnew_1}}
\end{figure*}

The 3D models described in Sect.~\ref{sec:3dmodels} were isotropically illuminated by a central AGN with different bolometric luminosities:
$(\frac{1}{9}, \frac{1}{3}, 1) \times 8 \times 10^{44}$~erg\,s$^{-1}$ for Circinus and 
$(\frac{1}{9}, \frac{1}{3}, 1, 3, 9) \times 8 \times 10^{44}$~erg\,s$^{-1}$ for NGC~1068.
For each bolometric luminosity we calculated the MIR and NIR luminosities with the radiative transfer model 
at $10~\mu$m and $2~\mu$m. 
To be consistent with Burtscher et al. (2015), we use monochromatic luminosities $L_{\rm IR}=\nu \times \int I_{\nu} {\rm d}A$, where $\nu$ corresponds to 
the central frequency of the band. To study the contribution of the magnetocentrifugal wind on the MIR and NIR luminosities, we calculated
three different kinds of models:
\begin{enumerate}
\item
wind with $1/z$ density profile,
\item
wind with $1/z^2$ density profile, and
\item
no wind.
\end{enumerate}
Since we observe only minor differences between the MIR/NIR emission of the models with a $1/z$ and a $1/z^2$ wind, we
show our results for the $1/z^2$ wind in Appendix~\ref{sec:mirnirz2}.
The model MIR luminosities as a function of the bolometric luminosity are presented in Fig.~\ref{fig:plots_SURVEY1bnew_1}.
Without a wind component, the model MIR luminosities are about a factor of $3$ smaller than the bolometric luminosities
for type~1 objects. Type~2 objects nevertheless show a smaller ratio between the MIR and the bolometric luminosity compared to type~1 objects
(right panel of Fig.~\ref{fig:plots_SURVEY1bnew_1}).
The situation changes with the addition of wind component:
in type~1 objects the ratio between the MIR and bolometric luminosity is about $\frac{2}{3}$, that of
type~2 objects about $\frac{1}{2}$ to $\frac{1}{10}$. Overall, the MIR luminosities of type~2 objects are about
a factor of $2$ lower than those of type~1 objects. The differences between the $1/z$ and $1/z^2$ models are minor
for the MIR luminosity ($\sim 0.2$~dex). The MIR luminosities are thus approximately proportional to the
bolometric luminosity, because the wind component extends to high latitudes where the absorption by the thick gas disk
is low even for an inclination angle of $60^{\circ}$. The comparison between the models with and without a wind component
shows that the MIR luminosities of the thick gas disk and the wind are comparable.
As expected, the exact location of the puff-up does not significantly modify the MIR luminosities (Fig.~\ref{fig:plots_SURVEY1bnew_1s}).
The model $L_{\rm MIR}/L_{\rm bol}$ ratios of $\leq 0.5$ are significantly higher than the observed ratios of $0.06-0.17$ (Gandhi et al. 2009, Asmus et al. 2015).
These ratios rely on the $L_{\rm X}/L_{\rm bol}$ relation found by Marconi et al. (2004).
On the other hand, based on high resolution IR observations of $3$ local type~1 AGN Prieto et al. (2010) found ratios of $L_{\rm MIR}/L_{\rm bol} = 0.12$-$0.28$.
Their bolometric luminosities are directly derived from the SEDs. Thus, our model $L_{\rm MIR}/L_{\rm bol}$ ratios are at least a factor of two higher than the ratio
derived from observations. It is expected that a slightly different geometry of
the inner wall of the thick gas disk (a convex instead of a plane surface) and a clumpy wind decrease $L_{\rm IR}/L_{\rm bol}$.
We note that an additional screen of optical depth $\tau_{\rm V}=20$ (Sect.~\ref{sec:sed}) leads to a decrease of the MIR luminosity by a factor of $2$.

\begin{figure*}
  \centering
  \resizebox{\hsize}{!}{\includegraphics{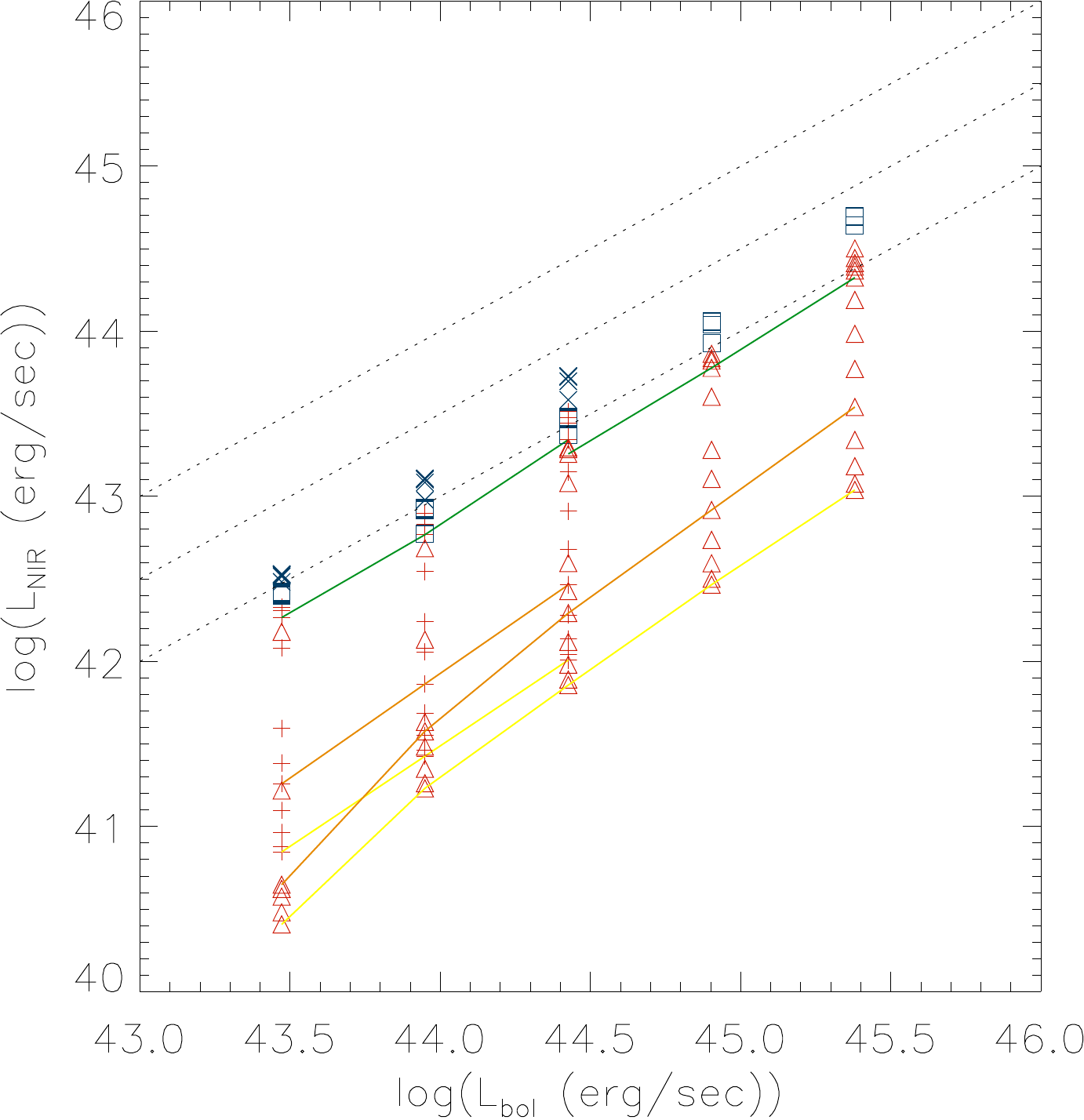}\includegraphics{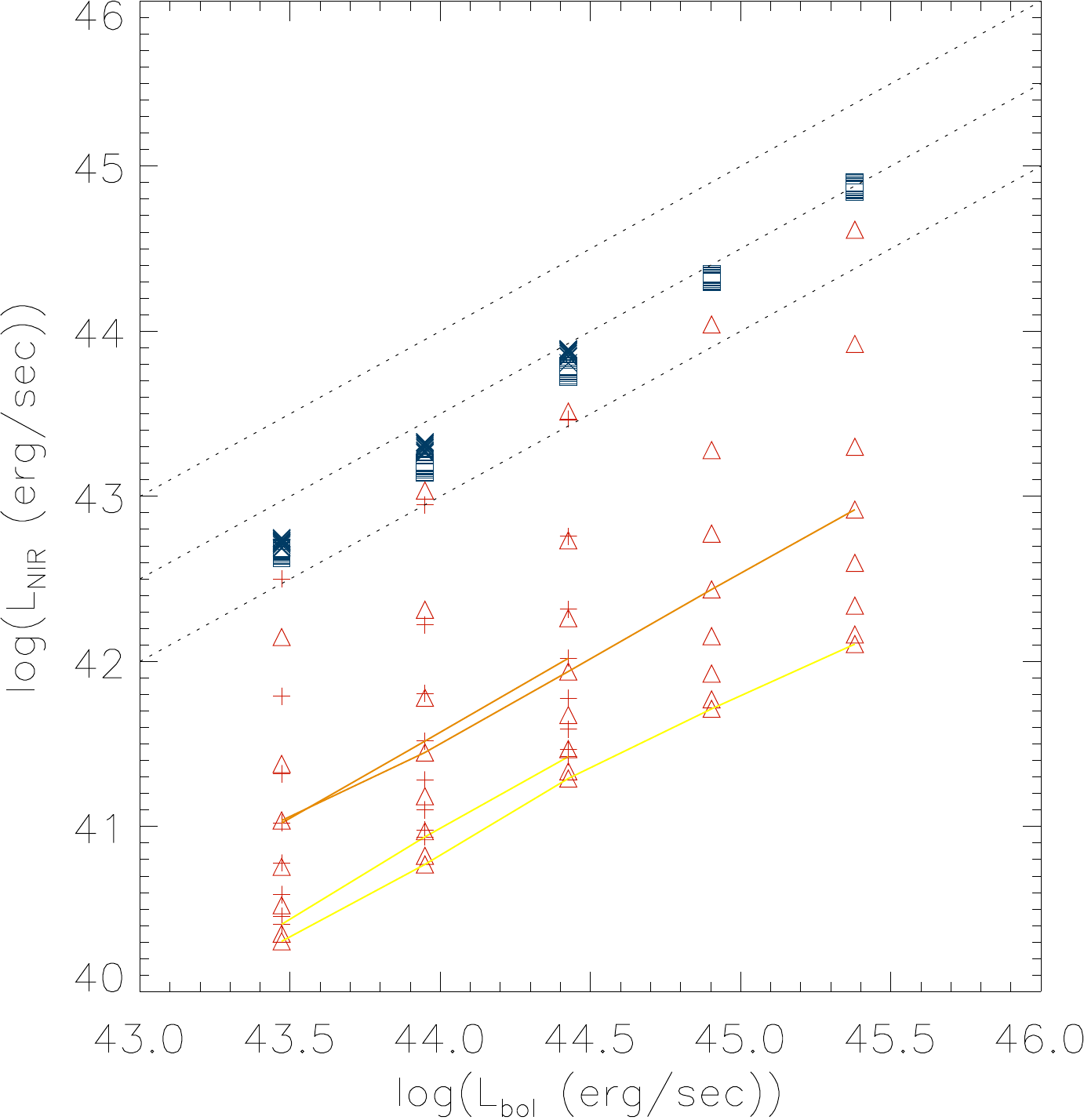}}
 \put(-450,240){\Large $1/z$ wind}
 \put(-200,240){\Large no wind}
  \caption{Model NIR luminosities as a function of the bolometric luminosity. Blue: type~1, red: type~2. Circinus model: 
    crosses/pluses; NGC~1068 model: squares/triangles. The inclination angles of type~2 objects are indicated: $i=50^{\circ}$ (green line),
    $i=70^{\circ}$ (orange line), $i=90^{\circ}$ (yellow line).
    Left panel: wind with $1/z$ density profile. Right panel: no wind component. The dotted lines correspond to $L_{\rm NIR}=\xi \, L_{\rm bol}$
    with $\xi=1,1/3,1/10$.
  \label{fig:plots_SURVEY1bnew_3}}
\end{figure*}

The model NIR luminosities as a function of the bolometric luminosity are presented in Fig.~\ref{fig:plots_SURVEY1bnew_3}.
The bulk of the NIR emission is produced close to the inner edge of the thin disk, i.e. the sublimation radius,
consistent with observations (continuum reverberation mapping: e.g. Suganuma et al. 2006; interferometry: e.g. Kishimoto et al. 2011).
Without a wind component, the model NIR luminosities are about $50$\,\% higher than those of the model with a wind component
for type~1 objects, because the wind provides a non-negligible NIR extinction. As expected, the NIR luminosities decrease significantly once the inner thin gas disk is
hidden by the thick gas disk. We observe a less significant drop of the NIR luminosities for high inclination angles ($i > 65^{\circ}$) in the presence of a wind.
We interpret the additional NIR emission as the contribution of the unobscured basis of the wind to the NIR emission.
Whereas the ratio between the NIR and bolometric luminosity is about $1/10$ for the basic wind model, it increases by a factor of $\sim 1.5$ and $\sim 2$
when the puff-up is located at a $\sqrt{2}$ and $2$ times smaller distance from the central black hole. The increase of the NIR luminosity due to higher 
dust temperatures is stronger than the decrease due to the smaller area where hot dust can be found.
\begin{figure*}
  \centering
  \resizebox{\hsize}{!}{\includegraphics{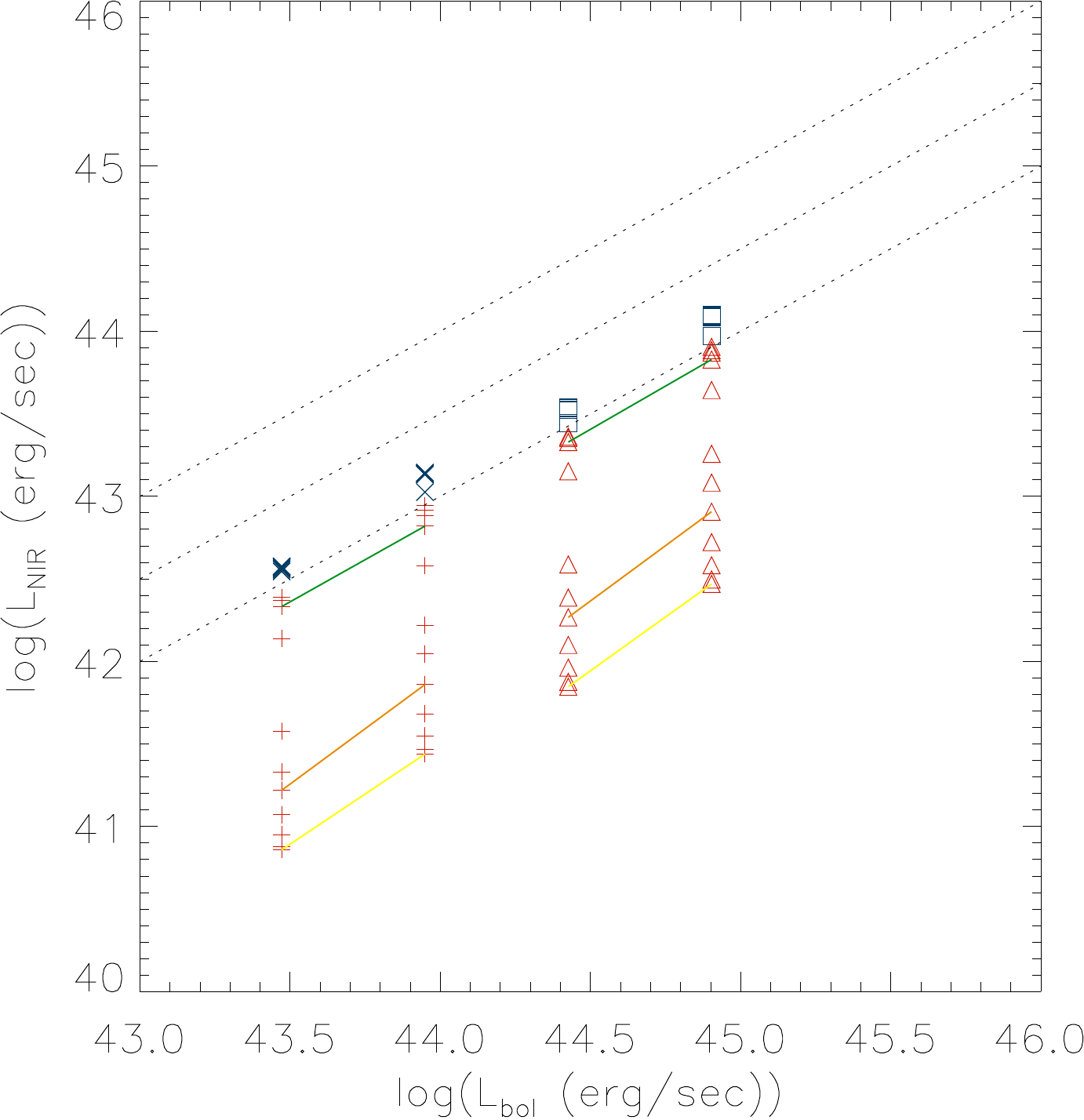}\includegraphics{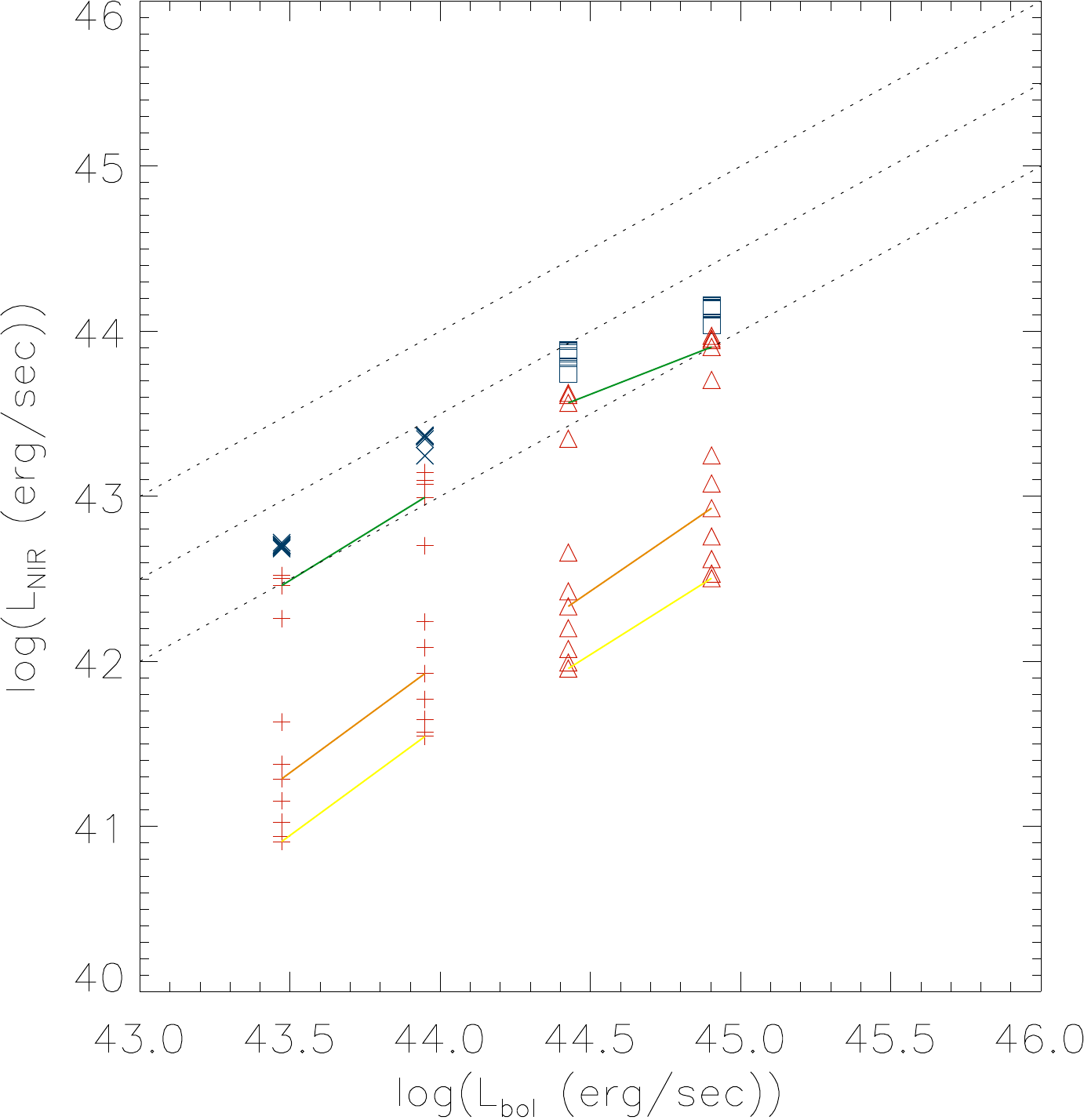}}
  \put(-450,240){\Large inner radius x 1/$\sqrt{2}$}
  \put(-200,240){\Large inner radius x 1/2}
  \caption{Model NIR luminosities as a function of the bolometric luminosity for the $1/z$ winds for a $\sqrt{2}$ (left panel) and $2$ (right panel) 
    times smaller inner radius of the thin gas disk. The dotted lines correspond to $L_{\rm NIR}=\xi \, L_{\rm bol}$
    with $\xi=1,1/3,1/10$. 
  \label{fig:plots_SURVEY1bnew_3s}}
\end{figure*}

\begin{figure*}
  \centering
  \resizebox{\hsize}{!}{\includegraphics{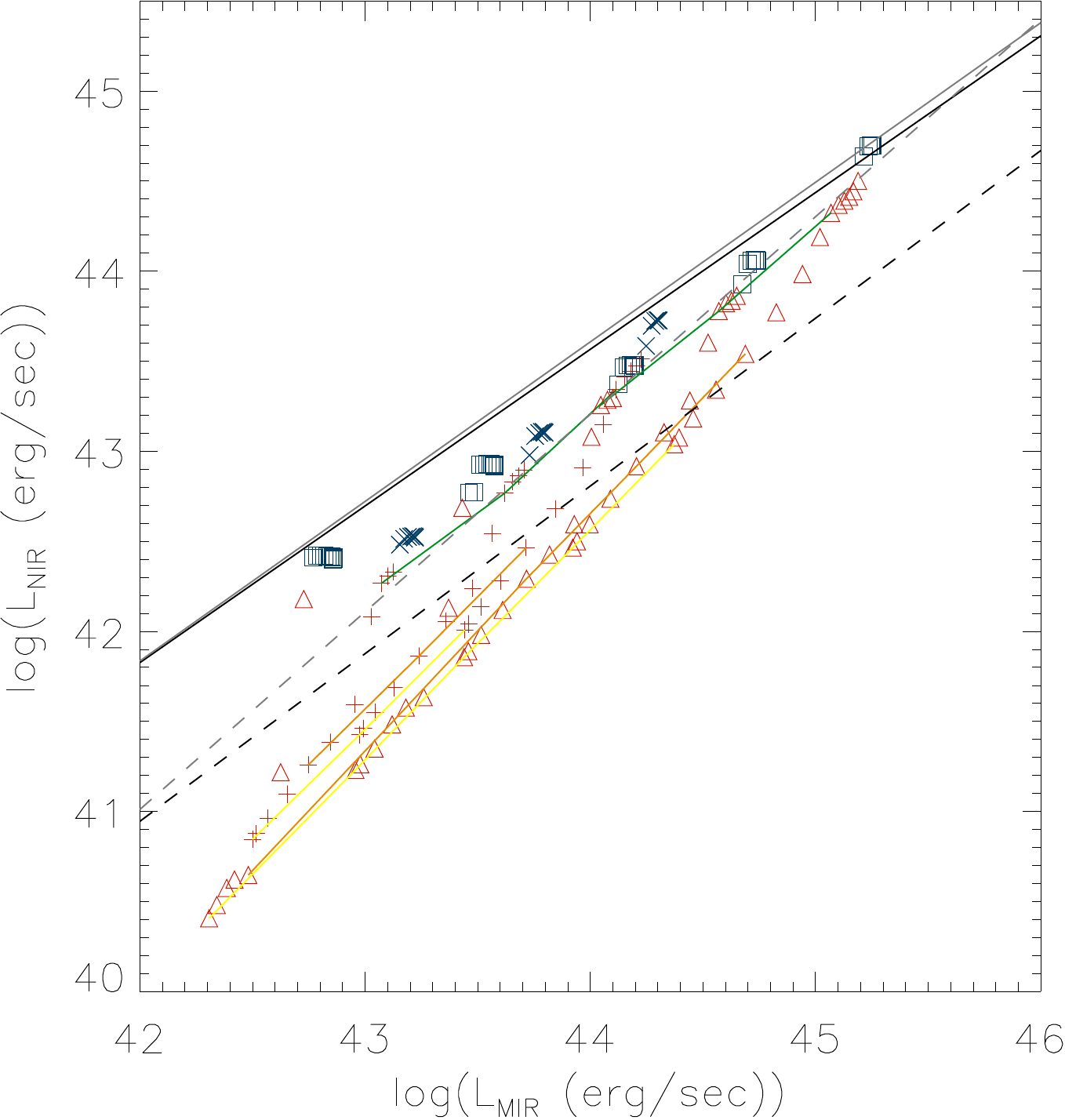}\includegraphics{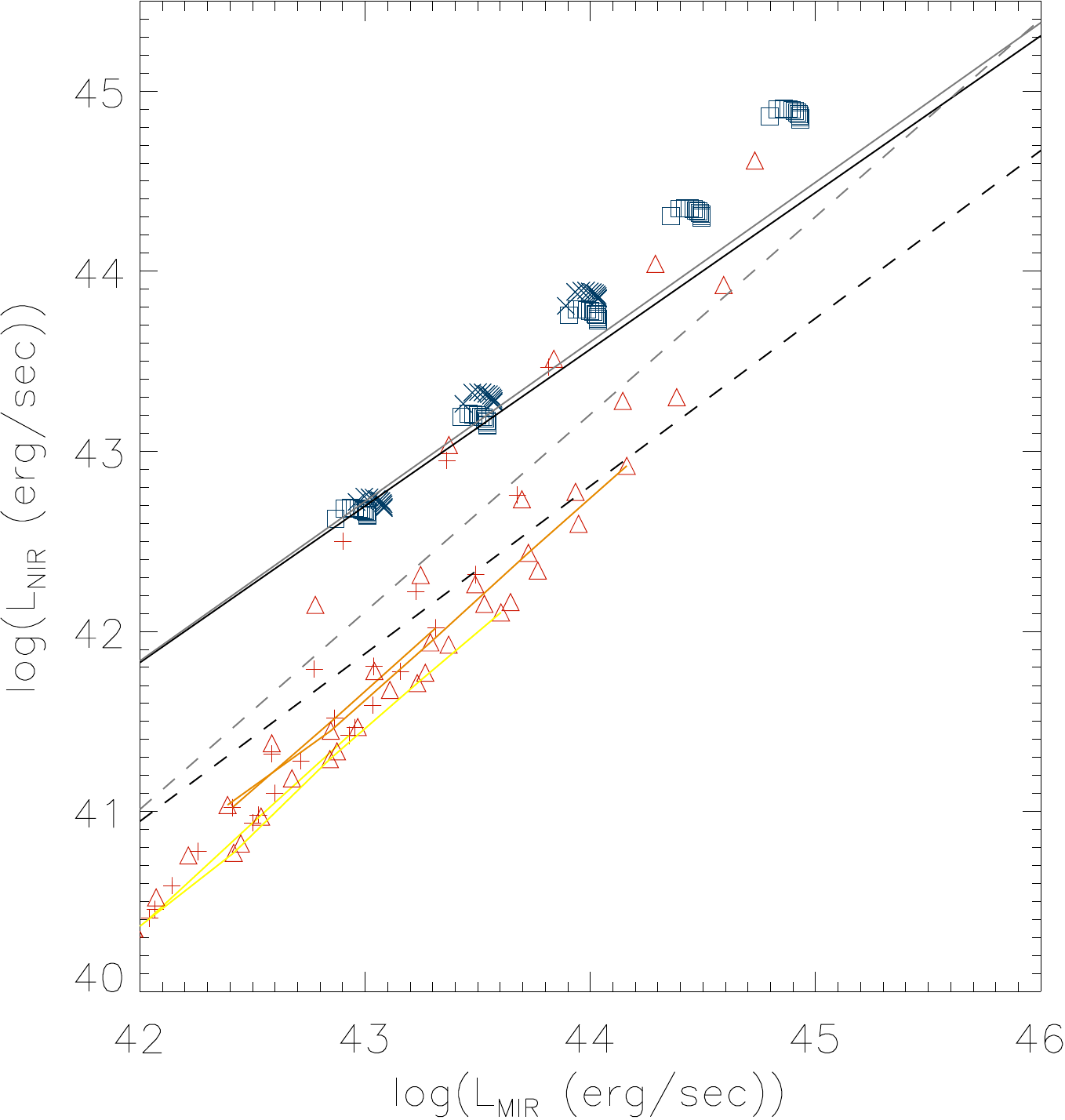}}
   \put(-450,245){\Large $1/z$ wind}
 \put(-200,245){\Large no wind}
  \caption{Model NIR luminosities as a function of the MIR luminosity. Blue: type~1, red: type~2. Circinus model: 
    crosses/pluses; NGC~1068 model: squares/triangles. The inclination angles of type~2 objects are indicated: $i=50^{\circ}$ (green line),
    $i=70^{\circ}$ (orange line), $i=90^{\circ}$ (yellow line).
    Left panel: wind with $1/z$ density profile. Right panel: no wind component. The solid and dashed lines are the relations found by Burtscher et al. (2015)
    for type~1 and type~2 objects, respectively.
  \label{fig:plots_SURVEY1bnew_5}}
\end{figure*}

\begin{figure*}
  \centering
  \resizebox{\hsize}{!}{\includegraphics{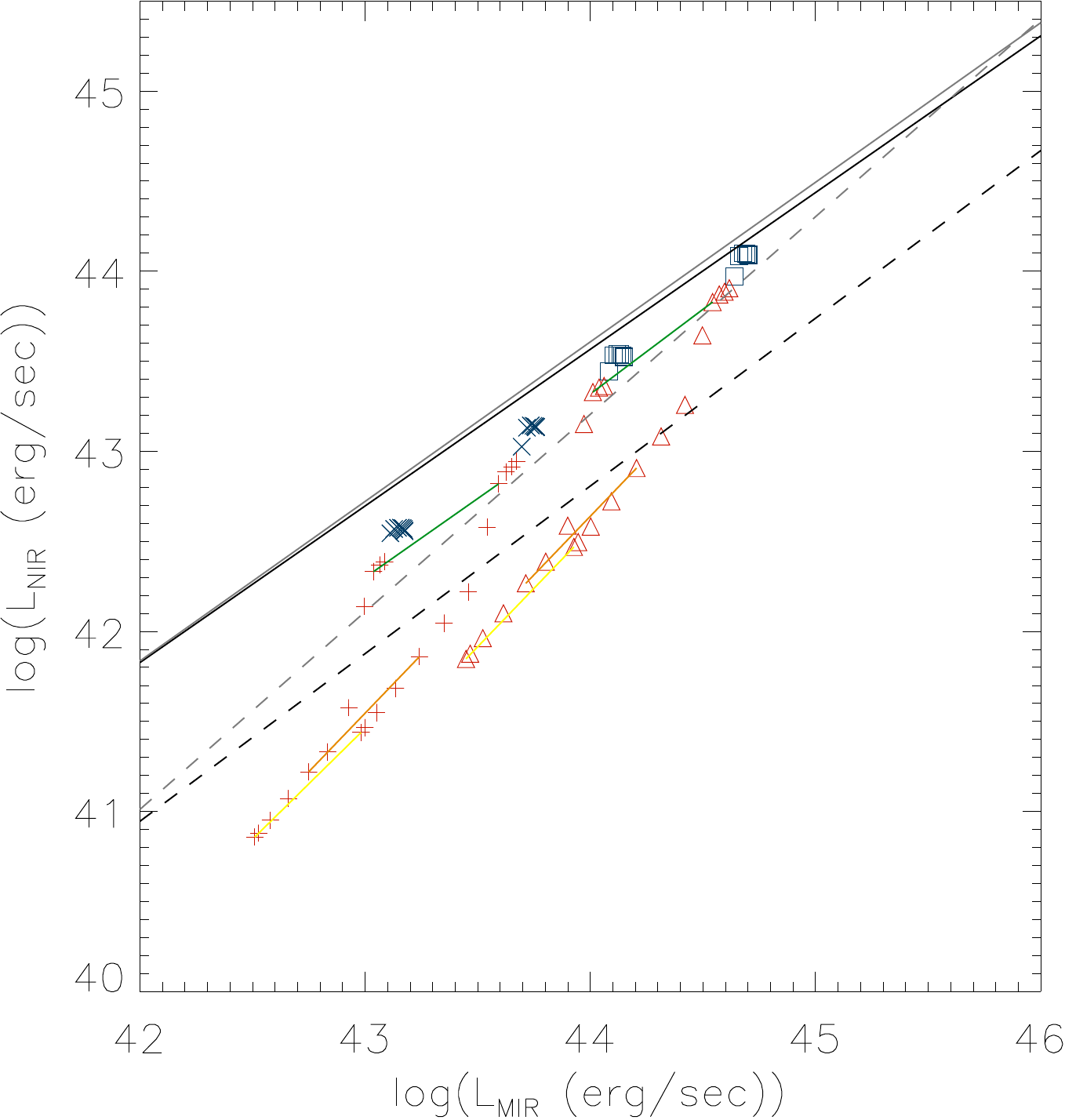}\includegraphics{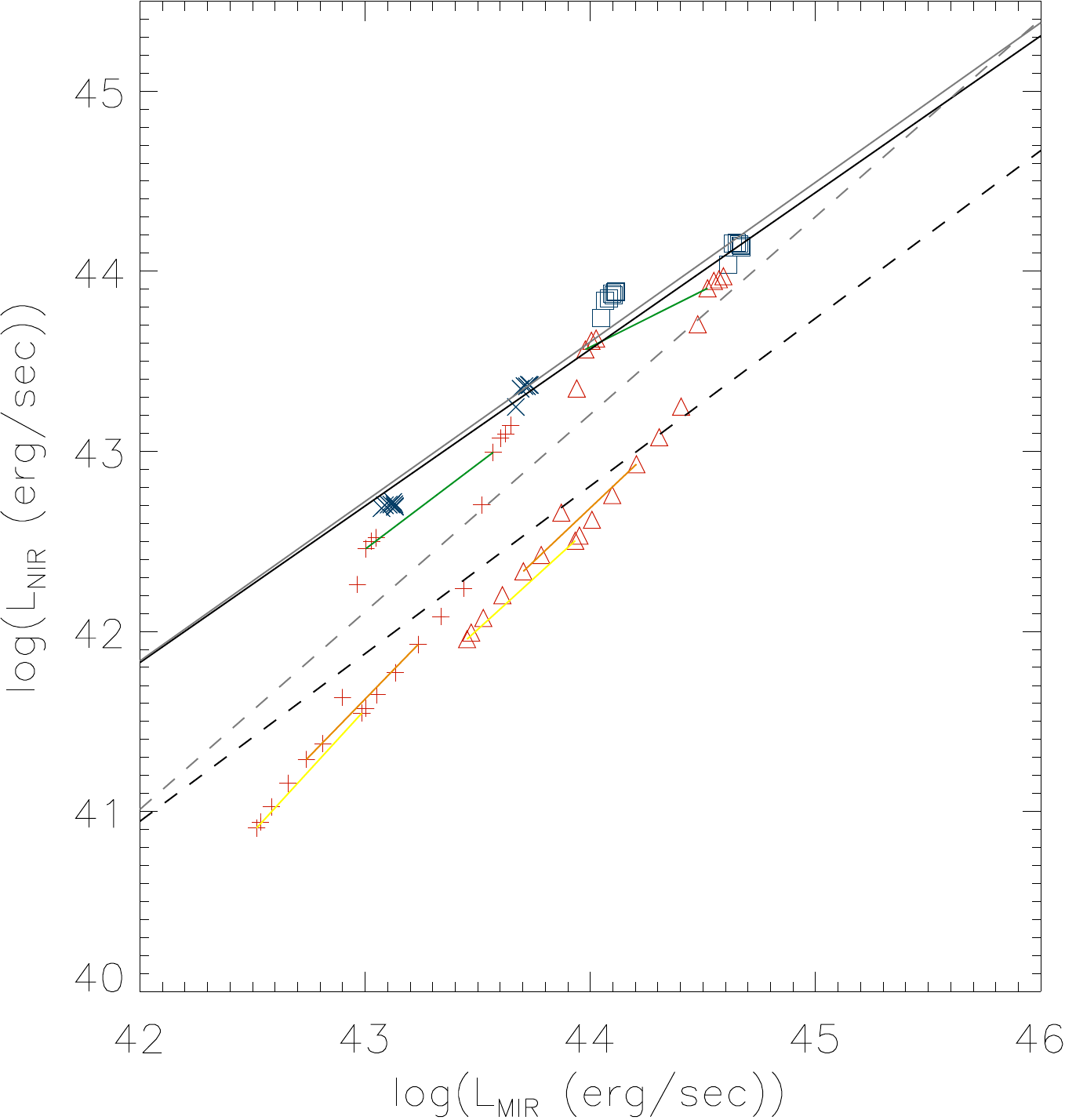}}
  \put(-450,240){\Large inner radius x 1/$\sqrt{2}$}
  \put(-200,240){\Large inner radius x 1/2}
  \caption{Model NIR luminosities as a function of the MIR luminosity for the $1/z$ winds for a $\sqrt{2}$ (left panel) and $2$ (right panel) 
    times smaller inner radius of the thin gas disk. The solid and dashed lines are the relations found by Burtscher et al. (2015)
    for type~1 and type~2 objects, respectively.
  \label{fig:plots_SURVEY1bnew_5s}}
\end{figure*}

The NIR luminosities as a function of the MIR luminosities are presented in Fig.~\ref{fig:plots_SURVEY1bnew_5}.
As for the previous correlations, the differences between the $1/z$ and $1/z^2$ models are minor.
Whereas type~1 objects show NIR/MIR ratios between $0.5$ and $1$, the NIR luminosities of type~2 models
are more than $5$ times smaller than the MIR luminosities. For luminosities smaller than $10^{44}$~erg\,s$^{-1}$ and inclination angles between $50^{\circ}$ and
$60^{\circ}$ the MIR/NIR luminosity ratio is about $10$.
The model with an inner radius located at $2 \times r_{\rm sub}$ (right panel of Fig.~\ref{fig:plots_SURVEY1bnew_5s}) reproduces the
observations of Burtscher et al. (2015) best.

\begin{figure*}
  \centering
  \resizebox{\hsize}{!}{\includegraphics{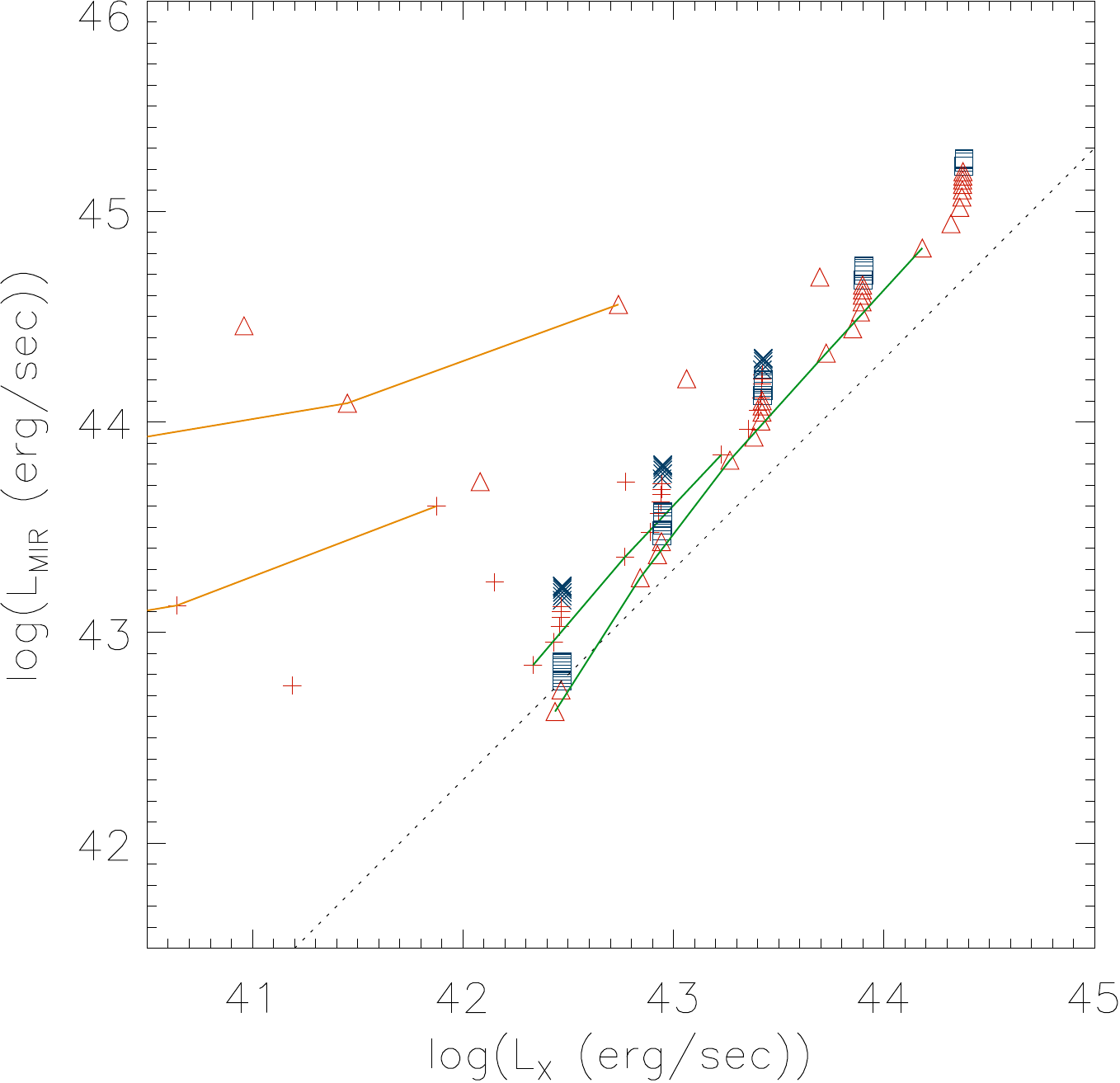}\includegraphics{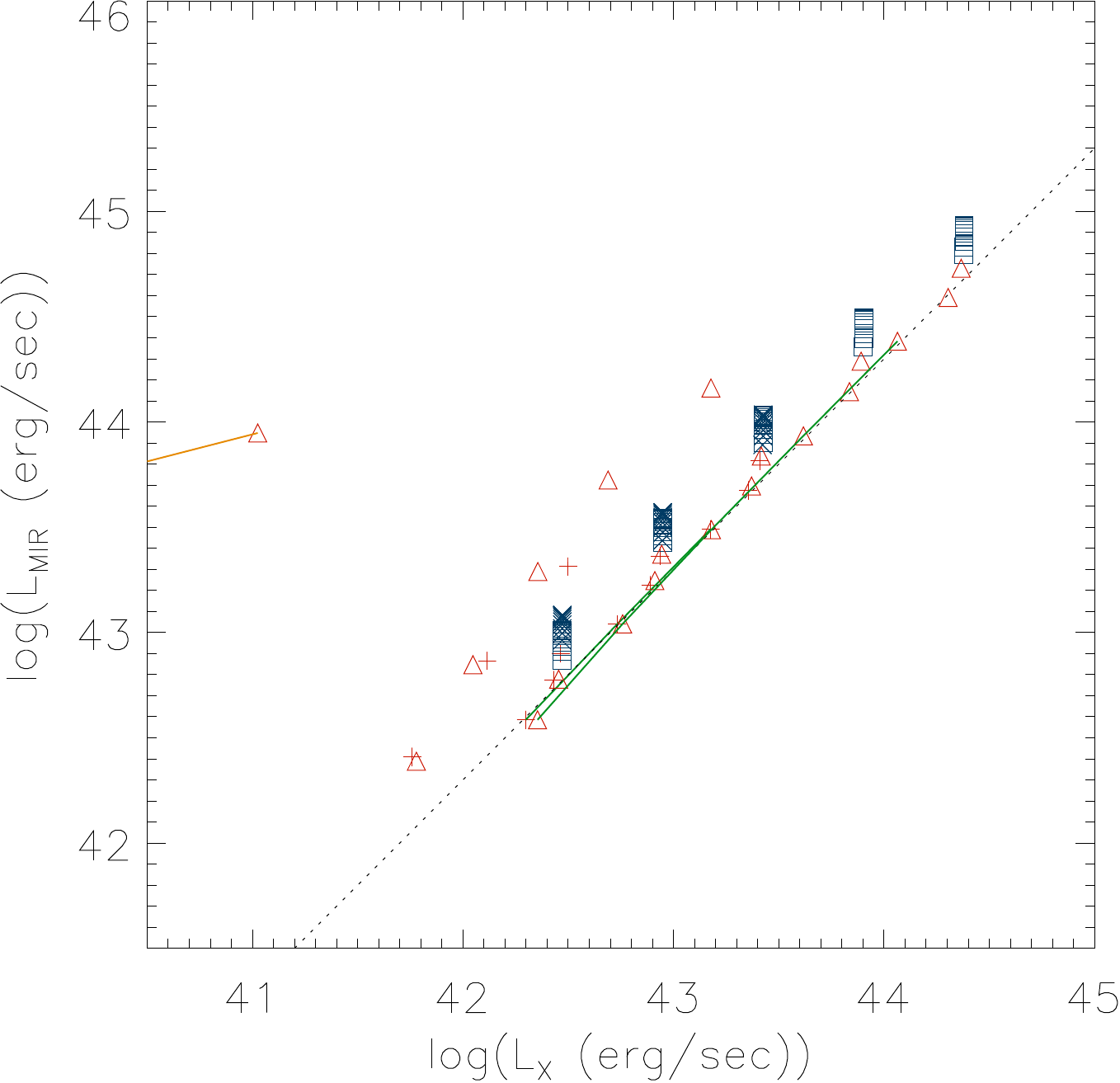}}
 \put(-450,220){\Large $1/z$ wind}
 \put(-200,220){\Large no wind}
  \caption{Model MIR--X-ray correlation.  Left panel: wind with $1/z$ density profile. Right panel: no wind component. Blue: type~1, red: type~2. Circinus model: 
    crosses/pluses; NGC~1068 model: squares/triangles. The inclination angles of type~2 objects are indicated: $i=65^{\circ}$ (green line),
    $i=75^{\circ}$ (orange line).
    The dotted line corresponds to the correlation found by Asmus et al. (2015): $\log(L_{\rm MIR})=\log(L_{\rm X}) + 0.33$.
  \label{fig:plots_SURVEY1bnew_2}}
\end{figure*}

As a last step, we compare our model results to the correlation between the MIR and the X-ray luminosities
(Asmus et al. 2015). These authors found that the MIR--X-ray correlation is nearly linear and within a factor of 2 independent of 
the AGN type and the wavebands used. The observed scatter of the correlation is $< 0.4$~dex.
We calculate the X-ray luminosity by assuming that the intrinsic X-ray luminosities of all models is $1/10$ of the bolometric luminosity (Marconi et al. 2004).
Assuming that the $1$-$10$~keV emission becomes optically thick at $N \sim 10^{24}$~cm$^{-2}$ (which corresponds to $\tau_{\rm V}=500$), 
the observed X-ray luminosity is then calculated via
\begin{equation}
L_{\rm X}=\frac{1}{10} \times L_{\rm bol} \exp{(-\frac{\tau_{\rm V}}{500})}\ .
\end{equation}
The results are presented in Fig.~\ref{fig:plots_SURVEY1bnew_2}.
Since the MIR luminosity integrated over a solid angle of $4 \pi$ is the reprocessed fraction of the bolometric accretion disk luminosity, it is proportional to the
covering factor. On the other hand, the observed MIR luminosity depends on the viewing angle (Fig.~\ref{fig:plots_SURVEY1bnew_1}) (a factor of $\ga 2$ for $i \ga 60^{\circ}$).
It is thus trivial that both wind models reproduce the observed MIR--X-ray correlation within a factor $2$. The scatter of the model correlation is 
determined by the viewing angle through the absorption of the MIR and X-ray emission. The scatter of the $L_{\rm MIR}/L_{\rm X}$ ratio is $0.3$~dex for the wind model
and $0.25$~dex for the model without a wind component. If we take into account that the probability of a galaxy that is observed with an inclination angle $i$ is 
proportional to the solid angle around that angle (in our case $i \pm 5^{\circ}$), we obtain a scatter of $0.36$~dex and $0.30$~dex, respectively.
With a variation/scatter of the $L_{\rm X}/L_{\rm bol}$ ratio of $0.3$~dex (Marconi et al. 2004), we obtain a total scatter of $0.42/0.39$~dex or $0.47/0.42$~dex
for the models with and without a wind, respectively. 
Surprisingly, all these values are comparable to that of the observations ($0.39$~dex; Asmus et al. 2015). The reason for
the tightness of the correlation even without a wind component is found in the high column densities of the absorbing disk material that also
decreases the X-ray emission together with the MIR emission.
Therefore, the wind component is not mandatory to reproduce the scatter of the $L_{\rm MIR}$--$L_{\rm X}$ correlation.

Only objects with $i \ge 70^{\circ}$ deviate significantly (more than $0.3$~dex) from the MIR--X-ray correlation,
i.e. they show much smaller X-ray luminosities due to X-ray absorption by the thick gas disk.
In terms of solid angle, this means that less than $6$\,\% of all objects deviate significantly from the correlation.
The same is found for the model without a wind component.
By comparing the models with and without wind, it becomes clear that the wind enhances the MIR emission of type~2 objects.
As expected, a smaller distance of the puff-up from the central black hole does not significantly modify the model MIR--X-ray correlation.

The tightness of the MIR--X-ray correlation can be of different origins:
\begin{itemize}
\item
Extended polar MIR emission caused by a dusty wind (see, e.g., Asmus et al. 2016) making the total MIR emission more isotropic;
\item
massive and dense thick gas disk as proposed here. In this case an extended polar MIR emission is not mandatory;
\item
clumpy models of the gas and dust distribution with and without a wind component are also able
to reproduce the observed MIR--X-ray correlation (H\"{o}nig et al. 2011, H\"{o}nig \& Kishimoto 2017). Clumpiness naturally increases the MIR isotropy;
\item
the X-ray emission might be mildly anisotropic (Liu et al. 2014; Sazonov et al. 2015; Yang et al. 2015).
\item
in the presence of a distribution of covering factors, type~1/2 sources will have a lower/higher covering factor, because the probability to
observe a certain type depends on the covering factor (Elitzur 2012). This naturally reduces the difference in IR emission between type~1 and type~2 sources; 
\end{itemize}

We conclude that the density profile of the wind ($1/z$ or $1/z^2$) has a minor influence on the NIR and MIR luminosities.
The existence of a wind component leads to MIR luminosities whose dependence on the inclination angle is relatively small.
The NIR component stems mostly from the thin gas disk and is thus prone to extinction by the thick gas disk.
The relation between the NIR and MIR luminosities (Fig.~\ref{fig:plots_SURVEY1bnew_5s}) is well comparable to
the observed relation (Burtscher et al. 2015; Fig.~9).

\subsubsection{Anisotropic illumination}

\begin{figure*}
  \centering
  \resizebox{\hsize}{!}{\includegraphics{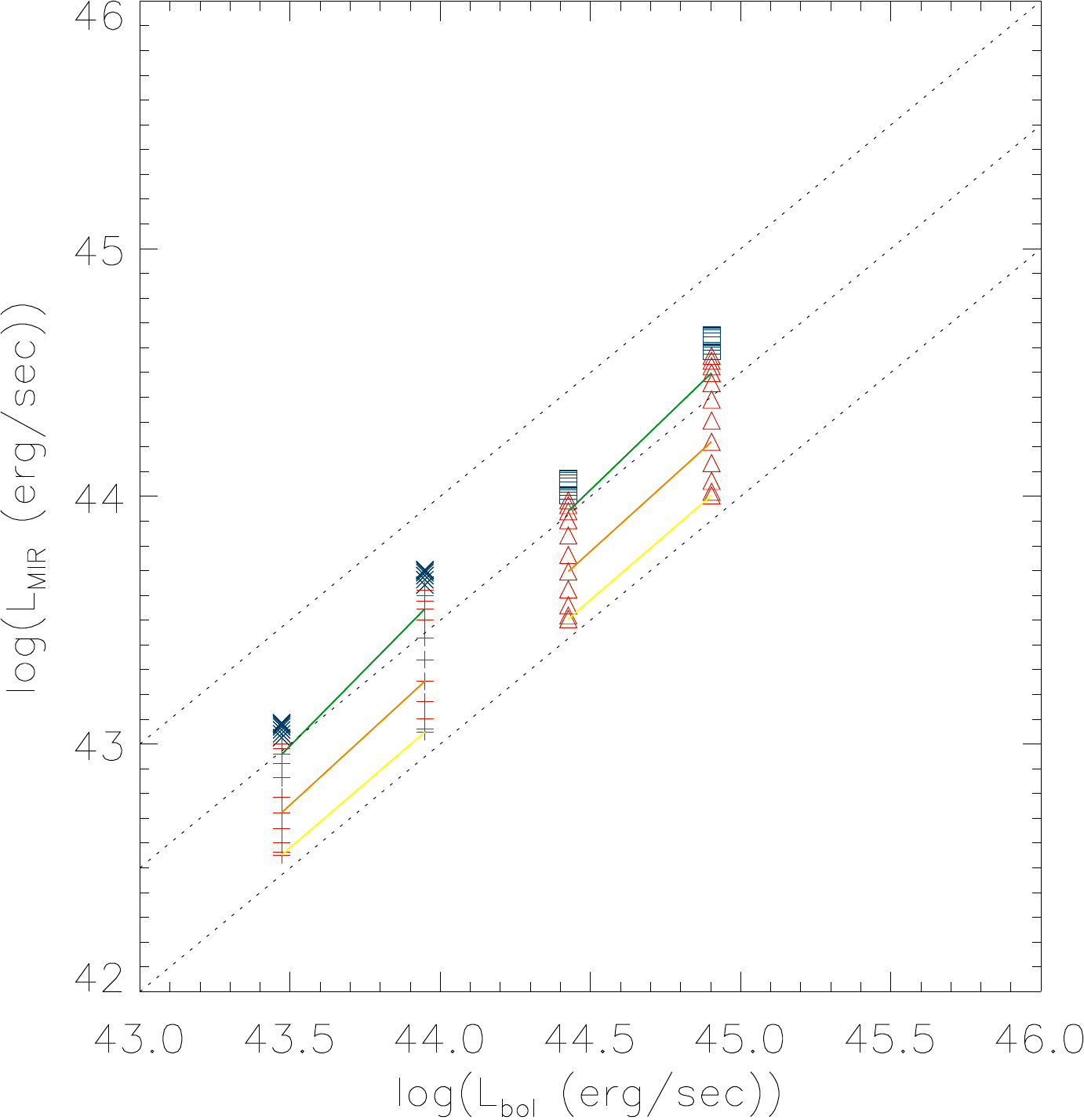}\includegraphics{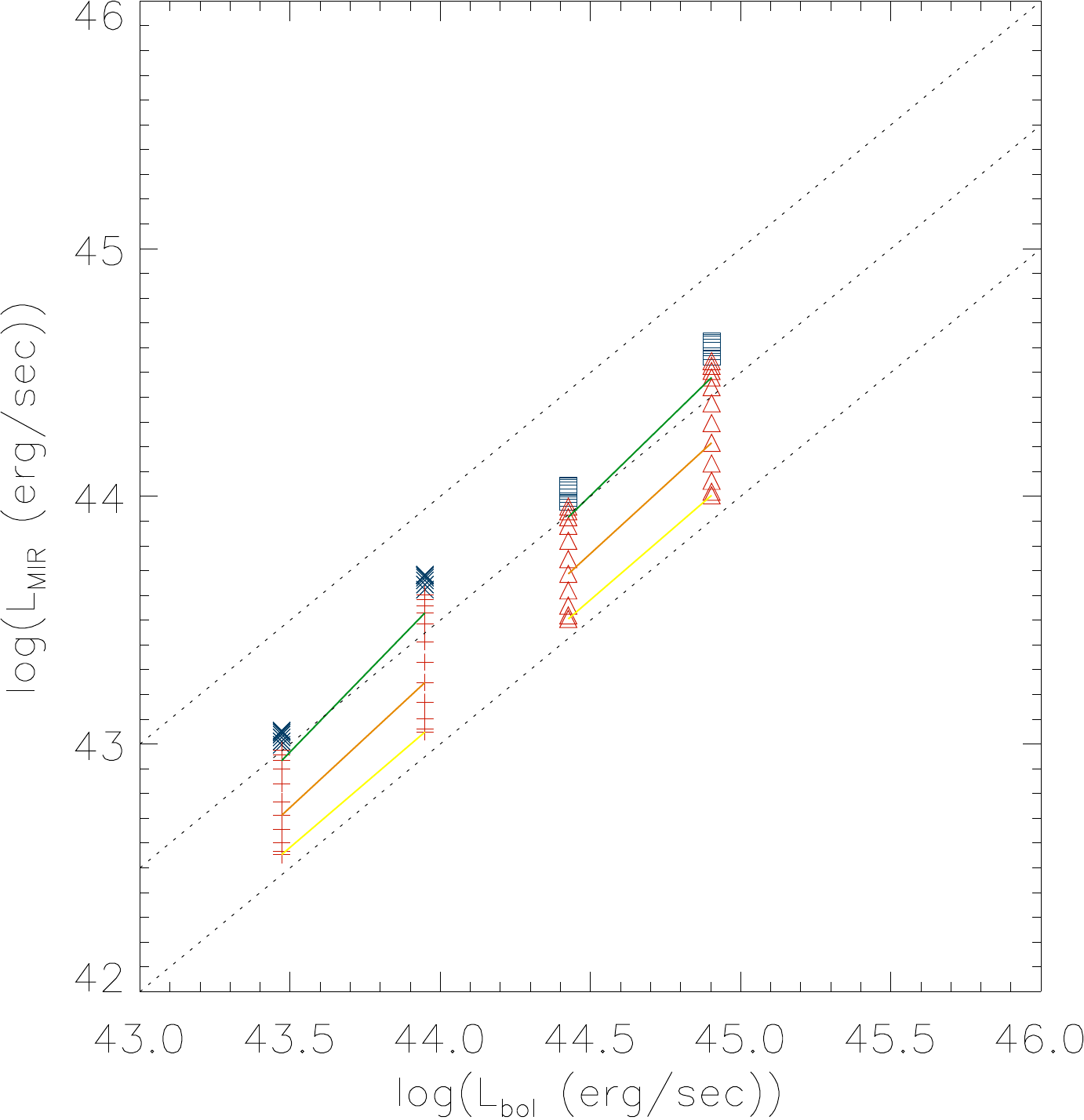}}
  \put(-450,220){\Large inner radius x 1/$\sqrt{2}$}
  \put(-450,240){\Large anisotropic illumination}
  \put(-200,220){\Large inner radius x 1/2}
  \put(-200,240){\Large anisotropic illumination}
  \caption{Model MIR luminosities as a function of the bolometric luminosity for the $1/z$ winds with {\it anisotropic illumination} 
    for a $\sqrt{2}$ (left panel) and $2$ (right panel) times smaller inner radius of the thin gas disk. The dotted lines correspond to $L_{\rm MIR}=\xi \, L_{\rm bol}$
    with $\xi=1,1/3,1/10$.
  \label{fig:plots_SURVEY1bnew_1ai}}
\end{figure*}

In a second step, we illuminate the gas distribution with a $\cos(\theta)$ pattern which is caused by limb darkening of the hot accretion disk
(Netzer 1987) located at a distance smaller than the dust sublimation radius. 
The natural consequence is that the thick gas disk and the basis of the wind receive less flux, whereas the upper wind regions
receive somewhat more flux. The net effect is a decreased MIR and NIR luminosity with respect to the isotropic illumination (Fig.~\ref{fig:plots_SURVEY1bnew_1ai}).
For these models, the MIR luminosity is about $3$ times lower than that of the models with isotropic illumination.
The ratio $L_{\rm MIR}/L_{\rm bol} \sim 1/5$ is still a factor of two higher than the value determined by Gandhi et al. (2009).
Again, a screen with $\tau_{\rm V}=20$ (Sect.~\ref{sec:sed}) leads to two times lower MIR luminosities.
Given that the AGN sample of Gandhi et al. (2009) includes also type~2 objects, there is reasonable agreement between our model and observations.

\begin{figure*}
  \centering
  \resizebox{\hsize}{!}{\includegraphics{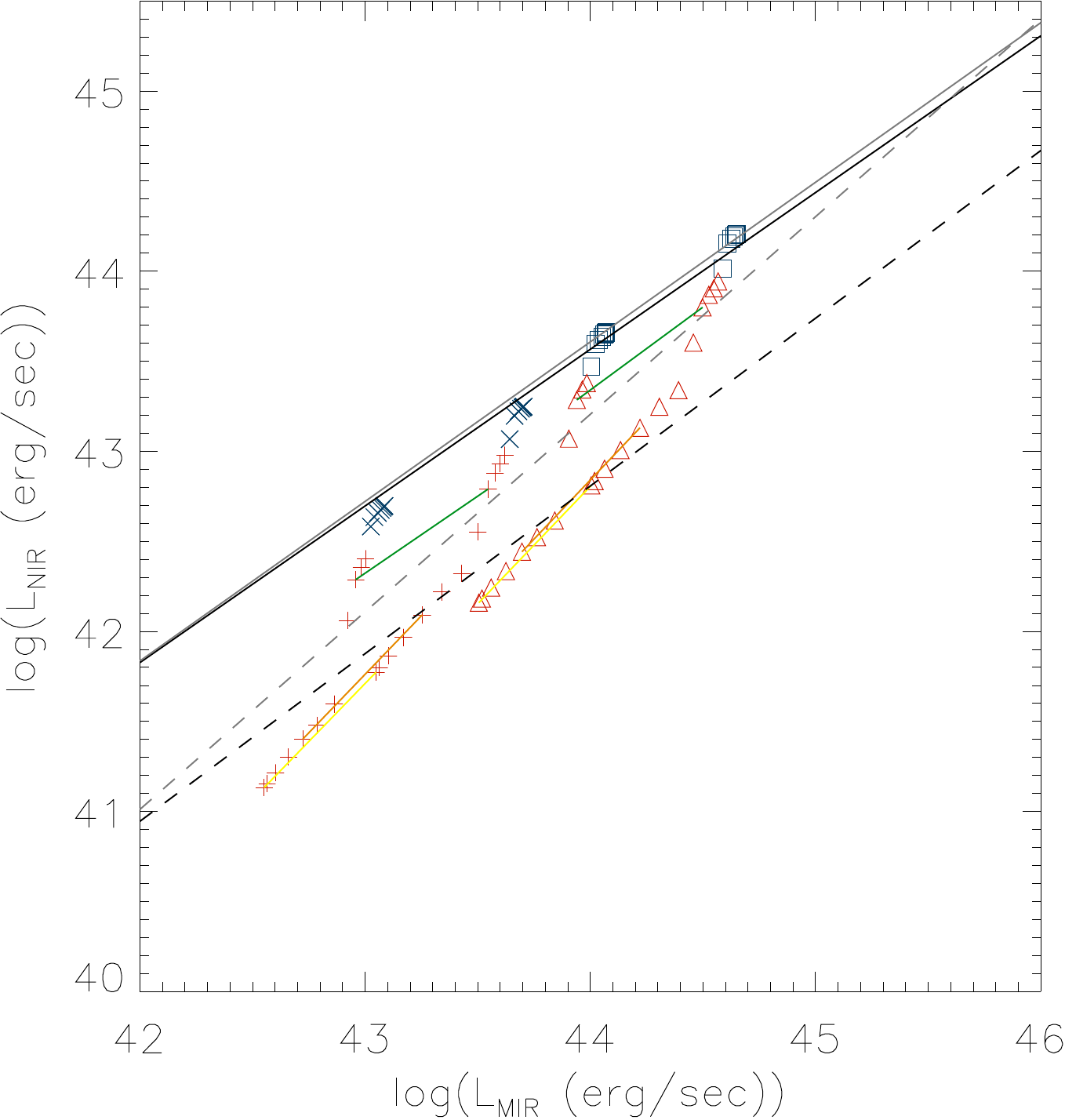}\includegraphics{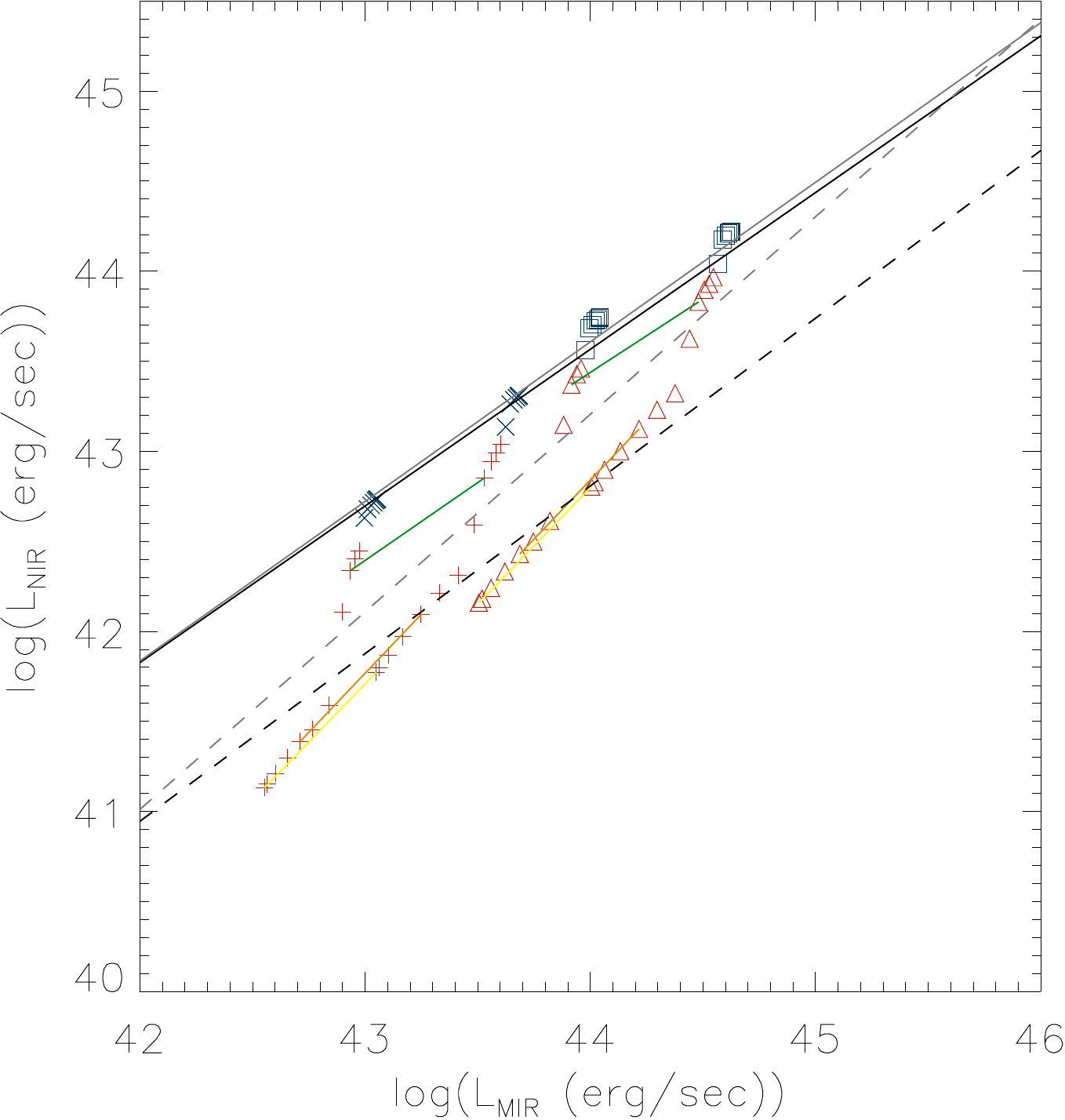}}
  \put(-450,220){\Large inner radius x 1/$\sqrt{2}$}
 \put(-450,240){\Large anisotropic illumination}
  \put(-200,220){\Large inner radius x 1/2}
  \put(-200,240){\Large anisotropic illumination}
  \caption{Model NIR luminosities as a function of the MIR luminosity for the $1/z$ winds with {\it anisotropic illumination} 
    for a $\sqrt{2}$ (left panel) and $2$ (right panel) times smaller inner radius of the thin gas disk.
    The solid and dashed lines are the relations found by Burtscher et al. (2015)
    for type~1 and type~2 objects, respectively.
  \label{fig:plots_SURVEY1bnew_5ai}}
\end{figure*}

Since the NIR is less affected by the change of the illumination pattern, the NIR to MIR luminosity ratio increases with respect
to the models with isotropic illumination (Fig.~\ref{fig:plots_SURVEY1bnew_5ai}). The model with the puff-up being located at
$3 \times r_{\rm sub}$ (left panel of Fig.~\ref{fig:plots_SURVEY1bnew_5ai}) reproduces the observations of Burtscher et al. (2015) best.

We conclude that for the population of local AGN observed by Burtscher et al. (2015) the $\cos(\theta)$ illumination seems to be preferred
over the isotropic illumination.

\subsection{The point source fraction \label{sec:psf}}

\begin{figure}
  \centering
  \resizebox{\hsize}{!}{\includegraphics{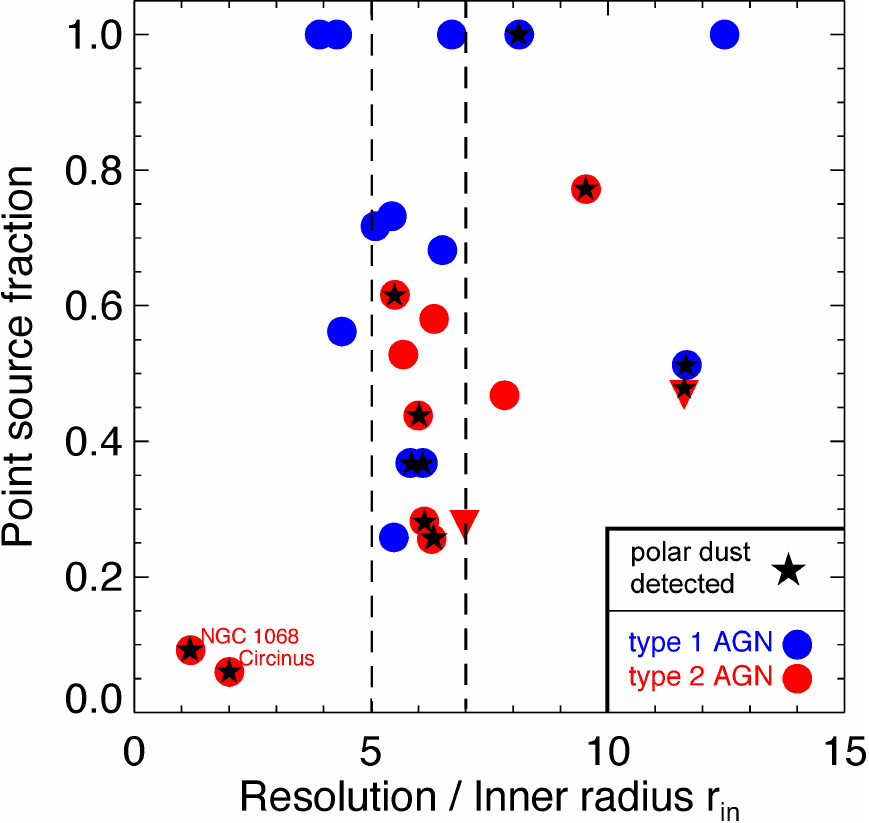}}
  \caption{Point source fraction for the MIR interferometric observations of Burtscher et al. (2013). Only objects around $6 r_{\rm in}$ should be compared
    to Fig.~\ref{fig:psf1}. Objects with detected extended dust emission are marked with a star. This emission is mostly in polar direction, i.e. the 
    direction of the narrow line region, and was detected on 10 pc-scale (Lopez-Gonzaga et al. 2016 and references therein) and on 100 pc-scale (Asmus et al. 2016), 
    but not for the same objects. Note that the detectability of polar dust emission depends on a number of observational parameters and non-detections do not 
    mean that these AGNs are intrinsically different.
  \label{fig:fp_rin_nocase1}}
\end{figure}

\begin{figure*}
  \centering
  \resizebox{\hsize}{!}{\includegraphics{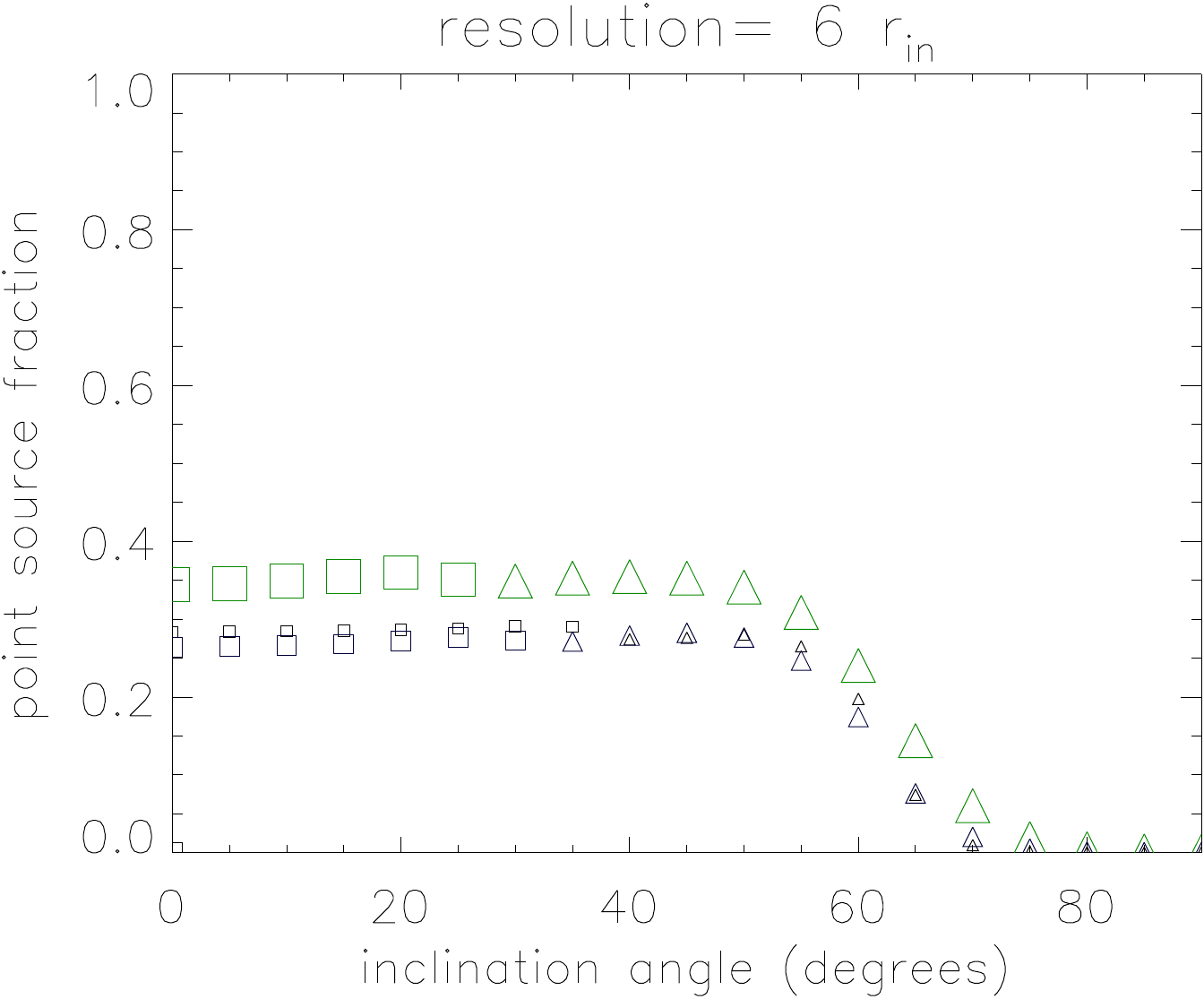}\includegraphics{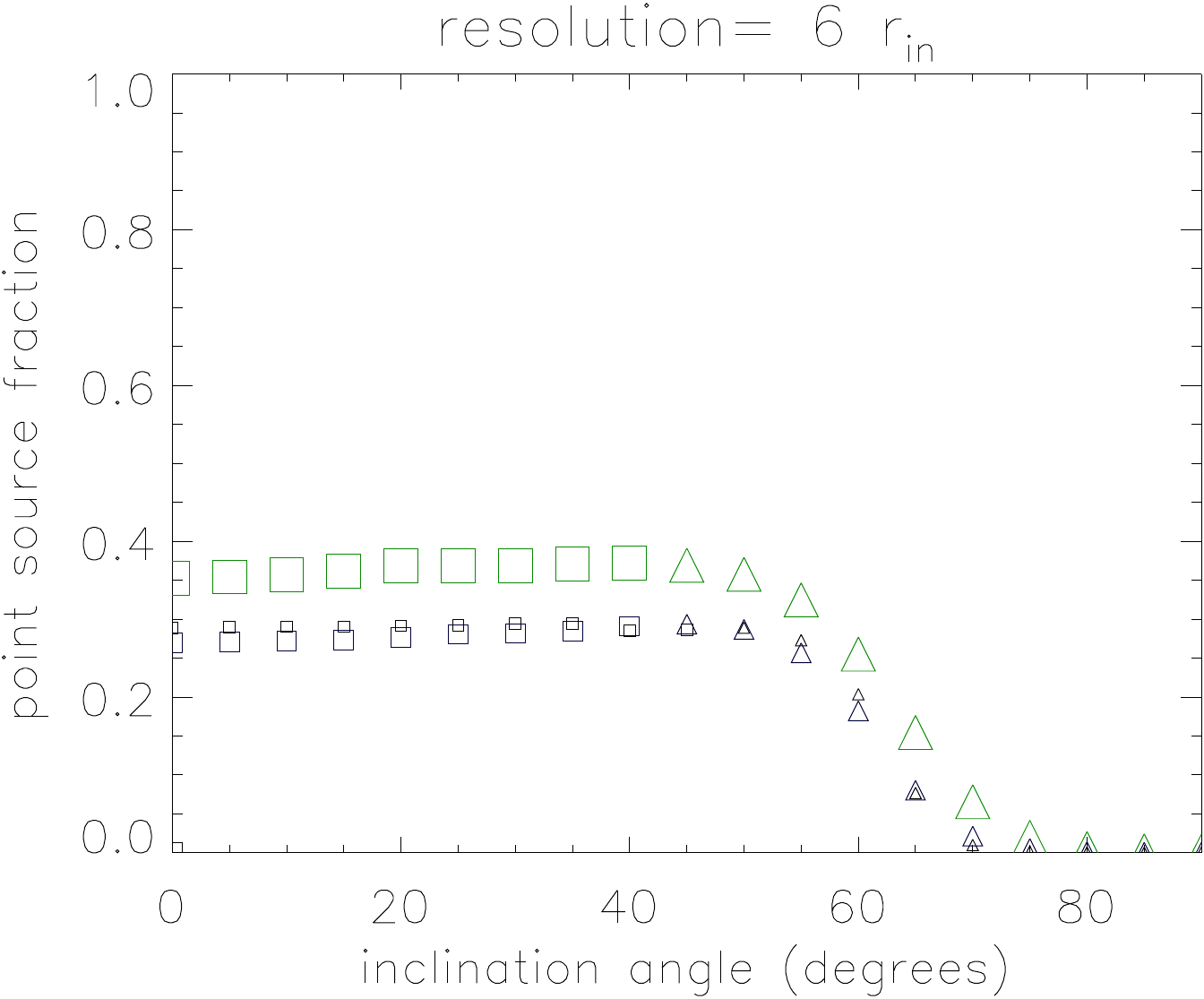}\includegraphics{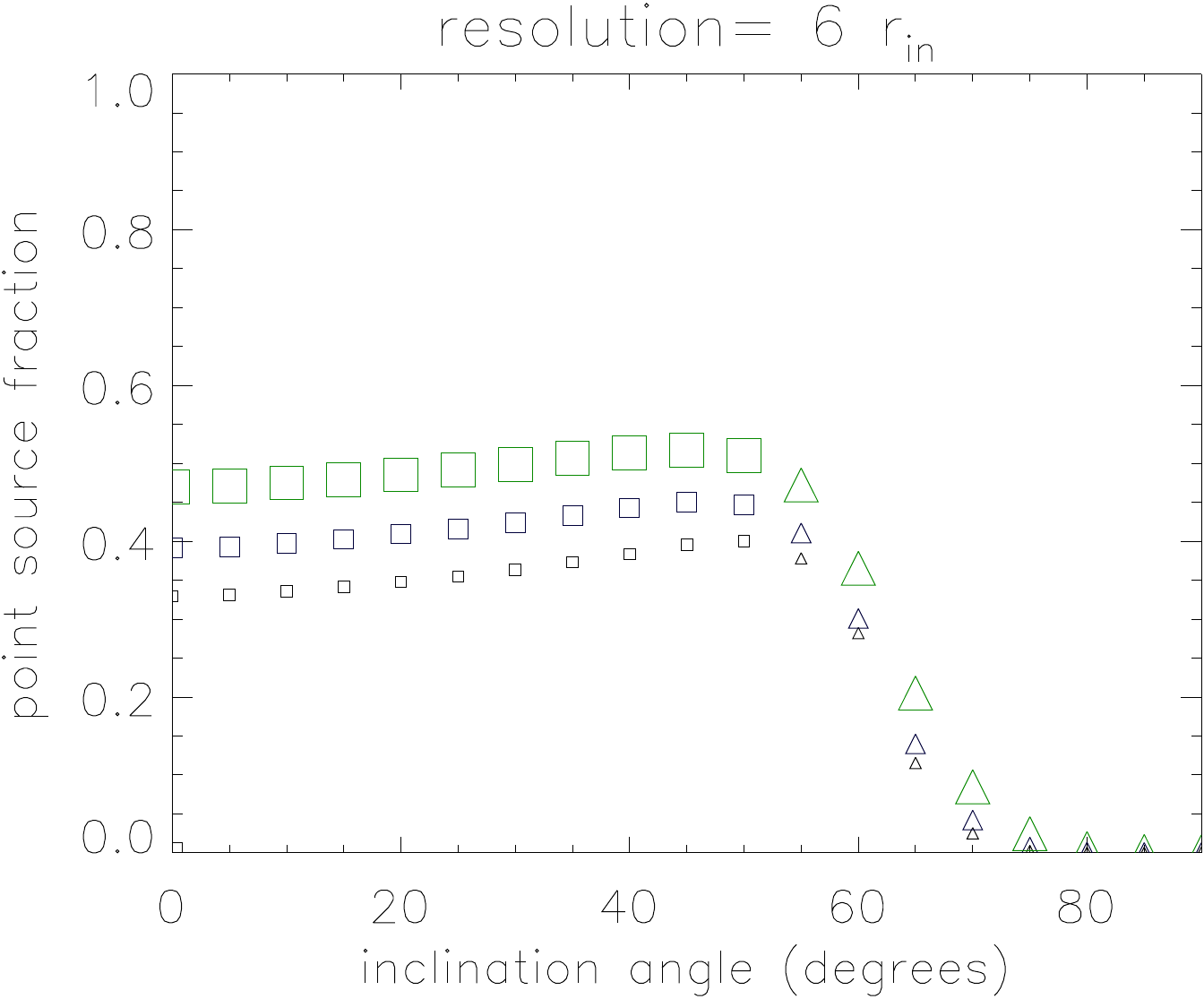}}
  \resizebox{\hsize}{!}{\includegraphics{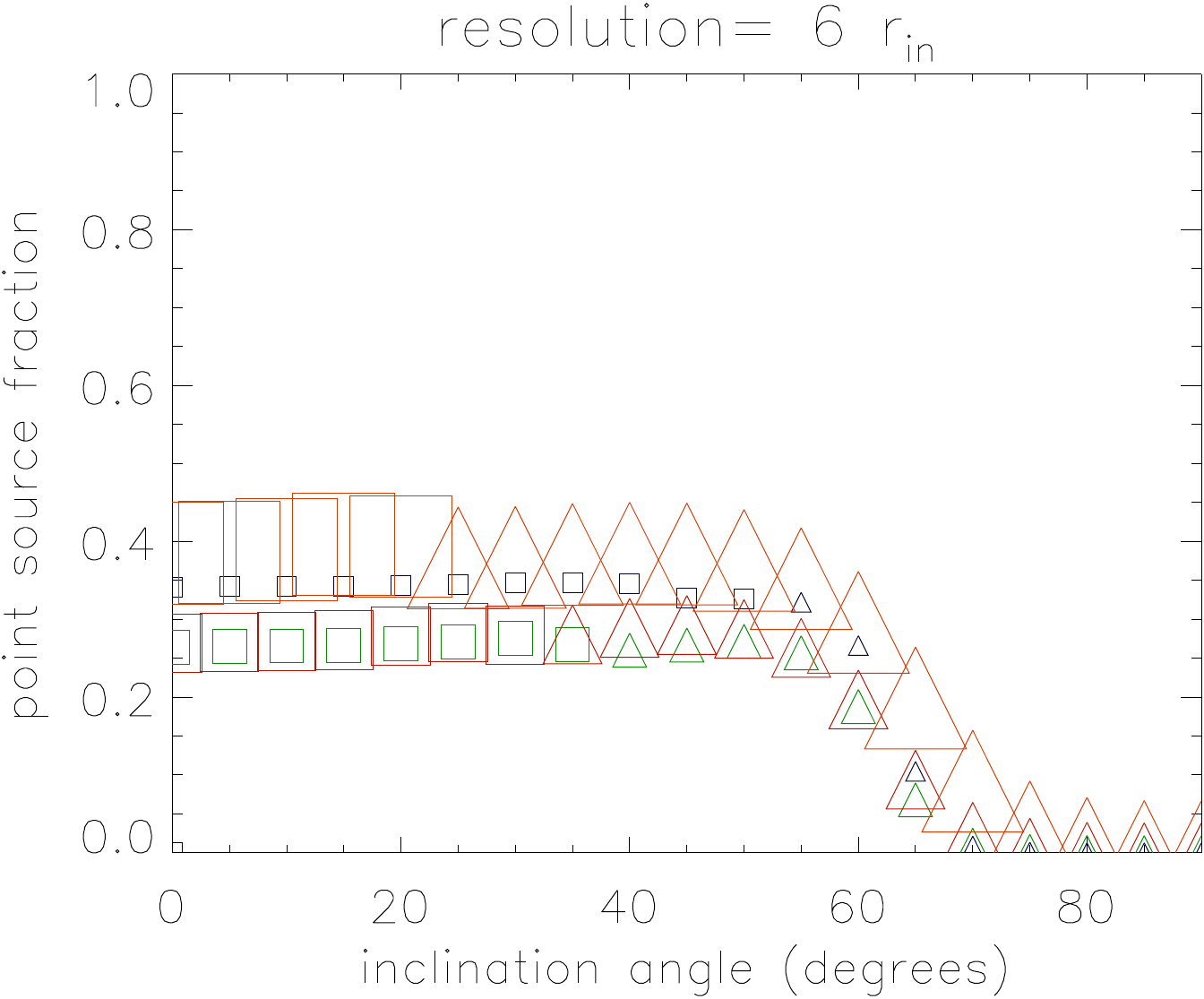}\includegraphics{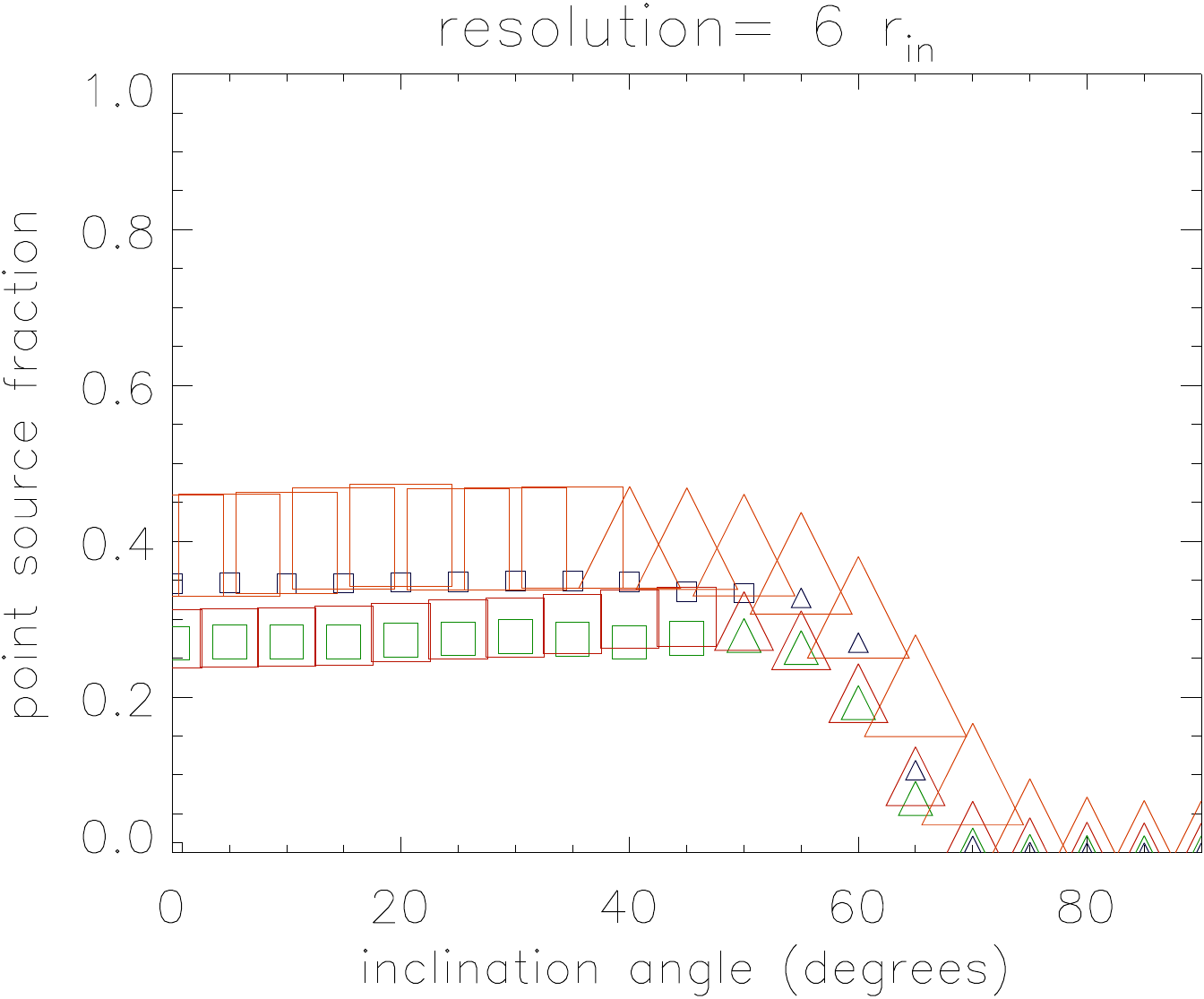}\includegraphics{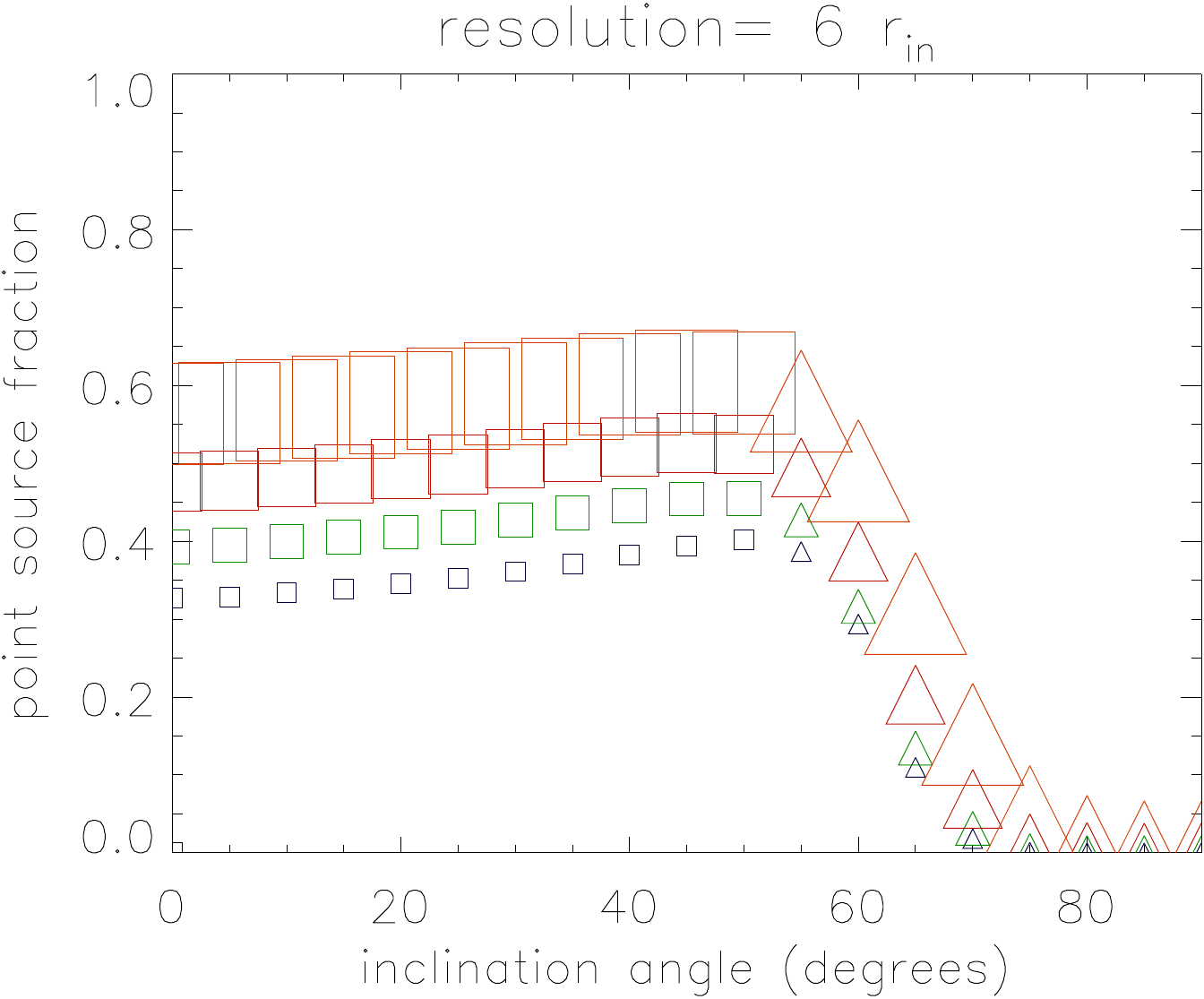}}
  \caption{Point source fraction ($6 r_{\rm in}$) for the Circinus model (upper row) and NGC~1068 model (lower row). Left column: $1/z$ wind (standard model); middle column: $1/z^2$ wind;
    right column: no wind. Boxes: type~1 objects; triangles: type~2 objects. The size of the symbols is proportional to the bolometric luminosity.
  \label{fig:psf1}}
\end{figure*}

The detailed geometry of the (sub-)parsec scale dust distribution, 
i.e. the multi-component structure as well as sizes, elongations and position angles of the 
components, can be observationally best constrained in the two mid-IR brightest objects, Circinus and NGC~1068. 
The basic nuclear dust structure has been determined in another two dozen objects, though (Burtscher et al. 2013, see also H\"{o}nig et al. 2012, 2013). 
The most straight-forward observable in these objects is the visibility at 
long baselines, indicating how well a source is resolved on scales of about $6 r_{\rm in}$ (Fig.~\ref{fig:fp_rin_nocase1}). This highest-resolution visibility or 
``point-source fraction'' is a robustly measured quantity (uncertainty $\sim$ 5 \%) and can be compared directly against our disk--wind model (Fig.~\ref{fig:psf1}).

The model point source fraction is defined as the flux density in the inner $N\,r_{\rm in}$ divided by the total flux density of the image.
In the observations of Burtscher al. (2013) $N=6$ for the majority of AGNs of their sample. 
For all models described in Sect.~\ref{sec:isotrop} we calculated the point source fraction for N=6. In addition, we determined
the point source fractions of the Circinus model at $i=70^{\circ}$ ($N=2$) and the NGC~1068 model at $i=60^{\circ}$ ($N=1$).
Our model point source fractions of Circinus and NGC~1068 agree with the observed point source fractions.
 
\begin{figure}
\resizebox{\hsize}{!}{\includegraphics{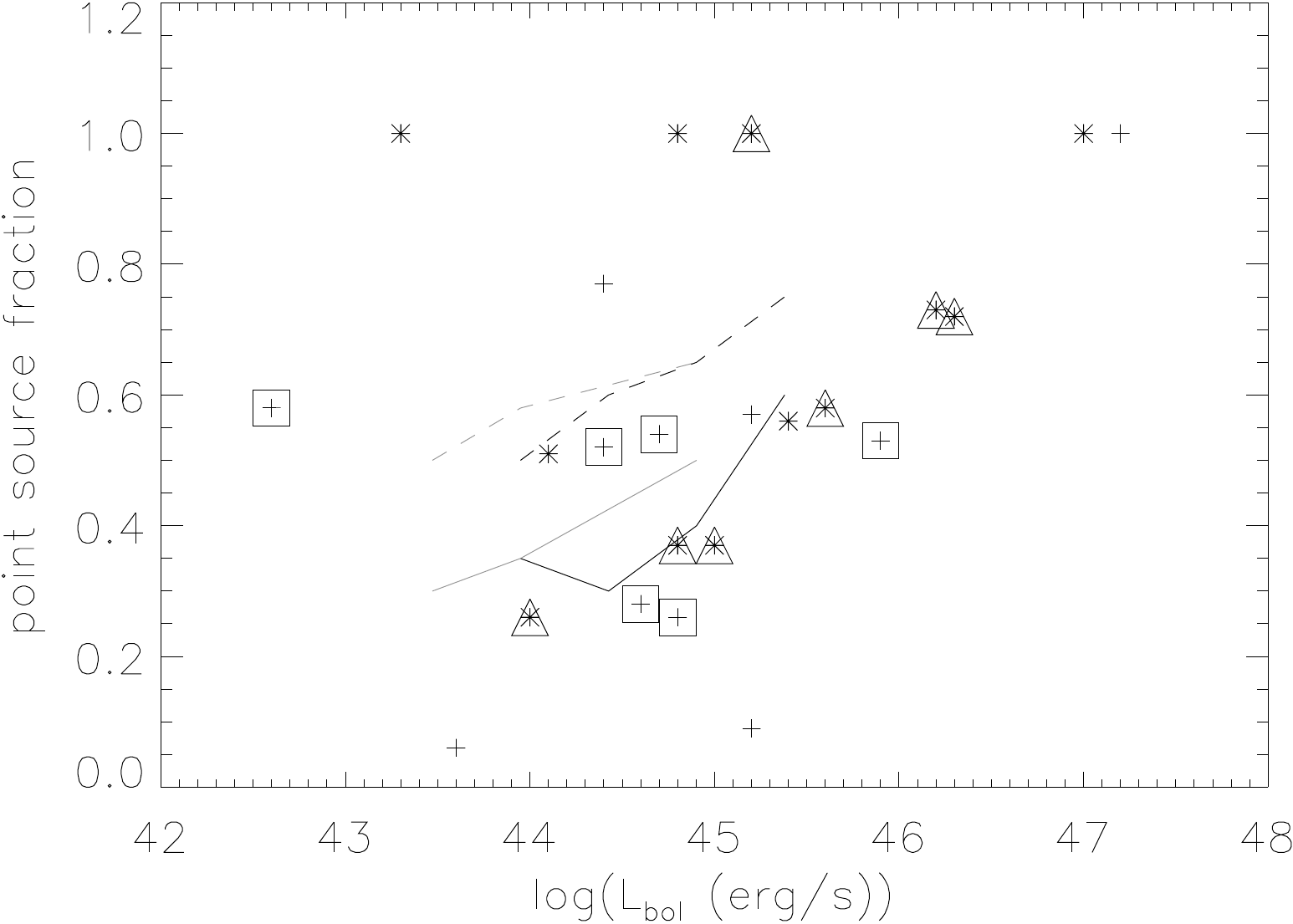}}
\caption{Observed point source fraction as a function of the bolometric luminosity. Pluses: type~2 objects. Stars: type~1 objects. Triangles: type~1 objects with point 
source fluxes within $(6 \pm 1) \times r_{\rm sub}$. 
Squares: type~2 objects with point source fluxes within $(6 \pm 1) \times r_{\rm sub}$}. Dark solid line: NGC~1068 model with 
$r=6 \times r_{\rm sub}$. Grey solid line: NGC~1068 model with $r=9 \times r_{\rm sub}$. Dark dashed line: Circinus model with $r=6 \times r_{\rm sub}$.
Grey dashed line: Circinus model with $r=9 \times r_{\rm sub}$.
\label{fig:pointsourcefrac_type}
\end{figure}

We see a clear dependence of the point-source fraction on the inclination of the model. In addition, the model point source fraction of type~1 objects
depends on the bolometric luminosity, i.e. the point source fraction increases with increasing luminosity. 
In order to see if such a trend also exists in the data of Burtscher et al. (2013), we show the point source fraction as a function of the
bolometric luminosity in Fig.~\ref{fig:pointsourcefrac_type}. The type~1 objects with $r=6 \times r_{\rm sub}$ (triangles) indeed 
reproduce the observed increase of the point source fraction with increasing bolometric luminosity.
This is caused by the increase of the sublimation radius with $\sqrt{L_{\rm bol}}$, whereas the inner illuminated edge of the thick disk and
the high-density part of the wind are located at a constant radius. If most of the IR luminosity is produced within an area that is close
to the inner edge of the thick disk, an increase of the beam width with a constant size of the major IR emitting region leads to an increasing
point source fraction. Once the physical size that corresponds to the resolution (FWHM) of the interferometric observations with the longest baselines
is about $4$ times the radius of the inner edge of the thick gas disk, the point source fraction becomes $\sim 0.5$.

Point source fractions $>0.7$ are only observed in models without a wind, high bolometric luminosities, and $r=9 \times r_{\rm sub}$ (Fig.~\ref{fig:psf16rin}).
In type~1 objects with point source fractions close to unity even the thick gas disk is absent.
The variety of point source fractions ($0.25 \le f_{\rm p} \le 0.75$) for type~1 and type~2 objects might thus be caused by luminosity effects
for type~1 objects and inclination effects for type~2 objects.
Moreover, we argue that observed point source fractions of $\sim 1$ (Fig.~\ref{fig:fp_rin_nocase1}) in type~1 AGNs indicate the absence of a polar 
wind. In quasars, where strong outflows are detected (e.g., Feruglio et al. 2010), the winds can also have an equatorial geometry (e.g., Elvis 2000).

\section{Infrared Interferometry \label{sec:irinterferometry}}

In this section we compare the model visibility amplitudes of the radiative transfer models to existing MIR observations.

\subsection{From images to visibility amplitudes \label{sec:visibilities}}

We compare our model images to observations of the actual geometry of the circum-nuclear region. At infrared wavelengths, especially 
in the mid-IR atmospheric $N$ band window (8--13 $\mu$m), the emission of radio-quiet AGNs is dominated by the thermal radiation of dust. We can therefore 
directly compare the radiative transfer of our model dust distributions with high-resolution IR observations of AGNs. Since the circum-nuclear dust 
distributions even in the most nearby AGNs are essentially unresolved with single-dish telescopes, IR long-baseline interferometry is required to 
probe their geometry. Successful observations of more than two dozen of nearby AGNs have been obtained in the near-IR and mid-IR (see Burtscher et al.
 2016 for a recent review). The most detailed studies have been possible with MIDI at the VLTI for the two brightest objects NGC~1068 
(Lopez-Gonzaga et al. 2014) and in the Circinus~Galaxy (Tristram et al. 2014).

The result of these observations are visibility amplitudes, which themselves need to be compared to model images to constrain the actual surface brightness 
of the (sub-)parsec region. To be closer to observations, we compare the MIDI observations directly to visibilities derived 
from the model images of our disk--wind model. The visibility amplitudes are computed from the model images by means of a Fourier transform with proper 
scaling. To facilitate the comparison between model images and observed visibility data, we have created a Python class, 
{\em img2vis}\footnote{available for download at \url{https://github.com/astroleo/img2vis}}. It takes a model 
image with a given pixel scale and wavelength, converts it into visibility amplitudes (on the so-called $(u,v)$ plane) and compares the results with 
observations given as an ``OIFITS'' file. With the uncertainties associated to the visibility amplitudes, the $\chi^2$ of each model is calculated. 

For each AGN we calculated the following model series
\begin{enumerate}
\item
symmetric RT model with $1/z$ wind and puff-ups at $2,3,4 \times r_{\rm sub}$ and isotropic illumination,
\item
symmetric RT model with $1/z$ wind and puff-ups at $2,3,4 \times r_{\rm sub}$ and $\cos(\theta)$ illumination,
\item
symmetric RT model with $1/z$ wind and a helical wind component (Sect.~\ref{sec:helical}).
\end{enumerate}
For each model series we varied the inclination angle ($0^{\circ}$-$90^{\circ}$) and the spatial scaling ($0.5,0.75,1.0,1.25,1.5$).

\subsection{Application of the model \label{sec:application}}

In the following we compare our model series of the Circinus~Galaxy and NGC~1068 with IR-interferometry observations.
All model images were produced at wavelengths of $8,9,10,11,12~\mu$m. Since our model does not contain an intrinsic
silicate absorption feature and this seems to pose a problem for NGC~1068 at $9~\mu$m (an exceedingly high $\chi^2$ compared to
the $\chi^2$ at the other wavelengths), we decided to calculate the corresponding $\chi^2$ only 
at $8,10,11,12~\mu$m.

\subsubsection{Circinus}

\begin{figure*}
  \centering
  \resizebox{\hsize}{!}{\includegraphics{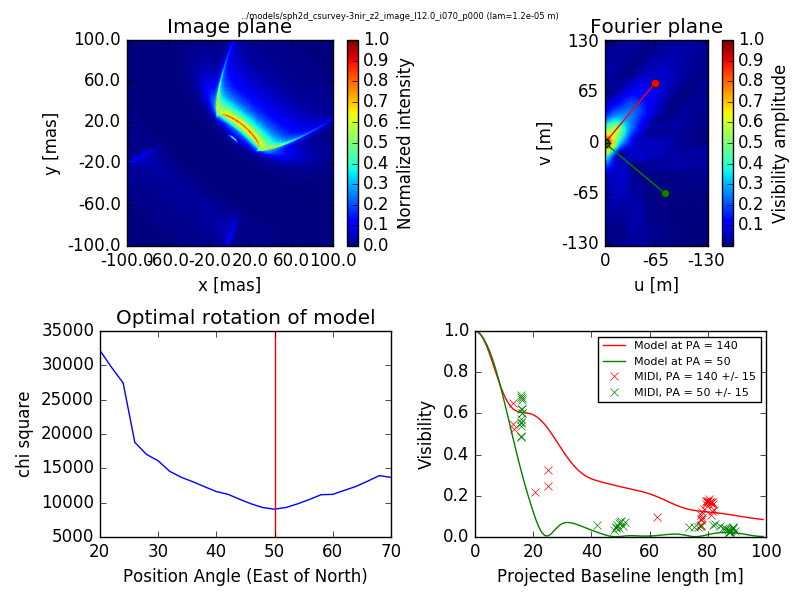}\includegraphics{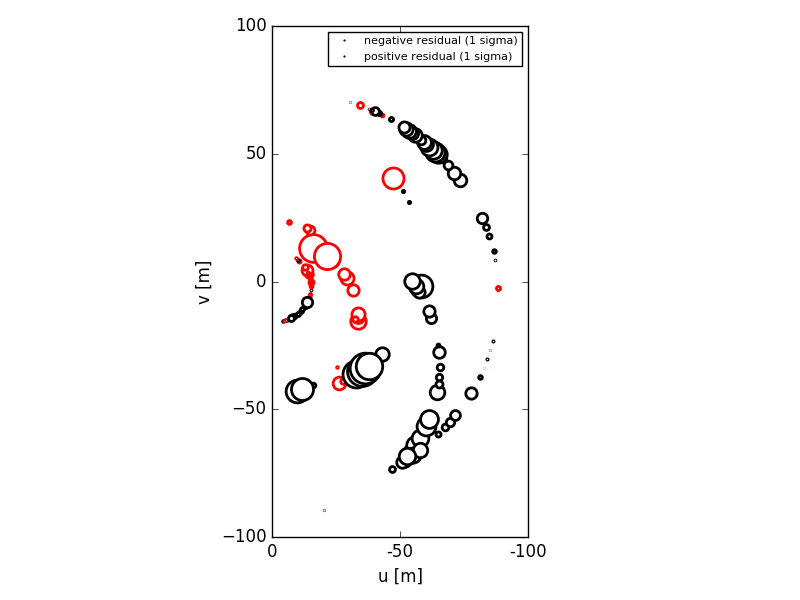}}
  \caption{Radiative transfer model for Circinus. Upper left panel: image at $12~\mu$m; upper middle panel: visibility amplitudes; lower
    left panel: $\chi^2$ as a function of position angle; lower middle panel: visibility amplitudes as a function of projected 
    baseline length (solid lines: model, crosses: observations); right panel: difference between the model and observed 
    visibility amplitudes (red: (model-observations) $> 0$, black: (model-observations) $< 0$). The size of the rings
    is proportional to the value of the residual.
  \label{fig:models_MARC_circinus_image_l12_i065_p000_1.00}}
\end{figure*}

For each model series, we selected the model with the lowest $\chi^2$ (Table~\ref{tab:interferometry}). 
All RT model $\chi^2$ are normalized with that of the Tristram et al. (2014) model: $\chi^2_{\rm rel}$. 
The ``best-fit'' model has an isotropic illumination, an inclination angle of $i=70^{\circ}$ and $\chi^2_{\rm rel}=2.5$.
In Fig.~\ref{fig:models_MARC_circinus_image_l12_i065_p000_1.00} we show the comparison for our standard model (including a $1/z$ wind).
For comparison, Fig.~\ref{fig:circinus_konrad_11.0_1.00} shows the result for the Tristram et al. (2014) model.
The corresponding comparisons at $8~\mu$m and $10~\mu$m are shown in Figs.~\ref{fig:models_MARC_circinus_image_l08_i065_p000_1.00} and 
\ref{fig:models_MARC_circinus_image_l10_i065_p000_1.00}.
We show only the negative $u$ axis of the Fourier plane since the Fourier transform of a real-valued image is axis-symmetric. To find the best-fitting 
solution, we rotate the model image in position angle (lower-left plot). Two cuts through the model image (``radial plots'') with according 
observations are shown in the lower-middle plot and a comparison of the residuals on the $(u,v)$ plane is shown on the right.
 
Overall, the model reproduces the observed visibility amplitudes in an acceptable way. 
In particular, the disk and wind model is able to explain the observed two-component structure 
which can be seen in the radial plots: the visibilities drop quickly up to a baseline length of about 20~m (indicating a large structure, here 
represented by the wind component) and then show some ``wiggles'' (modelled by the thin disk component).
The range of observed visibilities at a baseline length just short of 20~m corresponds to different observed position angles and indicates a significant elongation of 
the large-scale component.
\begin{table*}
\begin{center}
\caption{Circinus models. \label{tab:interferometry}}
\begin{tabular}{lcccccc}
\hline
Model & illumination & spatial scaling & inclination & $\chi^2$ & $\chi^2/ \chi^2_{\rm Tristram}$ & Fig.\\
\hline
Tristram et al. (2014) & - & 1.0 & - &  11681 & 1.0 & \ref{fig:circinus_konrad_11.0_1.00} \\
radiative transfer model & isotropic & 1.0 & $70^{\circ}$ &  29637 & 2.5 & \ref{fig:models_MARC_circinus_image_l12_i065_p000_1.00} \\
RT model with corotating helical wind & isotropic & 1.0 & $70^{\circ}$ & 25159 & 2.2 & \ref{fig:models_MARC_spiral_circinus_marc_circinus_bild_10_25L0.0370370size15winding1wfactor200azangle-120_1.00} \\
\hline
\end{tabular}
\end{center}
\end{table*}

We conclude that the symmetric RT model reproduces the available interferometric observations in an acceptable way. 
Our model has a physical background with four components (puff-up, thin disk, thick disk, and wind component)
whose extent and orientation cannot be varied. On the other hand, the Tristram et al. (2014) model contains three independent Gaussian components 
(but they also fitted the differential phases).
We believe that a factor of two between the $\chi^2$ of our model and that of Tristram et al. (2014) is acceptable.
We derive an inclination angle of $i=70^{\circ}$. Since the maser disk is most probably seen edge-on ($i=90^{\circ}$), it must be tilted or
warped with respect to the thick gas disk.

\subsubsection{NGC~1068 \label{sec:appn1068}}

\begin{figure*}
  \centering
  \resizebox{\hsize}{!}{\includegraphics{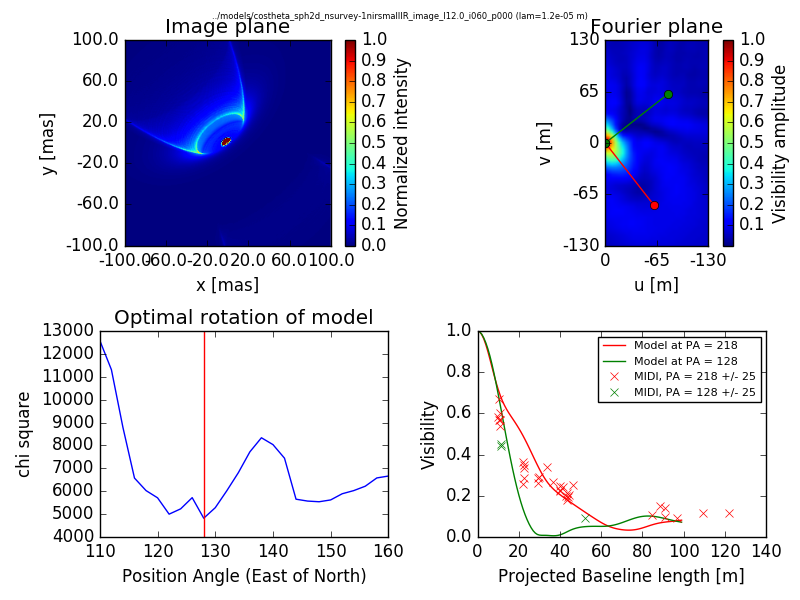}\includegraphics{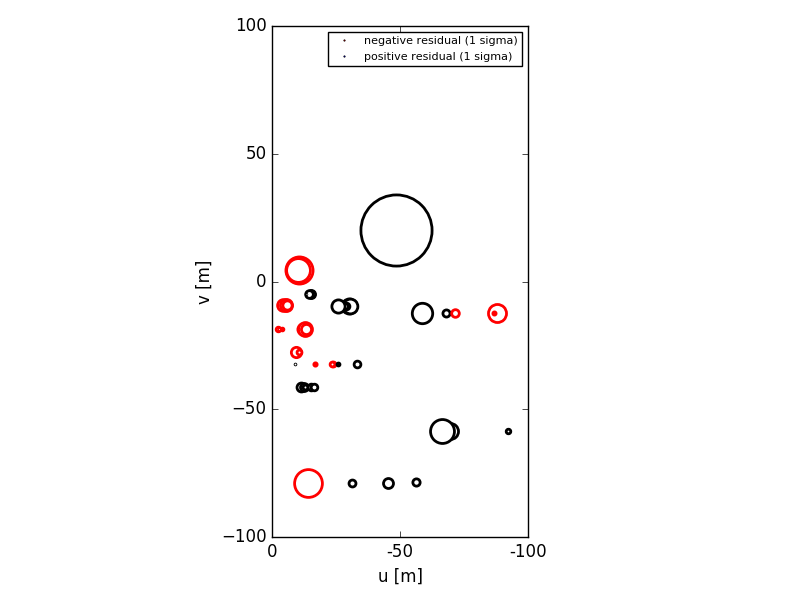}}
  \caption{Same as Fig.~\ref{fig:models_MARC_circinus_image_l12_i065_p000_1.00} for the NGC~1068 radiative transfer model at $12~\mu$m.
    The puff-up of the inner thin disk is located at $r=3 \times r_{\rm sub}$.
  \label{fig:costheta_sph2d_nsurvey-1nirsmallIR_image_l12.0_i060_p000_1.25}}
\end{figure*}

As for Circinus, we selected the model with the lowest $\chi^2$ (Table~\ref{tab:interferometry1}). 
All RT model $\chi^2$ are normalized with that of the Lopez-Gonzaga et al. (2014) model.
The ``best-fit'' model has a $\cos(\theta)$ illumination, an inclination angle of $i=60^{\circ}$ and $\chi^2_{\rm rel}=2.8$ 
(Fig.~\ref{fig:costheta_sph2d_nsurvey-1nirsmallIR_image_l12.0_i060_p000_1.25}).  
For comparison, Fig.~\ref{fig:n1068_lopez_12.0_1.00} shows the result for the Lopez-Gonzaga et al. (2014) model.
The corresponding comparisons at $8~\mu$m and $10~\mu$m are shown in Figs.~\ref{fig:costheta_sph2d_nsurvey-1nirsmallIR_image_l08.0_i060_p000_1.25} and 
\ref{fig:costheta_sph2d_nsurvey-1nirsmallIR_image_l10.0_i060_p000_1.25}.
The thick gas disk of NGC~1068 is thus less inclined than that of Circinus,
the relative $\chi^2$ is comparable to that of Circinus. The inclination angle derived from the model is significantly different from the inclination angle 
derived by the fitting of the IR SED with a clumpy torus model (H\"{o}nig et al. 2008; $i=90^{\circ}$).
Since the inner thin disk, and especially its inner puff-up is visible in the model image,
the exact location of the puff-up is important. It turned out that a radius of $r=3 \times r_{\rm sub}$ leads to the lowest $\chi^2_{\rm rel}$.
This is consistent with the location of the puff-up determined by the IR luminosities (Sect.~\ref{sec:luminosities}).

To investigate the influence of the wind component on the visibilities, we re-calculated the radiative transfer of our ``best-fit'' model
with a dust-free wind, which is devoid of IR emission. The $\chi^2$ of this model is $1.6$ times higher than that of the model with a dusty wind.
Most importantly, the emission distribution in the UV plane of the model with a dust-free wind is significantly more extended than that of the dusty wind model and observations, mainly in the polar direction but also in the direction of the disk plane.
Our model is close to model~1 of Lopez-Gonzaga et al. (2014):
their first component corresponds to the inner part of the thin disk disk, their second component to the inner rim of the thick disk and
the base of the wind. A part of their third component potentially corresponds to emission of the wind cone at higher altitudes.
Since our model is symmetric, we can only speculate that the bulk of the emission of the third component of the Lopez-Gonzaga et al. model
is caused by an asymmetric illumination of the hollow wind cone as modelled for Circinus by Stalevski et al. (2017).

We conclude that the symmetric RT model reproduces the available interferometric observations in an acceptable way.
The comparison with the model of Lopez-Gonzaga (2014) shows why our RT model is quite successful: the two almost 
parallel components with comparable sizes, the inner component being brighter are well reproduced by our models.
The inner component being closer to the central engine is naturally warmer than the outer component, as it is observed in NGC~1068 by Lopez-Gonzaga (2014).
\begin{table*}
\begin{center}
\caption{NGC~1068 models. \label{tab:interferometry1}}
\begin{tabular}{lccccccc}
\hline
Model & illumination & spatial scaling  & inclination & $\chi^2$ & $\chi^2/ \chi^2_{\rm Lopez-Gonzaga}$ & Fig. \\
\hline
Lopez-Gonzaga et al. (2014) & - & 1.0 & - & 6286 & 1.0 & \ref{fig:n1068_lopez_12.0_1.00} \\
radiative transfer model & cos($\theta$) & 1.25 & $60^{\circ}$ & 17871 & 2.8 & \ref{fig:costheta_sph2d_nsurvey-1nirsmallIR_image_l12.0_i060_p000_1.25} \\
RT model with corotating helical wind & cos($\theta$) & 1.25 & $60^{\circ}$ & 17239 & 2.7 & \ref{fig:n1068_bild_10_35L1.00000size15winding1wfactor50azangle-60_1.00} \\
\hline
\end{tabular}
\end{center}
\end{table*}

\subsubsection{Helical wind components \label{sec:helical}}

Motivated by the fact that the large-scale components of the $10$~$\mu$m interferometric observations
are not orthogonal to the smaller-scale (disk) components in the Circinus galaxy (Tristram et al. 2014)
and NGC~1068 (Lopez-Gonzaga et al. 2014), we set up an additional helical density distribution of the wind.
The basic picture is that disk clumps are elevated as entities by the wind and transported upwards.
During the clump ejection, the clump is rotating and sheared. This naturally leads to a helical structure.
In fact, a main characteristic of magnetocentrifugal winds is high rotation velocities. The wind corotates with
the disk until the Alfv\'en radius. Our prescription for the helical outflow has three open parameters:
(i) the thickness, (ii) the winding, and (iii) the azimuthal angle of the footpoint of the spiral.
We vary all three parameters to investigate their influence on the $10$~$\mu$m maps. We made
simplified radiative transfer models for these helical structures and added them to the full RT models.

\begin{figure*}
  \centering
  \resizebox{\hsize}{!}{\includegraphics{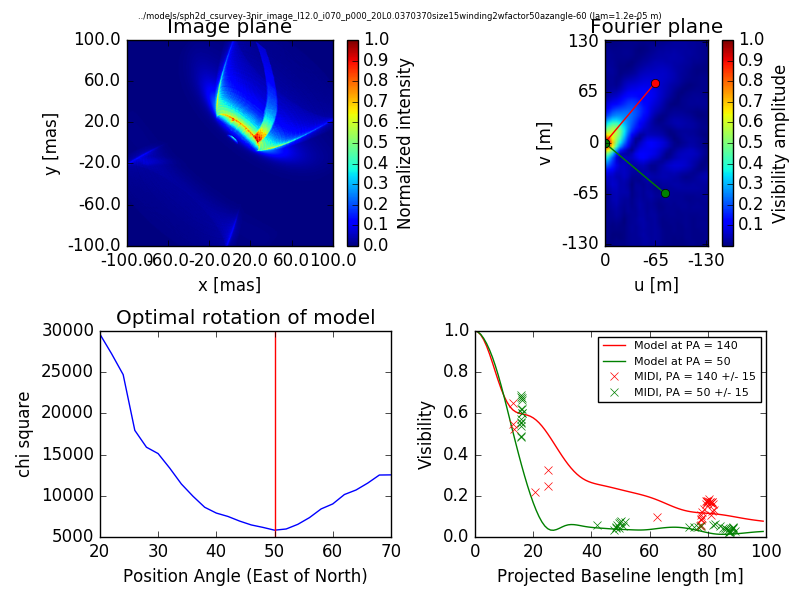}\includegraphics{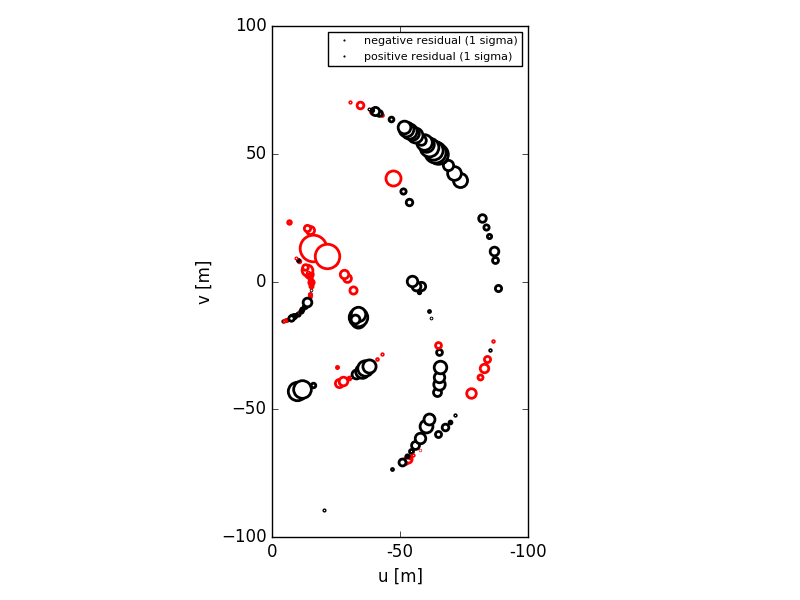}}
  \caption{Same as Fig.~\ref{fig:models_MARC_circinus_image_l12_i065_p000_1.00} for the Circinus radiative transfer model with an additional 
    corotating helical wind component.
  \label{fig:models_MARC_spiral_circinus_marc_circinus_bild_10_25L0.0370370size15winding1wfactor200azangle-120_1.00}}
\end{figure*}

Adding a helical wind component to break axis-symmetry decreases $\chi^2$ for both galaxies. The 
addition of a corotating helical wind to the RT model of Circinus leads to a decreased $\chi^2_{\rm rel}$ of $2.2$ 
(Fig.~\ref{fig:models_MARC_spiral_circinus_marc_circinus_bild_10_25L0.0370370size15winding1wfactor200azangle-120_1.00}).
This decrease is small, but significant.
Since $\chi^2$ depends on the exact wind geometry which is highly uncertain, one should not overinterpret the models with a helical wind component.
As shown by Tristram et al. (2014), such an asymmetric wind component is needed for Circinus.
Alternatively, a tilted $\cos(\theta)$ illumination might also lead to an asymmetric MIR emission distribution.

\begin{figure*}
  \centering
  \resizebox{\hsize}{!}{\includegraphics{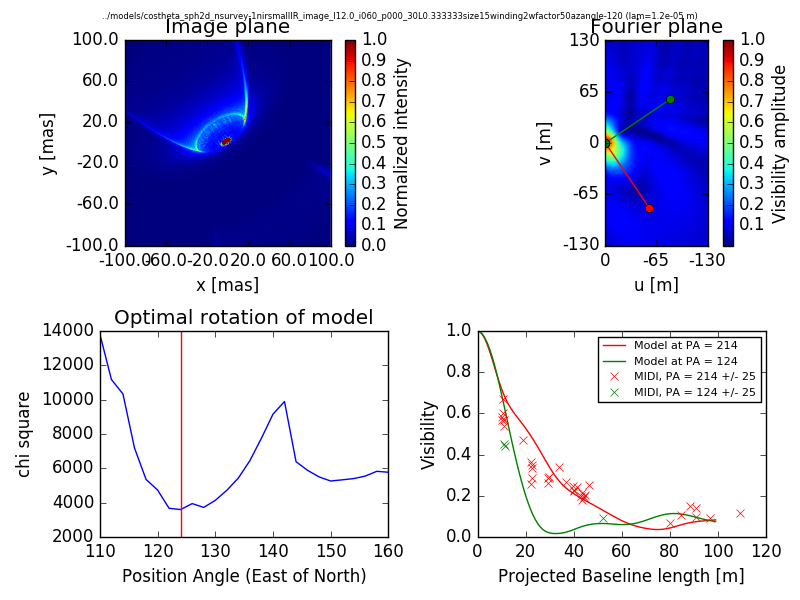}\includegraphics{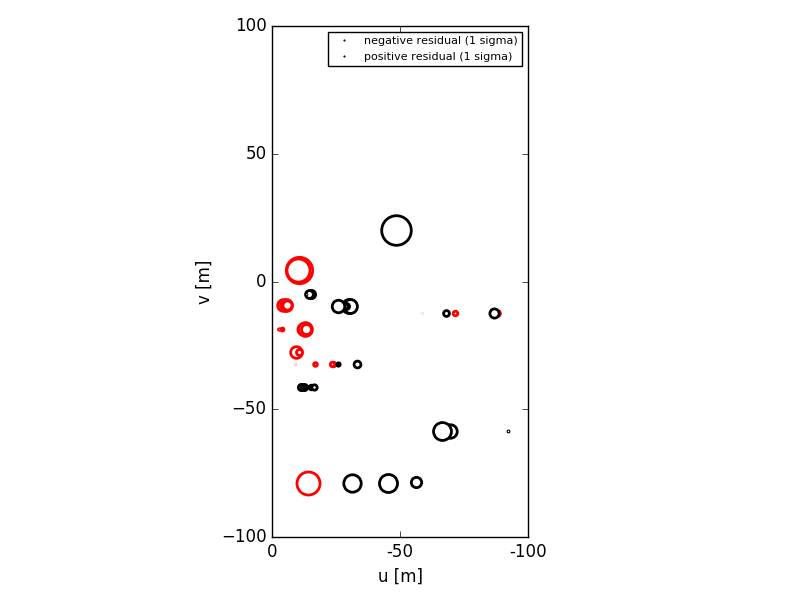}}
  \caption{Same as Fig.~\ref{fig:models_MARC_circinus_image_l12_i065_p000_1.00} for the NGC~1068 radiative transfer model with an 
    additional corotating helical wind component. 
  \label{fig:n1068_bild_10_35L1.00000size15winding1wfactor50azangle-60_1.00}}
\end{figure*}

We show the minimum $\chi^2$ RT model for NGC~1068 with an additional corotating wind component in 
Fig.~\ref{fig:n1068_bild_10_35L1.00000size15winding1wfactor50azangle-60_1.00}.
The addition of a corotating helical wind to the RT model leads to a modestly decreased $\chi^2_{\rm rel}$ of $2.7$. 
The comparison of the residuals of the symmetric to the helical wind models shows that mainly the most northern visibility decreased significantly.

We conclude that the addition of a helical wind improves the resemblance of the model with respect to the interferometric observations
by a small but significant amount in Circinus. For the NGC~1068, the addition of a helical wind does not lead to a significantly better resemblance.

\section{Optical polarization \label{sec:polarization}}

Since the unified model of AGN is based on the observation of optical polarization in nearby Seyfert galaxies (Antonucci \& Miller 1985, 
Antonucci 1993), we decided to run polarized radiative transfer simulations to investigate whether our models of the Circinus galaxy and 
NGC~1068 are consistent with archival polarization data. We confine ourselves to the near-infrared, optical and ultraviolet bands as: 
1) the code used to achieve our simulations only works from the X-ray band to the near-infrared (upper limit: $\sim$1~$\mu$m), and 
2) most of the past spectropolarimetric measurements were taken between the U and J bands. We used the Monte Carlo radiative transfer 
code {\sc stokes} (Goosmann \& Gaskell 2007, Marin et al. 2012, Marin et al. 2015) and built our 3D models of the Circinus galaxy and 
NGC~1068 according to Sect.~\ref{sec:ddist}. In {\sc stokes}, the 3D RT model is made of a finite number of discrete geometrical wedges of constant 
density with sharp edges. Therefore, the vertical density distribution of the thick gas disk is assumed to be constant due to numerical 
limitations. To mimic a Gaussian vertical density distribution, we decided to increase the thick disk opening angle within the model 
uncertainties by $30$\,\% for Circinus and NGC~1068. This indeed lead to a better agreement with observations. 
We accounted for the screen of cold dust with $\tau_{\rm V} = 20$ presented in Sect.~\ref{sec:sed} that is used to reproduce the silicate feature for both AGN. 
To do so, we included in the three-dimensional model a physical slab of dust which was placed beyond the border of the thick disk,
at a distance of $10$~pc from the central black hole for Circinus and NGC~1068. Since the screen is optically thick, a larger distance to the
central black hole does not influence our results.
The projected screen size was $5$~pc for NGC 1068 and $2$~pc for Circinus. Due to the large optical thickness of the slab, photons are mainly absorbed. 
The polarization of the few photons that travel through the entire dust screen is not expected to be strongly altered since forward scattering has a minimum 
impact on the polarization of optical light.
For the electron density distribution we assumed that 
the electrons are co-spatial with the dust grains, with an optical depth of $\tau_{\rm V} < 1$. We make sure that the electron densities 
are in agreement with observations, i.e., the polar electron density lies between $10^3$ and $10^6$~cm$^{-3}$, and the equatorial electron 
density is at least $10^7$~cm$^{-3}$ (Blandford et al. 1990). The spectral band of investigation was set from 1125 to 9775~\AA, and we looked 
at the resulting polarization for all inclinations.

\subsection{Circinus}
\label{Polarization:Circinus}

\begin{figure}
   \centering
   \resizebox{\hsize}{!}{\includegraphics{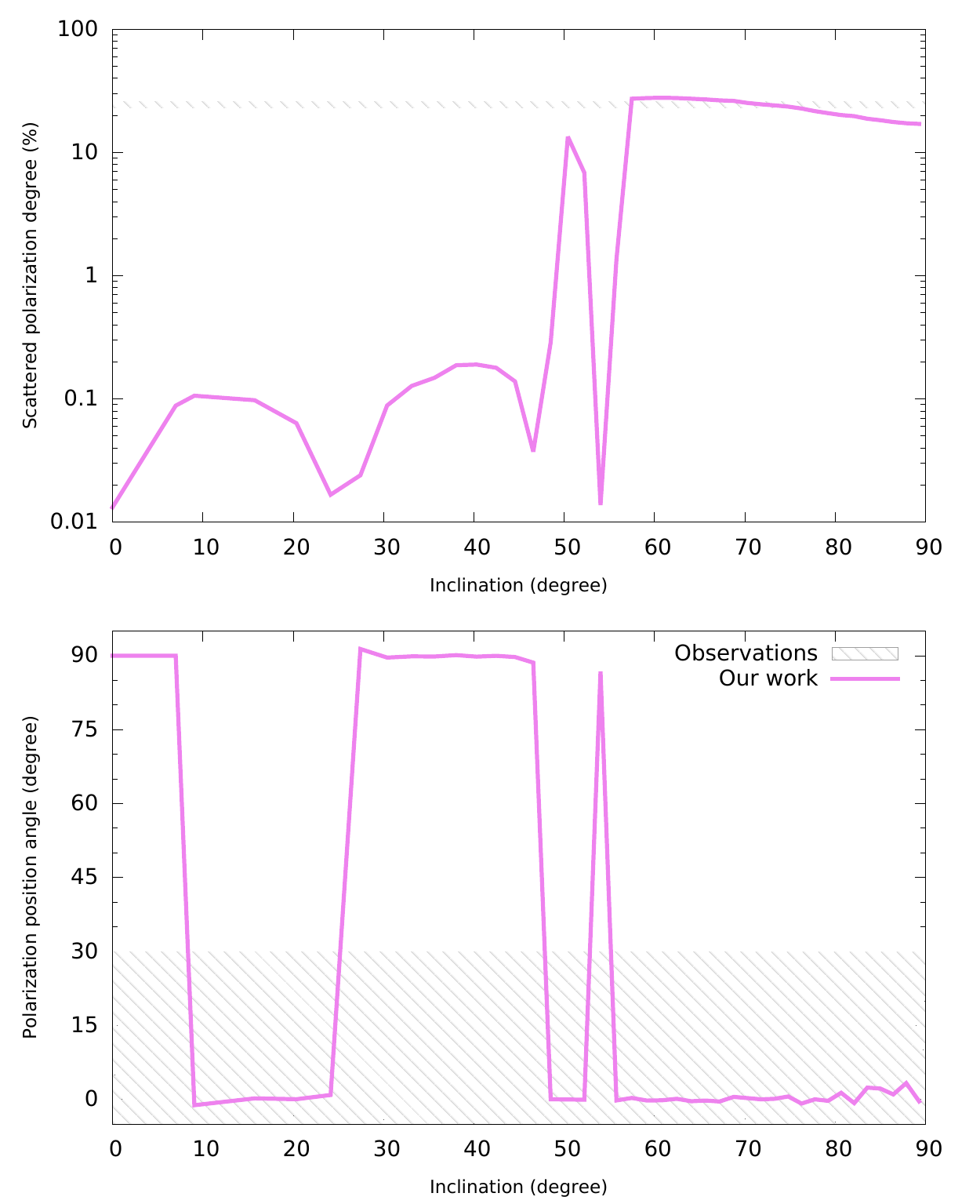}}
   \caption{Wavelength-integrated near-infrared and optical polarization of the 
   Circinus 3D model as a function of inclination. The observed properties are 
   indicated as hatched regions. Upper panel: degree of linear polarization as a 
   function of the inclination angle of the gas disk. Lower panel: polarization 
   position angle. The polarization results from electron and Mie scattering in 
   the Circinus 3D model.}
  \label{Fig:POL_Circinus_angles}
\end{figure}

\begin{figure}
   \centering
   \resizebox{\hsize}{!}{\includegraphics{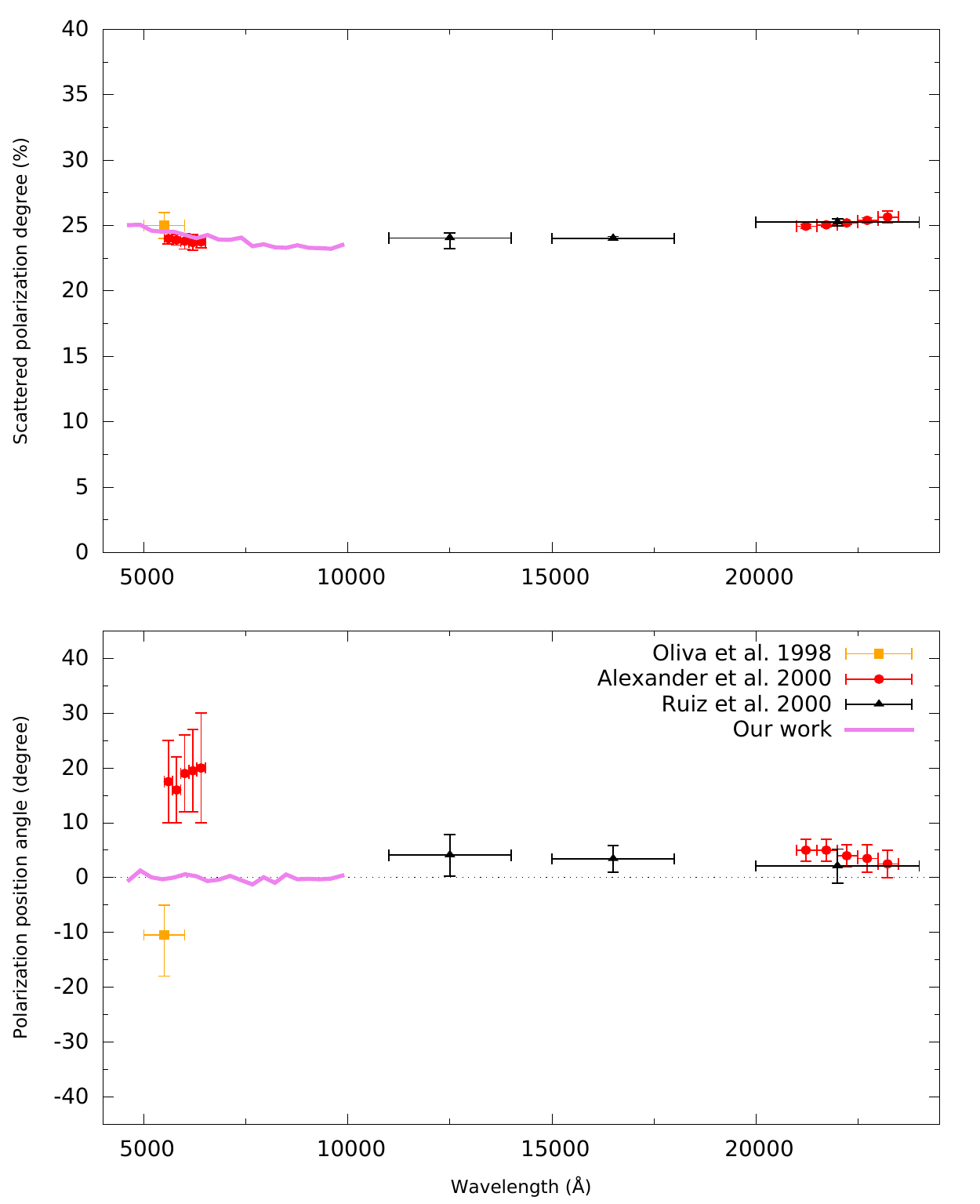}}
   \caption{Wavelength dependence of the optical polarization
    properties of the Circinus 3D model ($i=68^{\circ}$; magenta) compared 
    to observations.}
  \label{Fig:POL_Circinus}
\end{figure}

Our wavelength-integrated results for the Circinus 3D model are shown in Fig.~\ref{Fig:POL_Circinus_angles}. The magenta lines represent the 
polarization (top: polarization degree; bottom: polarization position angle) as a function of inclination angle and the shaded areas correspond to 
the observed polarization regardless of inclination. The degree of linear polarization is rather low at type-1 inclinations due to the dominant 
amount of unpolarized radiation coming from the continuum source and seen in transmission through the polar region. The net degree of polarization
is thus less than 0.2\% and shows a polarization position angle mainly equal to 90$^{\circ}$, as expected from atlases of type-1 AGN (e.g., 
Smith et al. 2002). However, the polarization angle rotates from 90$^{\circ}$ to 0$^{\circ}$ when the observer's line-of-sight matches the wind 
half-opening angle ($i \sim 15^{\circ}$). Multiple scattering within the medium induces a variation in polarization angle as photon reprocessing
happens in the polar region, decreasing the net polarization in this range of inclinations. Once the line-of-sight of the observer is below the 
wind half-opening angle ($i > 20^{\circ}$), the degree of polarization decreases with increasing inclination until a depolarization effect happens 
at $i \sim$~52$^{\circ}$, where the polarization position angle rotates again from 90$^{\circ}$ to 0$^{\circ}$. This transition is due to the 
predominance of polar scattering as the equatorial thin disk polarization becomes obscured by the optically-thick base of the wind. The fraction 
of photons that undergo equatorial scattering in the inner regions becomes weaker and reprocessing in the polar wind dominates the total emission. 
The rotation of the polarization angle indicates the transition from equatorial-scattering dominated (type-1 AGN) to polar-scattering dominated
emission (type-2 AGN). A sudden change of the polarization position angle, associated with a local minima in polarization degree, happens 
at an inclination of 53$^\circ$; this feature is only due to the finite edges of the models. With a Gaussian vertical density distribution this
feature would disappear. At inclination angles larger than $65^{\circ}$, the observer's line-of-sight is completely obscured by the equatorial 
thick disk and electron/Mie scattering in the polar wind dominates. The polarization degree is then plateauing at $\sim$ 20\% until a 
90$^\circ$ inclination due to the dust screen that obscures the base of the polar winds. Compared to observations, our model is able to 
reproduce both, the observed degree and angle of polarization, at inclinations $i \ge$ 68$^\circ$. This is in agreement with the nucleus inclination 
derived from MIR interferometric observations (Sect.~\ref{sec:application}), and also in agreement with the inclination angle derived by 
Fischer et al. (2013) for the same object. 

We thus fixed the inclination of the observer to 68$^\circ$ and plotted in Fig.~\ref{Fig:POL_Circinus} the wavelength dependence of the 
near-infrared and optical polarization properties of the Circinus model. We also report the observations made by Oliva et al. (1998), Alexander et al. (2000), 
and Ruiz et al. (2000) on the plot. The observed polarization degree has been corrected for host galaxy depolarization and starburst light dilution
following the method presented in Marin (2014). Our results are consistent within 1-$\sigma$ with the estimated amount of scattered 
polarization\footnote{About 25\% at all optical wavelengths, see Oliva et al. (1998).} from the modeling achieved by Oliva et al. (1998). 
Our model is in agreement with the observation of the scattered continuum of the Circinus galaxy in the 5000 -- 6500~\AA~band, where 
spectropolarimetry is available. The dependence of the polarization degree, decreasing from the optical to the near-IR band, is due to the
important contribution of dust obscuration by the screen. It would be necessary to extend the wavelength coverage of the code to investigate 
whether our results also agree with infrared data, but the tendency of the model curve seems to be in agreement with observations. The 
polarization position angle is also consistent with observations and we do not expect any rotation of the angle with wavelength upward 1~$\mu$m, 
because Mie and Thomson scattering will remain the main reprocessing processes until 2~$\mu$m.

\subsection{NGC~1068}
\label{Polarization:NGC1068}

\begin{figure}
   \centering
   \resizebox{\hsize}{!}{\includegraphics{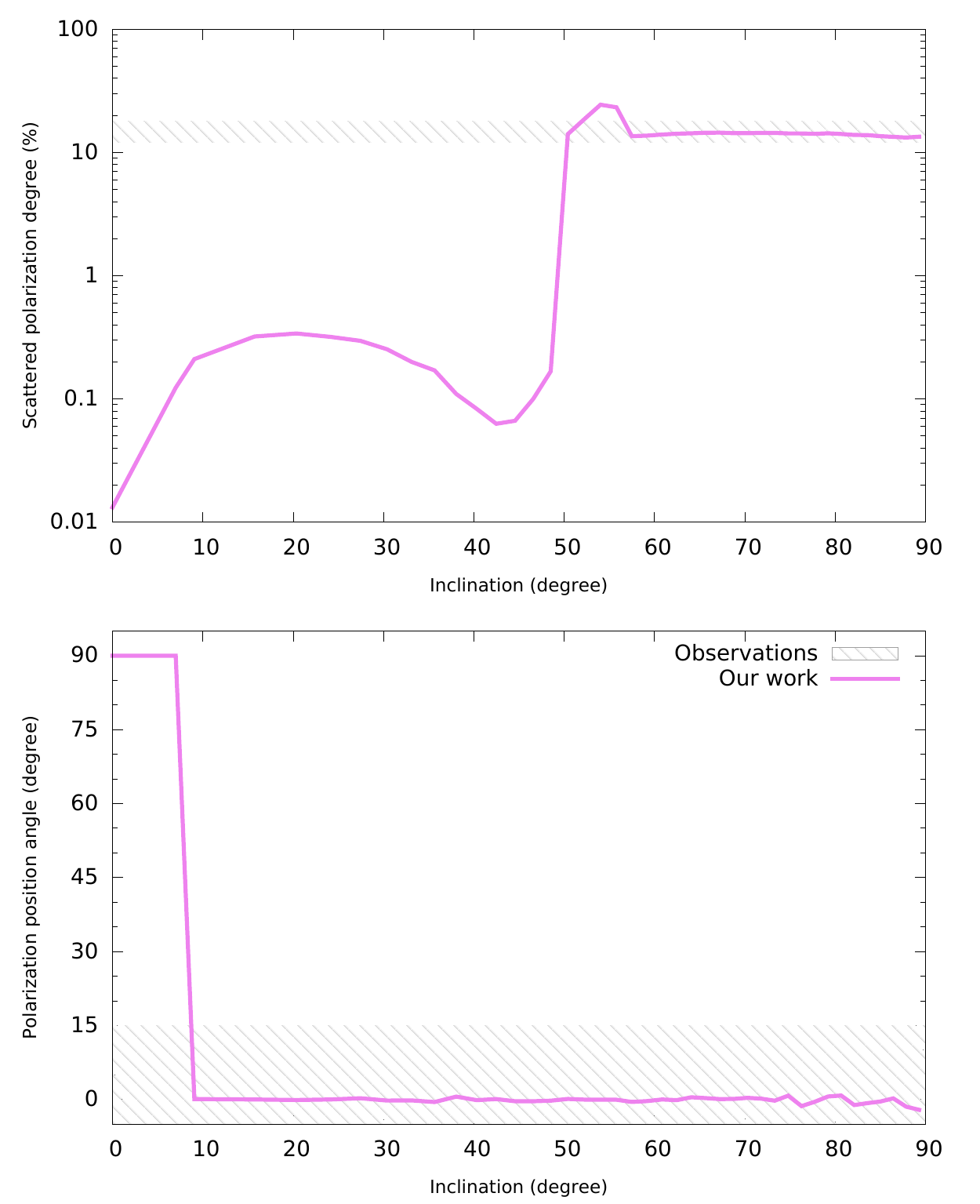}}
   \caption{Same as Fig.~\ref{Fig:POL_Circinus_angles} for the NGC~1068 3D model , with the addition of the ultraviolet 
   polarization.}
  \label{Fig:POL_NGC1068_angles}
\end{figure}

\begin{figure}
   \centering
   \resizebox{\hsize}{!}{\includegraphics{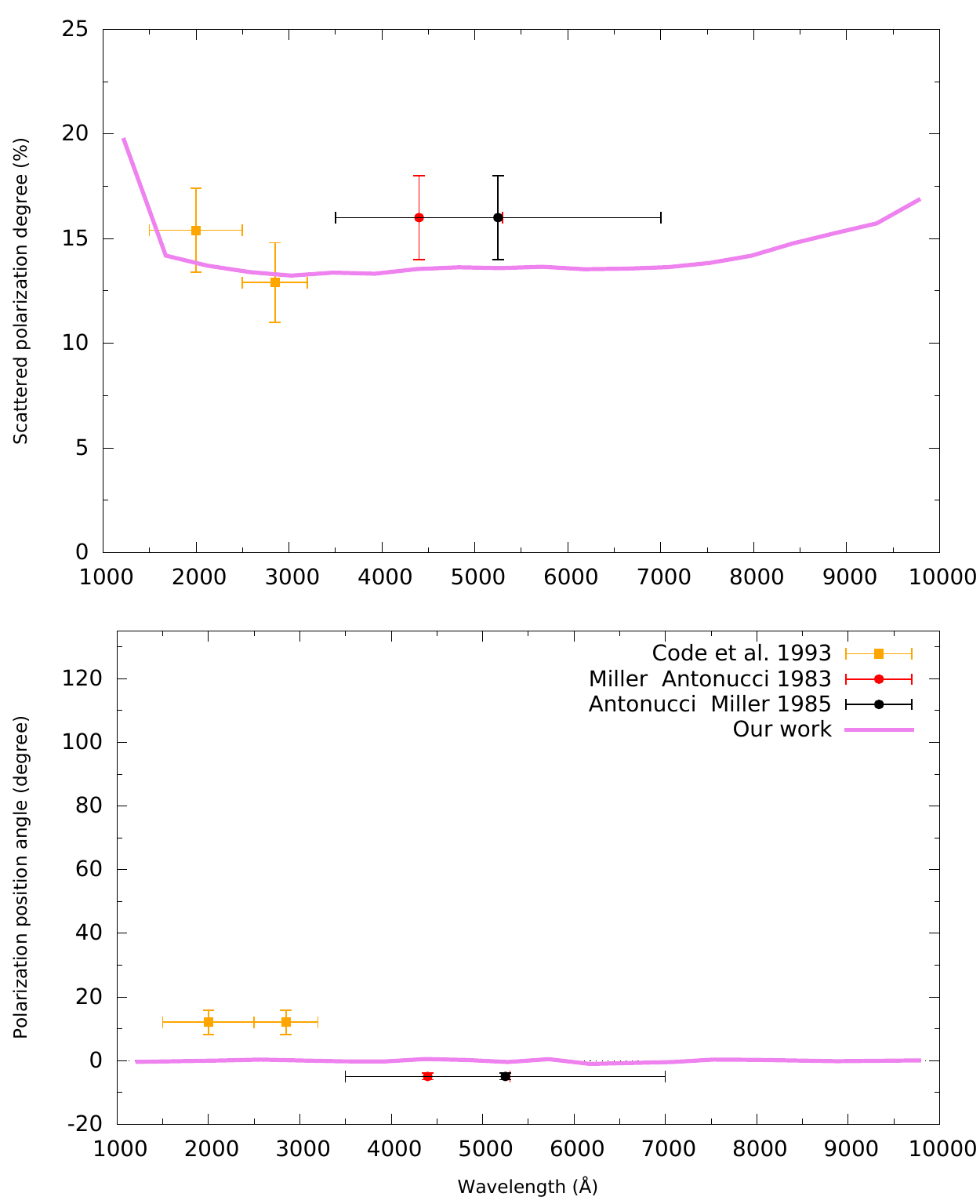}}
   \caption{Same as Fig.~\ref{Fig:POL_Circinus} for the NGC~1068 3D model ($i=60^{\circ}$), including the ultraviolet band.}
  \label{Fig:POL_NGC1068}
\end{figure}

The case of NGC~1068 is more complex. The determination of the inclination angle of the thick disk with different methods
led to different results: kinematical modelling of the NLR bicone yielded $i=85$-$90^{\circ}$ (Das et al. 2006, M\"{u}ller-Sanchez et al. 2011);
MIR interferometry (Lopez-Gonzaga et al. 2014) yielded an axis ratio of the compact component 1 of 7~mas/20~mas = 0.35, which translates to
an inclination angle of $70^{\circ}$ if one assumes that the emission of the MIR interferometry component 1 comes from a thin disk. 
If the emission of the MIR interferometry component 1 stems from the inner rim of the thick gas disk, the inclination angle can be lower, 
i.e. $i \sim 60^{\circ}$. We prefer an inclination angle of the thick gas disk of $i \sim 60^{\circ}$, because this naturally explains the
first two components found by Lopez-Gonzaga et al. (2014): component 1: inner thin disk with puff-up; component 2: inner rim of the thick disk.

The results for NGC~1068 are presented in Fig.~\ref{Fig:POL_NGC1068_angles}. Similarly to our previous polarization modeling, the magenta line 
is the inclination-dependent polarization and the shaded area corresponds to the observed polarization. Regarding the type-1 polarization 
signatures, we find a similar behavior of both the degree and angle of polarization with inclination with respect to the Circinus galaxy 3D 
model; the differences are due to the moderately different geometry and density profile of the polar wind. The polarization position angle is 
equal to 90$^{\circ}$ at polar inclinations, then rotates when the observer's viewing angle coincides with the half-opening angle of the wind. 
At this particular inclination range, the degree of polarization rises up to 0.3\%, which is usual for type-1 AGNs. Once the viewing angle is below 
the wind's lower boundary, obscuration by the extended dust screen covers the signature of the equatorial thin disk and the polarization position angle 
remains 0$^{\circ}$. The transition between type-1 and type-2 signature occurs between $i \sim 45^{\circ}$, a lower angle than that found for 
Circinus. It is only at an inclination of $\sim$~50$^{\circ}$ that the model reproduces both, the observed polarization degree and position angle. 
This value is in agreement with our interferometric results ($i$ = 60$^{\circ}$; Sect.~\ref{sec:application}), yet significantly
different from the inclination derived from the methods that consist of mapping and modeling the radial velocities of the [O\ion{III}] emission 
region in AGN (Das et al. 2006, M\"{u}ller-Sanchez 2011) and of IR SED fitting (H\"{o}nig et al. 2008).

To be consistent with our results from the MIR interferometric observations, we fixed the inclination of the observer to $60^\circ$ 
and plotted in Fig.~\ref{Fig:POL_NGC1068} the wavelength dependence of the ultraviolet, optical and near-infrared polarization properties 
of the NGC~1068 model. Archival Lick~3m and HST polarimetric observations of NGC~1068, corrected for starburst light, are reported on the 
plot (Miller \& Antonucci 1983, Antonucci \& Miller 1985, Code et al. 1993). For an inclination angle of $i=60^{\circ}$ the model is within 
the expected polarization levels and its polarization position angle is similar to what was reported by observations for NGC~1068. The 
wavelength-dependence of the scattered polarization indicates a dust origin and the gradient of the degree of polarization with respect to 
wavelength is due to the dust mixture itself. We used a standard Milky Way composition for the dust grains (Mathis et al. 1977)
but the real mineralogy and size distribution of extragalactic dust grains is poorly constrained. Assuming a different size distribution of 
silicates and graphite would lead to variations of the polarization degree in the UV-optical band. A more rigorous exploration of the 
polarized signal of AGN is thus mandatory to better constrain our dust prescriptions.

Tension still persists between the NLR bicone axis inclination, which translates into a disk inclination of $\sim 80^{\circ}$, and our 
inferred inclination of the inner thick gas disk  of $\sim 60^{\circ}$. In the following we will show that both inclination angles have 
not to be identical. The NLR bicone has a height of $\sim 100$~pc with an outflow velocity of $\sim 1000$~km\,s$^{-1}$ (M\"{u}ller-Sanchez 2011). 
This gives a timescale of $10^{5}$~yr. The dynamical timescale of the inner edge of the thick gas disk which determines the inclination 
angle of the wind is $\Omega=1.5~{\rm pc}/(170~{\rm km/s}) \sim 10^4$~yr (see Table~\ref{tab:input}). Hence, given the different timescales,
the inclination angle of the bicone has not necessarily to be that of the present thick gas disk in such a lively environment.

We conclude that our models of the Circinus galaxy and NGC~1068 are able to reproduce both, the polarization dichotomy between type-1 and type-2 AGN 
and the observed polarization levels. The model is slightly degenerated as variations in the line-of-sight or in optical depth of the wind
will change the final degree of polarization, but the values used here are in very good agreement with observational results from the literature.

\section{Conclusions \label{sec:conclusions}}

Recent IR high-resolution imaging and interferometry showed that the dust distribution is frequently elongated along the polar
direction of an AGN (H\"{o}nig et al. 2012, 2013; Tristram et al. 2014; Lopez-Gonzaga et al. 2014, 2016; Asmus et al. 2016).
In addition, recent interferometric CO(6--5) observations revealed a bipolar outflow in a direction nearly perpendicular to the nuclear disk
(Gallimore et al. 2016). It thus appears that a nuclear molecular and dusty outflow or wind plays an important role for the overall gas
flows in the vicinity ($r < 10$~pc) of the central black hole. 
 
We developed a model scenario for the inner $\sim 30$~pc of an AGN which takes into account the recent observational progress (Fig.~\ref{fig:modelsketch}).
Our view of an AGN is from outside in. The structure of the gas within this region is entirely determined by the gas inflow from larger scales.
We assume a rotating gas disk between about one and ten parsec. External gas accretion adds mass and injects energy via gas compression into this gas disk.
Since all observed gas disks are thick (Davies et al. 2007; Vollmer et al. 2008), we assume that the energy injection via external accretion drives
turbulence. If the energy injection or gas compression timescale is shorter than the turbulent dissipation timescale, the gas compression is adiabatic,
gas clouds are overpressured, star formation is suppressed, and the disk becomes thick (Vollmer \& Davies 2013). 
The behavior of the gas within $\sim 10$~pc is set by the gas mass and the mass accretion rate of the massive thick disk which is located
at radii beyond $\sim 1$~pc. 
Our thick gas disks are assumed to be strongly magnetized via equipartition between the turbulent gas pressure and the energy density of the magnetic field.
In our massive and strongly magnetized thick disks the outflow rate due to magnetocentrifugal forces dominates that due to radiation 
pressure (Sect.~\ref{sec:radpress}).
The strong magnetic field associated with this thick gas disk plays a major role in driving a magnetocentrifugal wind 
(Blandford \& Payne 1982) at a distance of $\sim 1$~pc from the central black hole (Sect.~\ref{sec:rw}). 
Once the wind is launched, it is responsible for the transport of angular momentum and
the gas disk can become thin. A magnetocentrifugal wind is also expected above the thin disk (Sect.~\ref{sec:thindisk}).
Radiation pressure might play a significant role
for the launching of the wind by bending the field lines to an angle $\ga 30^{\circ}$  from the polar axis within the inner edge of the thick disk, 
but is not included in our model.

We identify the thin disk at radii $< 1$~pc with the observed maser disks in AGN (e.g., Greenhill et al. 2003).
The inner edge of the dusty maser disk is determined by the dust sublimation radius (Barvainis 1987). The dust-free gas disk continues right to
the central black hole. The mass accretion rate decreases when the gas approaches the central black hole: it is about $1$~M$_{\odot}$yr$^{-1}$
for the thick gas disk, $< 0.5$~M$_{\odot}$yr$^{-1}$ for the thin disk, and $\sim 0.1$~M$_{\odot}$yr$^{-1}$ for the accretion disk very close to the
central black hole. This implies the existence of a strong BLR wind. The luminosity of the AGN is then set by the mass accretion rate of the
central accretion disk $\dot{M}_{\rm final}$ via $L_{\rm bol}=0.1\,\dot{M}_{\rm final} c^2=6 \times 10^{44} (\dot{M}_{\rm final}/(0.1~{\rm M_{\odot}yr^{-1}}))$~erg\,s$^{-1}$.

We extended the description of a turbulent thick gas disk developed by Vollmer \& Davies (2013) by adding a magnetocentrifugal wind which
starts at a given radius $r_{\rm wind} \simeq 0.5$-$2$~pc (Sect.~\ref{sec:model}). 
The structure and outflow rate of this wind is determined by the properties of the thick gas disk
assuming energy equipartition between the turbulent and magnetic energy densities. We assumed that the outflow rate of the wind above the thin disk is
comparable to that of the wind starting from the thick disk. Since angular momentum of the thin disk is removed by the magnetocentrifugal wind, the
mass accretion rate of the thin disk is directly linked to the wind outflow rate (Eq.~\ref{eq:mdotthin}). In addition, we assume conservation of 
mass flux (Eq.~\ref{eq:mm}). Viable wind models are calculated by varying turbulent velocity dispersion of the thick disk and the wind radii $r_{\rm wind}$
(Sect.~\ref{sec:viable}). 

In a second step, we built three dimensional density cubes based on the analytical model of a thick disk, magnetocentrifugal wind, and thin disk.
We added a puff-up to the thin disk, close to the dust sublimation radius, as observed in young stellar objects. All density distributions are smooth.
The model parameters were adjusted to reproduce available observations of the Circinus~Galaxy and NGC~1068. 
These two AGN bracket the range of local AGNs in terms of black hole mass, rotation velocity, and bolometric luminosity.

We assumed $1/z$ and $1/z^2$ density profiles
for the wind. In addition to the axis-symmetric model, we added a helical wind component to the model cubes (Sect.~\ref{sec:helical}).
These structures were illuminated by a central source (isotropic or $\cos(\theta)$ illumination) of different bolometric luminosities and 
2D radiative transfer calculations were performed. 
We calculated the MIR and NIR luminosities, the central extinctions, spectral
energy distributions, and point source fractions of these model series for varying inclination angles (Sect.~\ref{sec:results}).

In a third step, we calculated MIR visibility amplitudes and compared them to available observations (Sect.~\ref{sec:irinterferometry}).

All models assume smooth gas and dust distributions. We are mainly interested in the thick gas disk and the transition 
between the thick and thin gas disks involving a magnetocentrifugal wind. The detailed geometry of the inner thin gas disk is not subject of this article.

We conclude that within our model assumptions
\begin{enumerate}
\item
magnetocentrifugal winds starting from a thin and thick gas disk are viable in active galaxy centers (Sect.~\ref{sec:viable});
radiation pressure is expected to only play an important role above the inner thin disk making the wind more equatorial there (Sect.~\ref{sec:radpress}),
\item
thick gas disks with high Toomre parameters ($Q \ga 40$) and low mass accretion rates ($\dot{M} \la 0.1$~M$_{\odot}$yr$^{-1}$)
cannot have a magnetocentrifugal wind (e.g., the Circumnuclear Disk in the Galactic Center; Sect.~\ref{sec:viable}),
\item
the outflow scenario can account for the elongated dust structures, outer edges of the thin maser disks, and molecular outflows observed in local AGN;
it helps to decrease the mass accretion rate from the outer thick gas disk to the innermost accretion disk around the central black hole,
\item
the model terminal velocities are consistent with observations (Sect.~\ref{sec:speed}), 
\item
a key ingredient of the model is the transition region between the thick and the thin disk which creates a directly illuminated inner 
wall of the thick gas and dust disk (Sect.~\ref{sec:ddist}; Fig.~\ref{fig:modelcutcircinus}),
\item
based on the comparison between the MIR and NIR luminosities to observations of local AGN (Burtscher et al. 2015), a $\cos(\theta)$ illumination
is preferred over an isotropic illumination,
\item
our model $L_{\rm MIR}/L_{\rm bol}$ ratios are at least a factor of two higher than the ratio derived from observations; it is expected that a slightly different geometry of
the inner wall of the thick gas disk and a clumpy wind decrease $L_{\rm MIR}/L_{\rm bol}$,
\item
about half of the MIR luminosity is emitted by the wind (Sect.~\ref{sec:luminosities}); the inclination angle has thus a minor impact on the MIR luminosity;
this makes the MIR/intrinsic X-ray correlation possible,
\item
the wind or outflow can in principle account for a significant fraction of the central optical extinction (Sect.~\ref{sec:central}), 
\item
for a realistic comparison with observations of Circinus and NGC~1068, 
a puff-up of the inner edge of the thin disk near the dust sublimation radius and a local foreground screen of $A_{\rm V} \sim 20$ are needed;
\item
the IR SEDs of Circinus and NGC~1068 are reproduced by our models in a satisfactory way (Sect.~\ref{sec:sed}); 
the exact form of the SED depends on the structure of the inner thin disk (puff-up, warp, tilt),
\item
the point source fraction of type~2 objects mainly depends on inclination, that of type~1 objects on bolometric luminosity (Sect.~\ref{sec:psf});
a point source fraction $>0.7$ in type~1 AGNs indicates the absence of a wind/outflow; for point source fraction of about one even the thick disk is absent,
\item
our Circinus and NGC~1068 models reproduce available MIR interferometric observations in an acceptable way (Sect.~\ref{sec:irinterferometry}); 
the basic geometry of our model is thus consistent with observations; an asymmetric wind component or a tilted $\cos(\theta)$ illumination
as suggested by Stalevski et al. (2017) is needed for Circinus to better reproduce observations,
\item
the second, disk-like component identified from MIR interferometric observations by Tristram et al. (2014) might 
correspond to the inner wall of the thick gas disk; the first, small-scale component of Lopez-Gonzaga et al. (2014) to the thin maser disk with an inner puff-up,
\item
we derive inclination angles of $i=70^{\circ}$ and $i=60^{\circ}$ for Circinus and NGC~1068, respectively; the inner thin maser disk
are probably tilted/warped with respect to the thick gas disks,
\item
our Circinus and NGC~1068 models are consistent with available optical polarization data. 
\end{enumerate}

Our thick disk, wind, thin disk model is thus a promising scenario for local Seyfert galaxies.
The model is completed by an inner puff-up of the thin disk, as observed in YSOs, and a local foreground screen with $\tau_{\rm V} \sim 20$.
In a subsequent work we will have a look at the differential phases derived from the MIR interferometric observations.
These differential phases represent important additional constraints on the model. Moreover, we plan to go from our
smooth to a clumpy model to investigate the influence of clumpiness on our results.


\begin{acknowledgements}
We would like to thank the anonymous referee for helping to significantly improve this article.
L.B. was supported by the DFG grant within the SPP~1573 ``Physics of the interstellar medium''.
\end{acknowledgements}

\clearpage

\begin{appendix}

\section{Thick disk model without a wind}

\begin{figure}
  \centering
  \resizebox{\hsize}{!}{\includegraphics{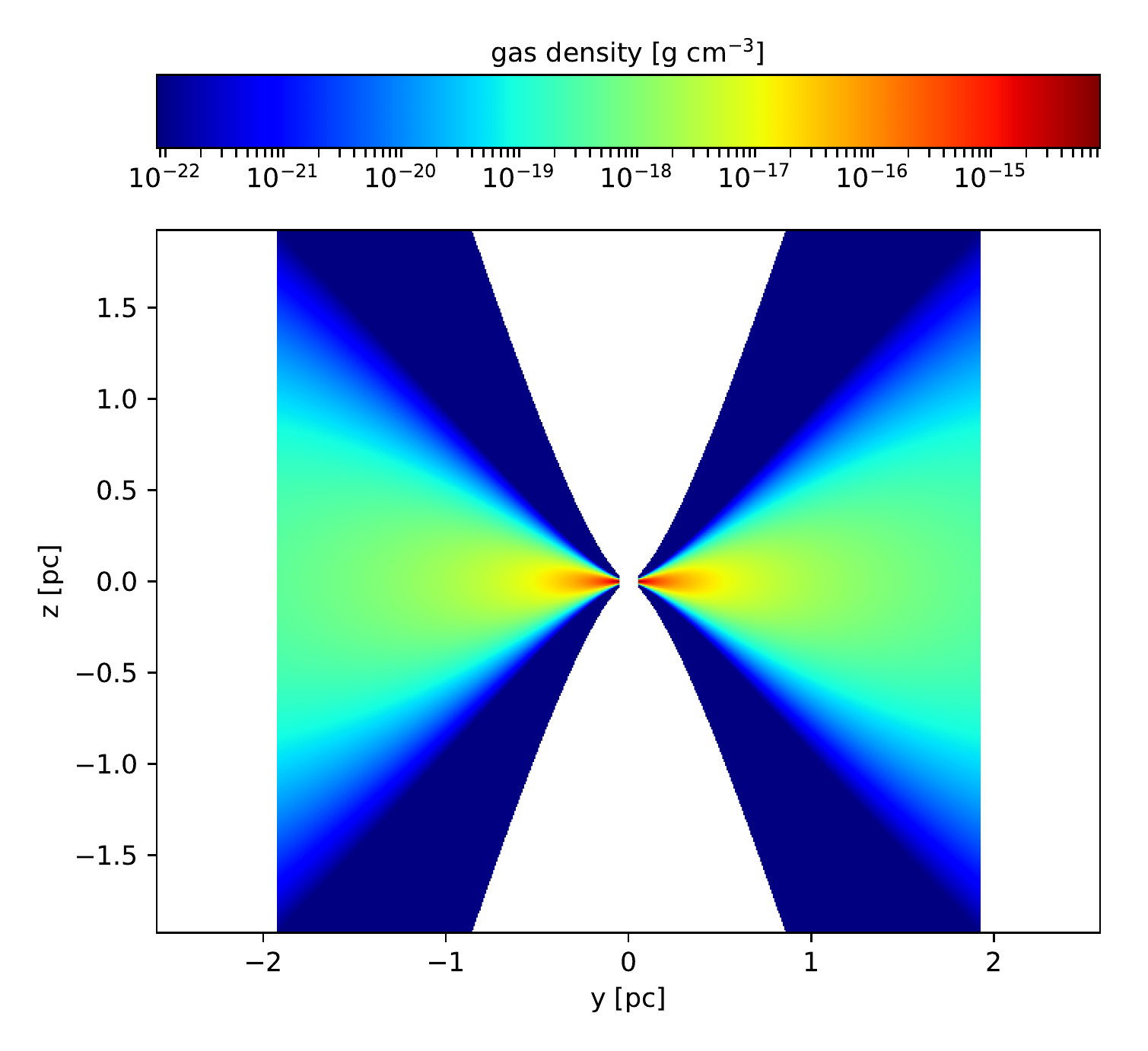}}
  \resizebox{\hsize}{!}{\includegraphics{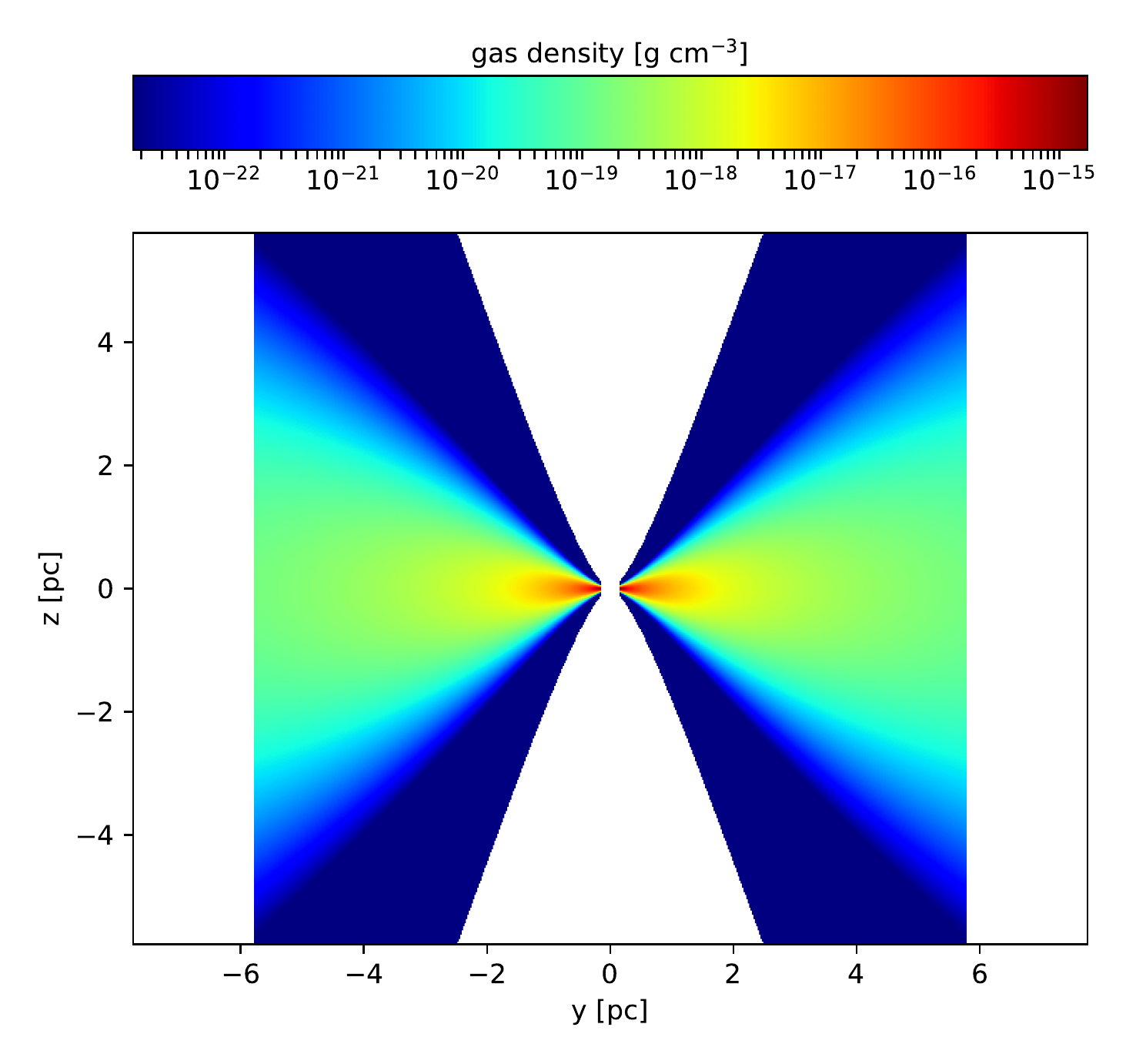}}
  \caption{Thick disk model without a wind. Density cut through the model cube of the Circinus galaxy (upper panel) and NGC~1068 (lower panel). The transfer function is logarithmic.
  \label{fig:modelcuts_nowind}}
\end{figure}

\section{MIR and NIR emission of the $1/z^2$ wind model \label{sec:mirnirz2}}

\begin{figure}
  \centering
  \resizebox{\hsize}{!}{\includegraphics{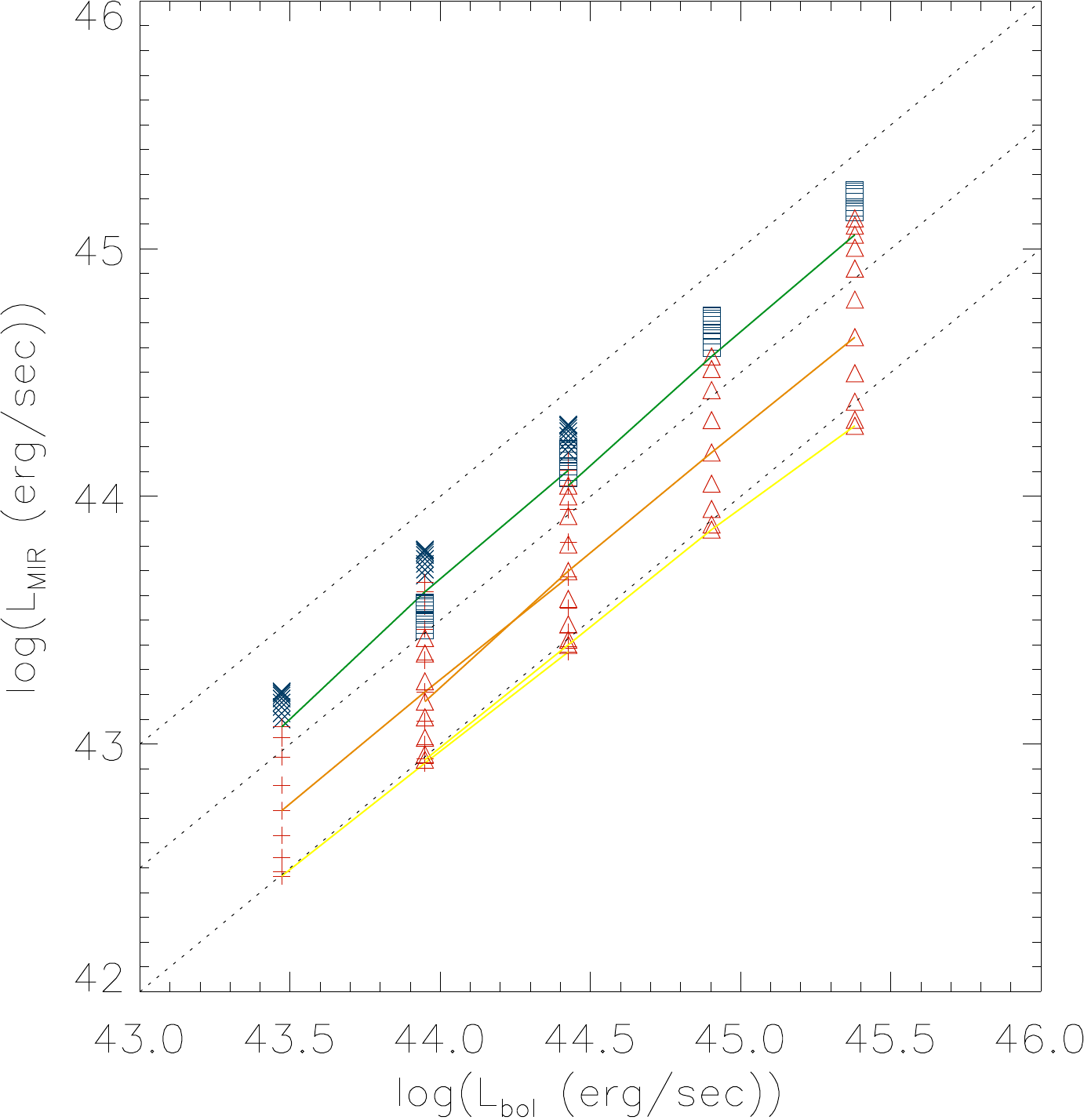}}
  \put(-200,200){\Large $1/z^2$ wind}
  \caption{Model MIR luminosities as a function of the bolometric luminosity. Blue: type~1, red: type~2. Circinus model: 
    crosses/pluses; NGC~1068 model: squares/triangles. The inclination angles of type~2 objects are indicated: $i=50^{\circ}$ (green line),
    $i=70^{\circ}$ (orange line), $i=90^{\circ}$ (yellow line).
  \label{fig:plots_SURVEY1bnew_1a}}
\end{figure}

\begin{figure*}
  \centering
  \resizebox{\hsize}{!}{\includegraphics{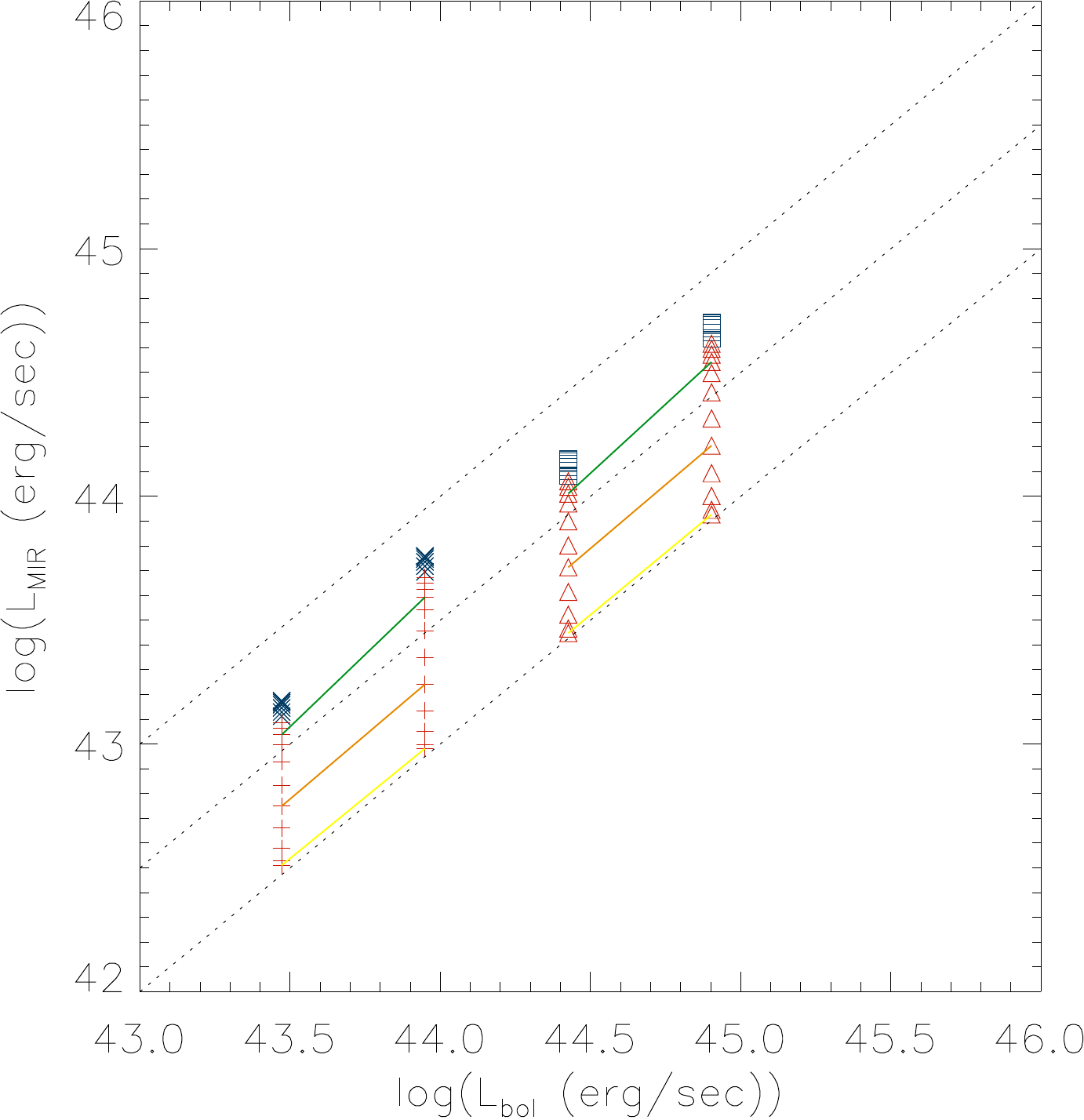}\includegraphics{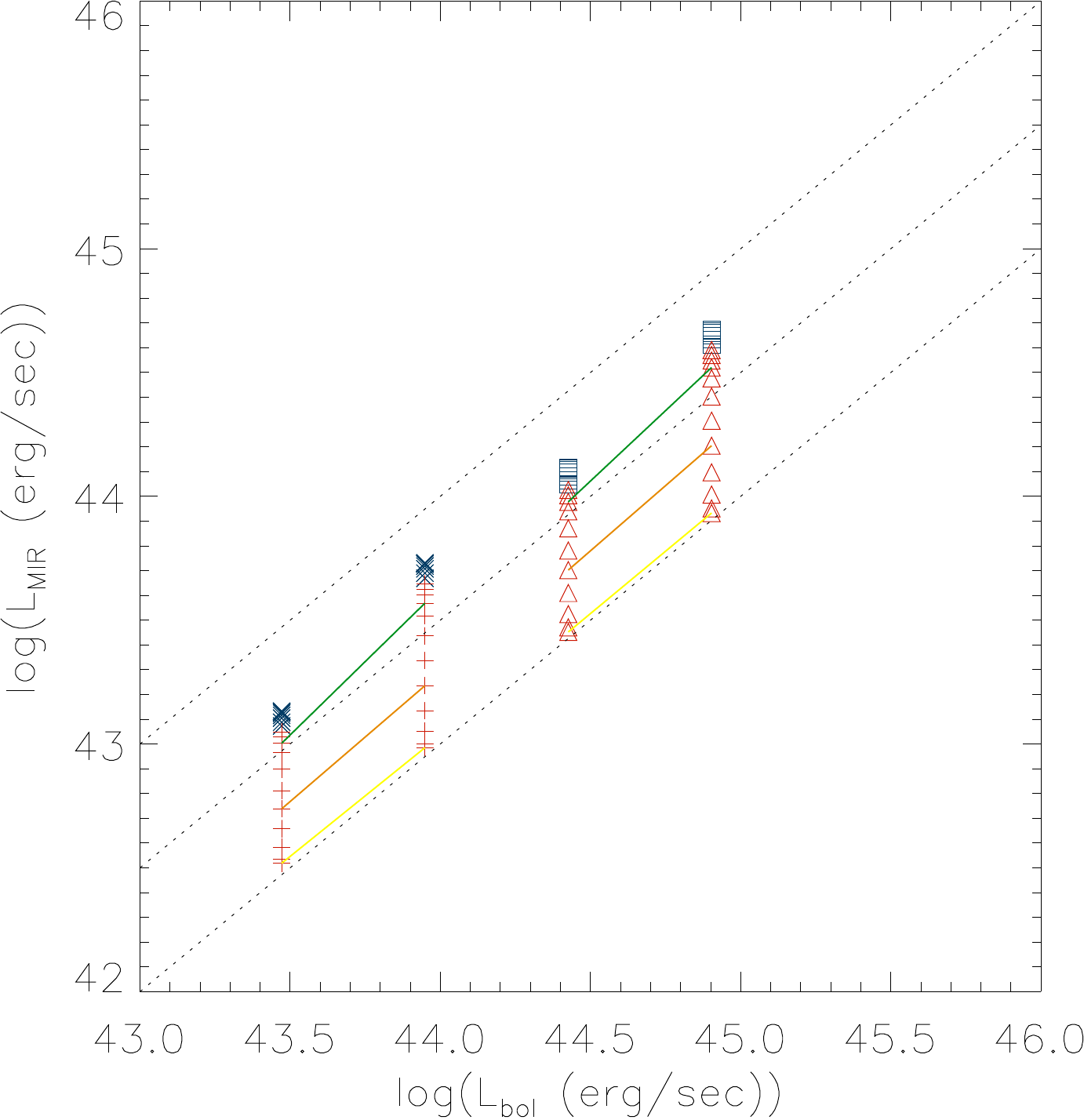}}
  \put(-450,240){\Large inner radius x 1/$\sqrt{2}$}
  \put(-200,240){\Large inner radius x 1/2}
  \caption{Model MIR luminosities as a function of the bolometric luminosity for the $1/z$ winds for a $\sqrt{2}$ (left panel) and $2$ (right panel) 
    times smaller inner radius of the thin gas disk (compare to Fig.~\ref{fig:plots_SURVEY1bnew_1}).
  \label{fig:plots_SURVEY1bnew_1s}}
\end{figure*}

\begin{figure}
  \centering
  \resizebox{\hsize}{!}{\includegraphics{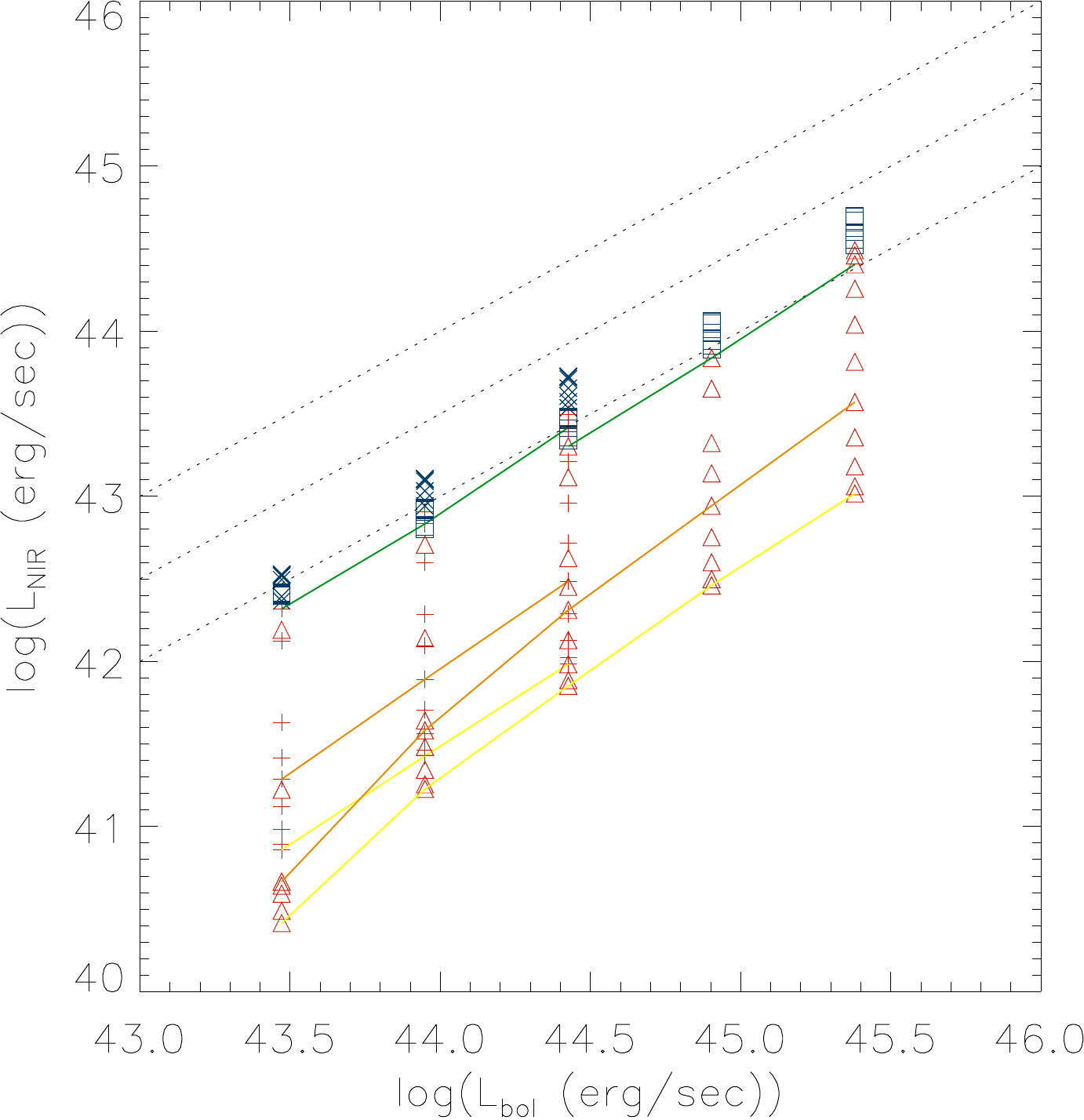}}
  \put(-200,230){\Large $1/z^2$ wind}
  \caption{Model NIR luminosities as a function of the bolometric luminosity. Blue: type~1, red: type~2. Circinus model: 
    crosses/pluses; NGC~1068 model: squares/triangles. The inclination angles of type~2 objects are indicated: $i=50^{\circ}$ (green line),
    $i=70^{\circ}$ (orange line), $i=90^{\circ}$ (yellow line).
  \label{fig:plots_SURVEY1bnew_3a}}
\end{figure}

\begin{figure}
  \centering
  \resizebox{\hsize}{!}{\includegraphics{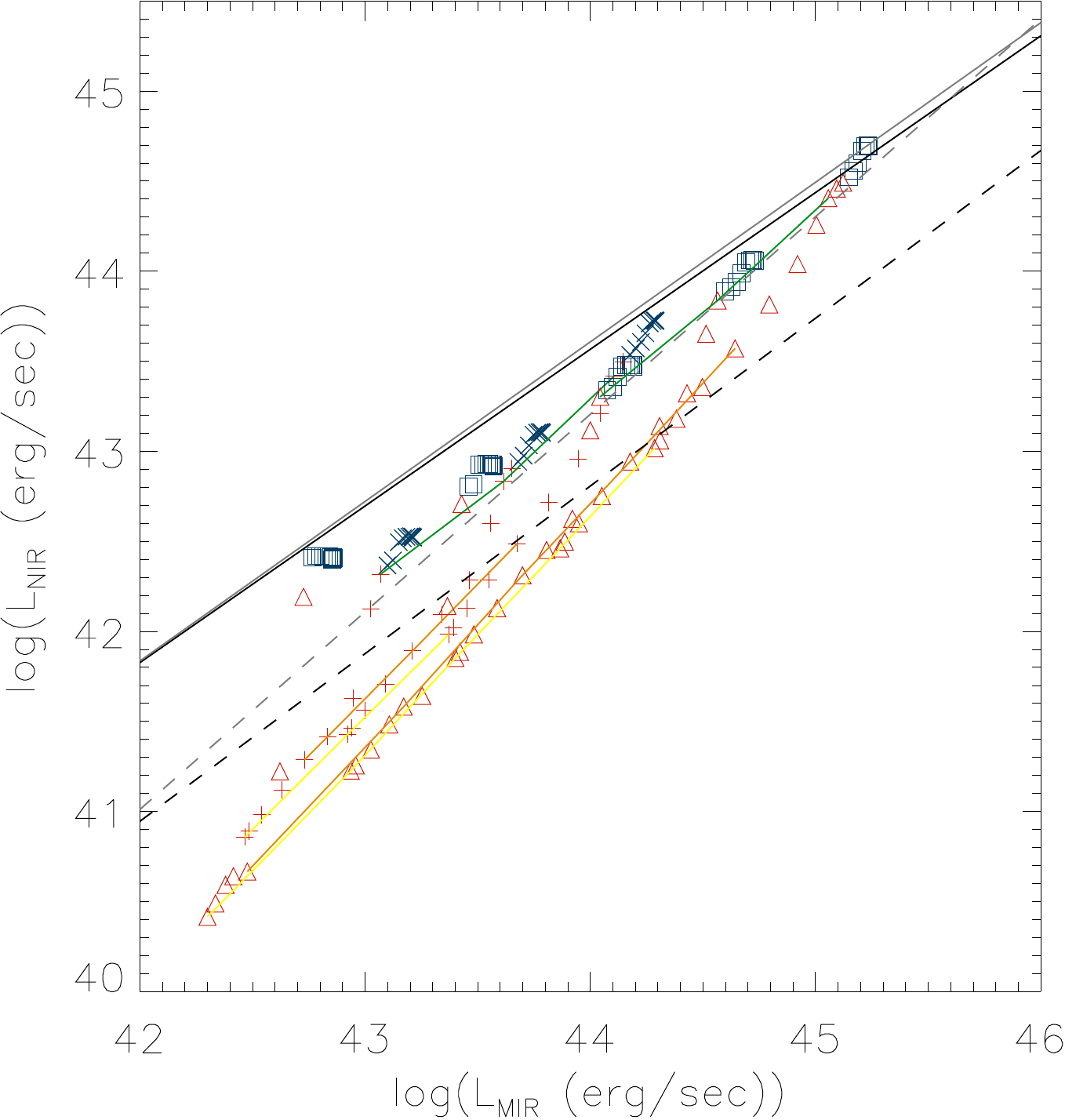}}
  \put(-200,230){\Large $1/z^2$ wind}
  \caption{Model NIR luminosities as a function of the MIR luminosity. Blue: type~1, red: type~2. Circinus model: 
    crosses/pluses; NGC~1068 model: squares/triangles. The inclination angles of type~2 objects are indicated: $i=50^{\circ}$ (green line),
    $i=70^{\circ}$ (orange line), $i=90^{\circ}$ (yellow line).
  \label{fig:plots_SURVEY1bnew_5a}}
\end{figure}

\begin{figure}
  \centering
  \resizebox{\hsize}{!}{\includegraphics{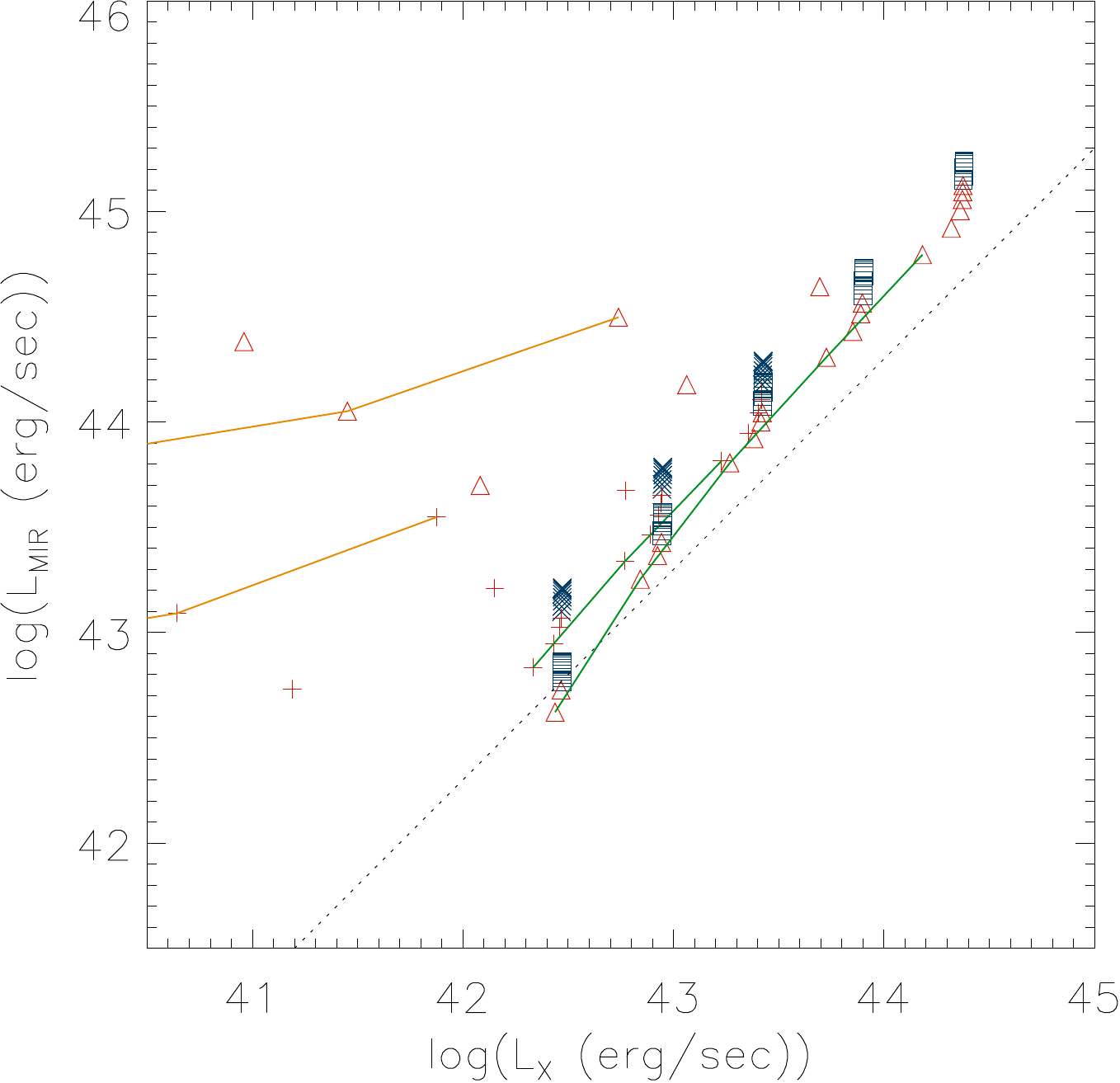}}
  \put(-200,220){\Large $1/z^2$ wind}
  \caption{Model MIR--X-ray correlation. Blue: type~1, red: type~2. Circinus model: 
    crosses/pluses; NGC~1068 model: squares/triangles. The inclination angles of type~2 objects are indicated: $i=65^{\circ}$ (green line),
    $i=75^{\circ}$ (orange line).
    The dotted line corresponds to the correlation found by Asmus et al. (2015): $\log(L_{\rm MIR})=\log(L_{\rm X}) + 0.33$.
  \label{fig:plots_SURVEY1bnew_2a}}
\end{figure}

\begin{figure*}
  \centering
  \resizebox{\hsize}{!}{\includegraphics{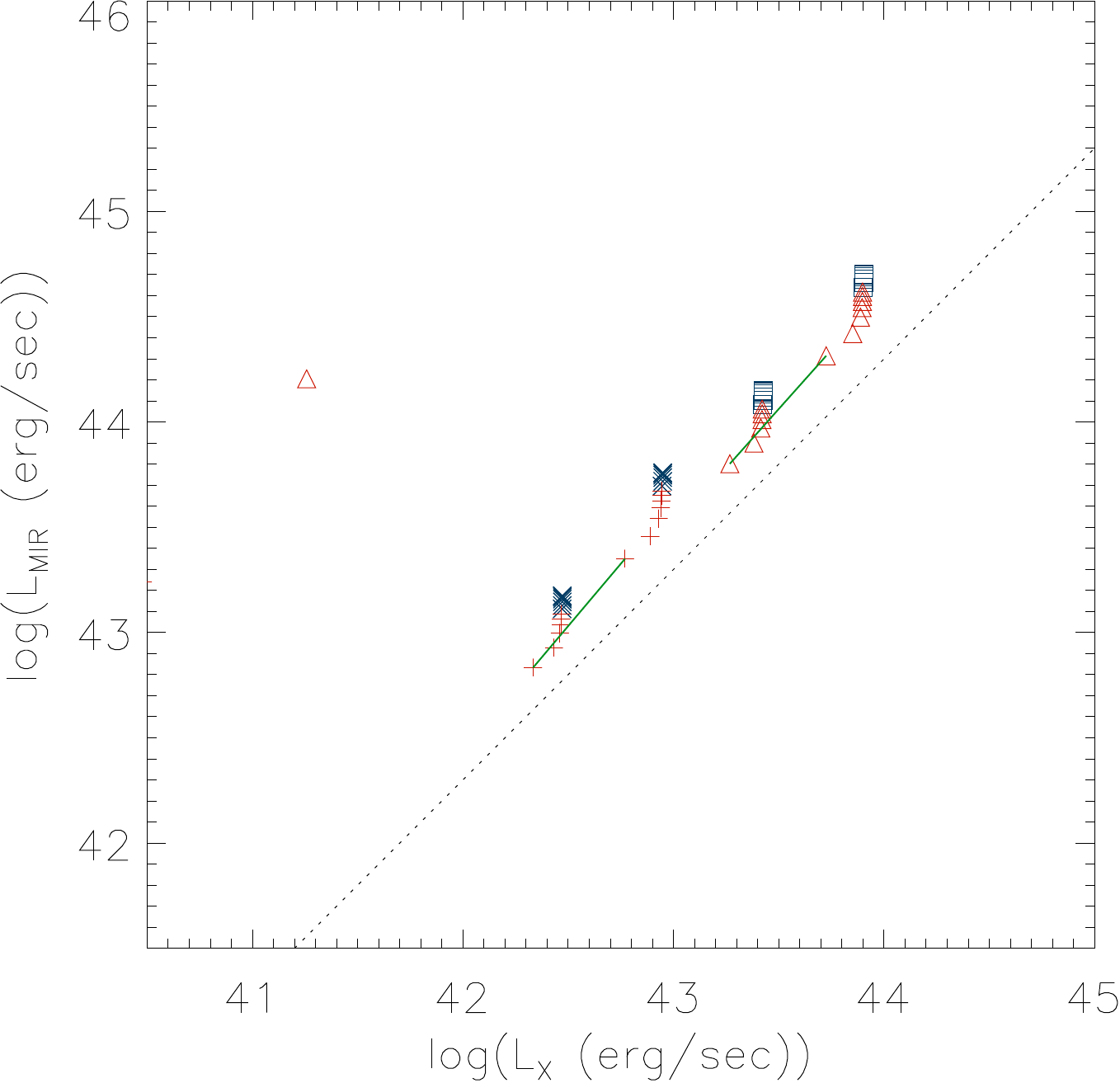}\includegraphics{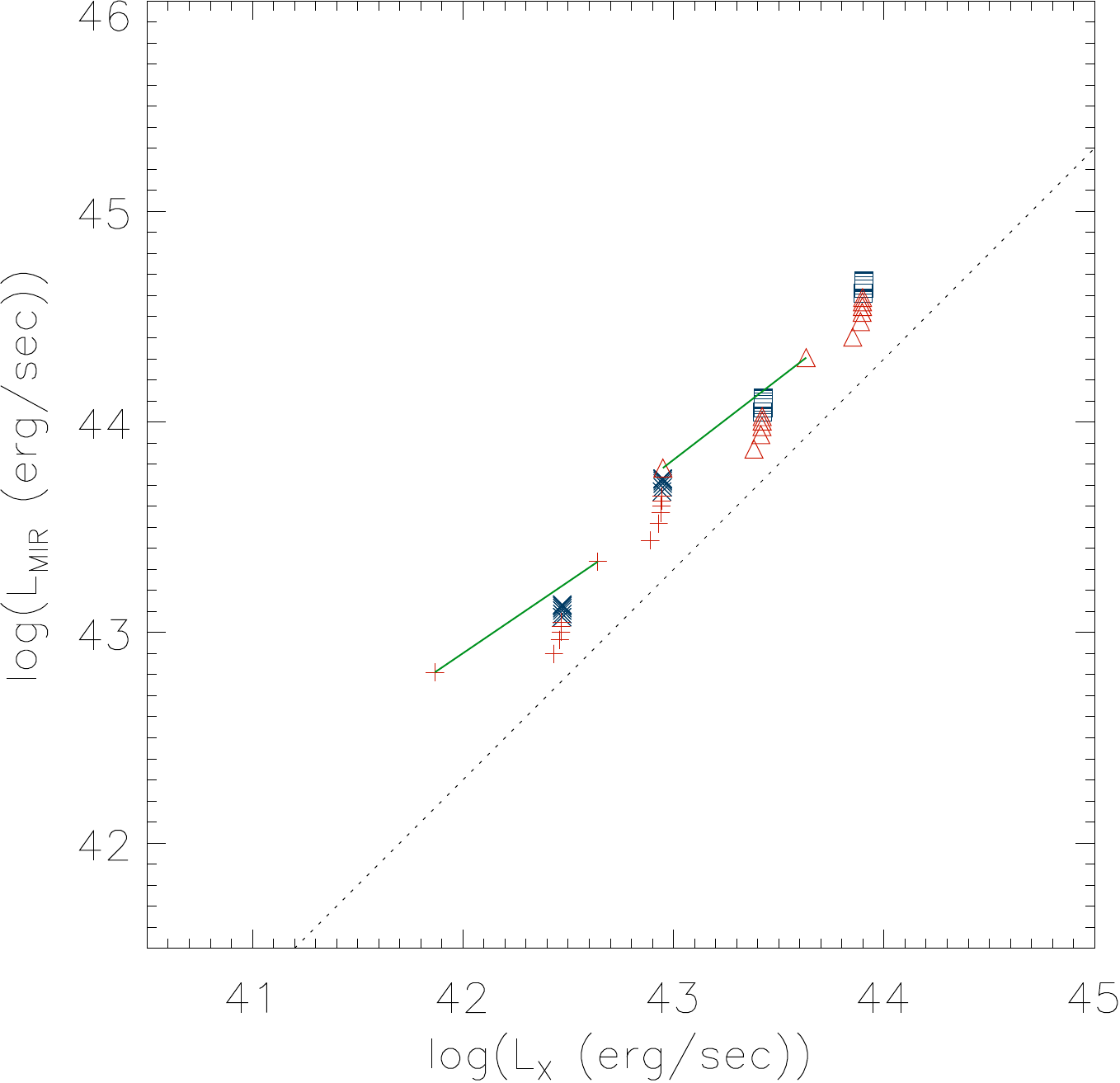}}
  \put(-450,220){\Large inner radius x 1/$\sqrt{2}$}
  \put(-200,220){\Large inner radius x 1/2}
  \caption{Model MIR--X-ray correlation for the $1/z$ winds for a $\sqrt{2}$ (left panel) and $2$ (right panel) 
    times smaller inner radius of the thin gas disk (compare to Fig.~\ref{fig:plots_SURVEY1bnew_2}).
  \label{fig:plots_SURVEY1bnew_2s}}
\end{figure*}

\clearpage

\section{Point source fraction}

\begin{figure*}
  \centering
  \resizebox{\hsize}{!}{\includegraphics{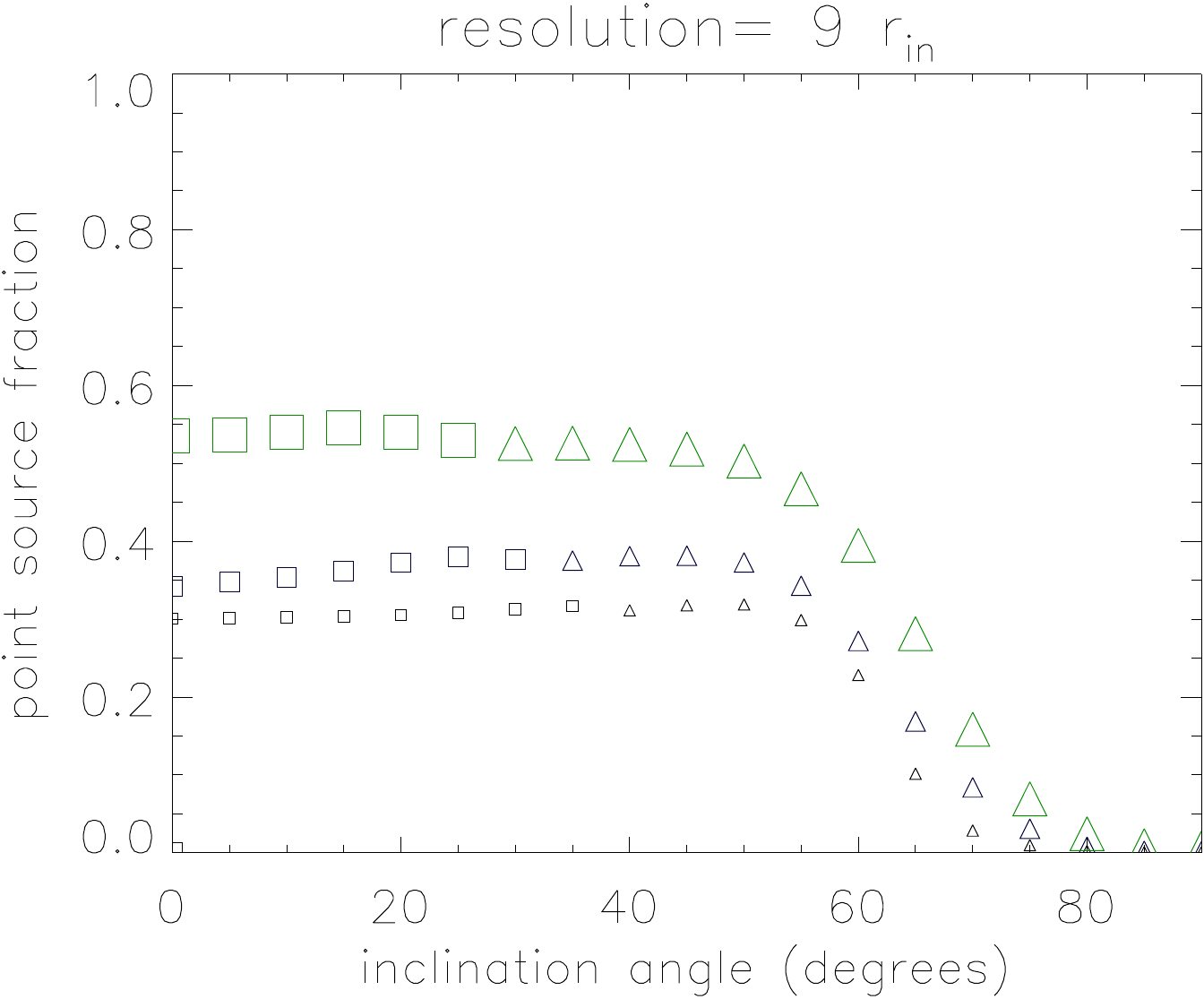}\includegraphics{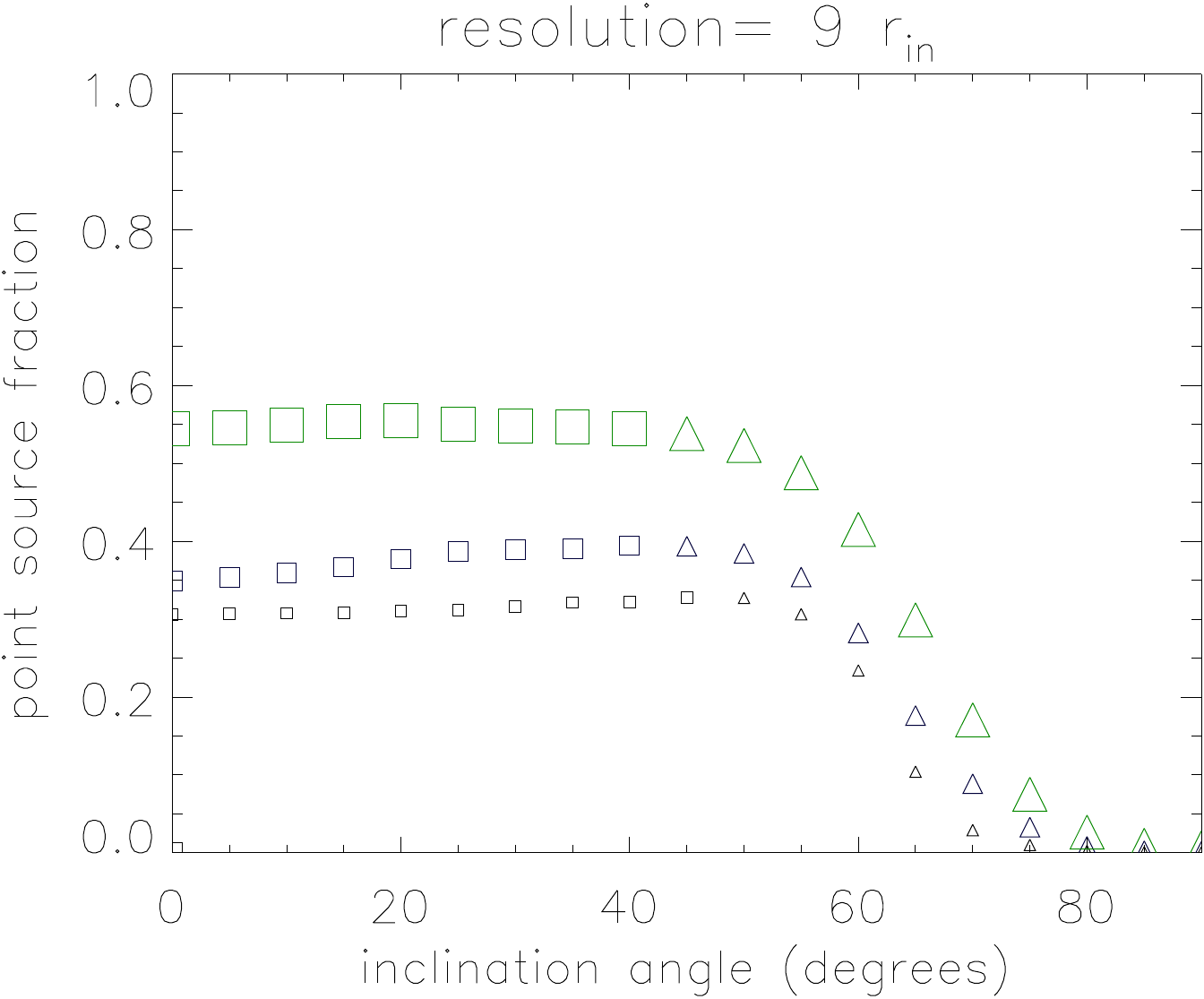}\includegraphics{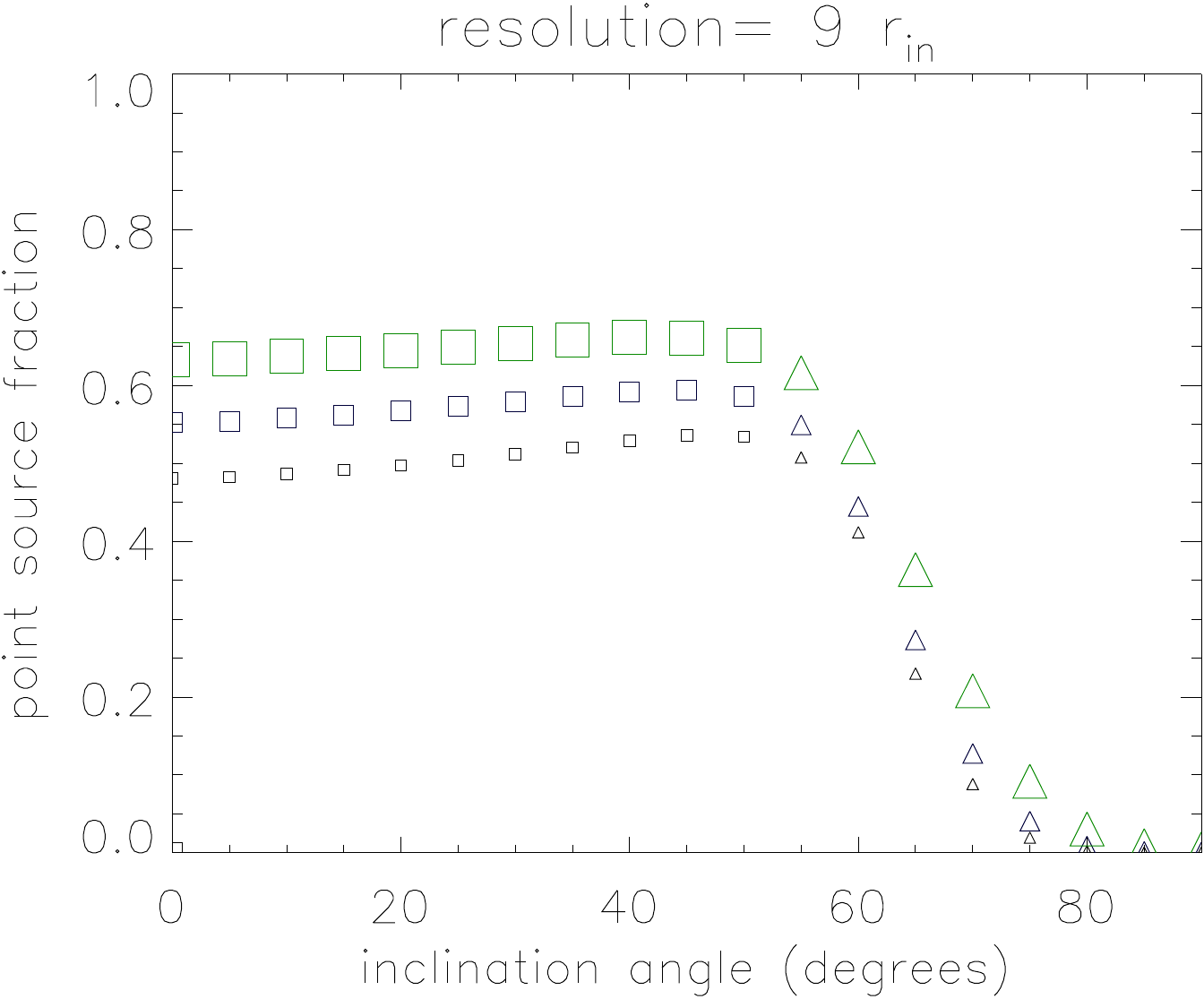}}
  \resizebox{\hsize}{!}{\includegraphics{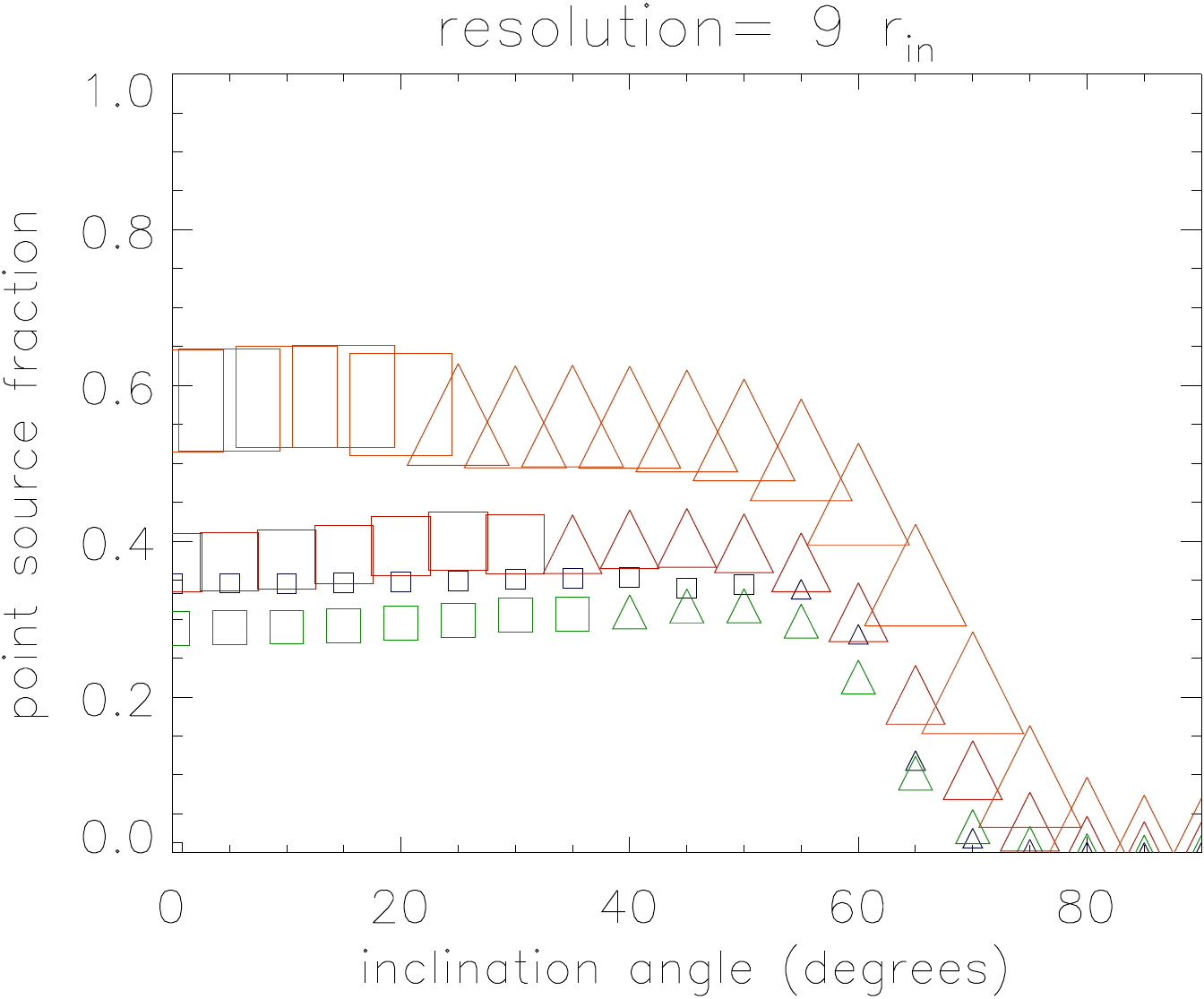}\includegraphics{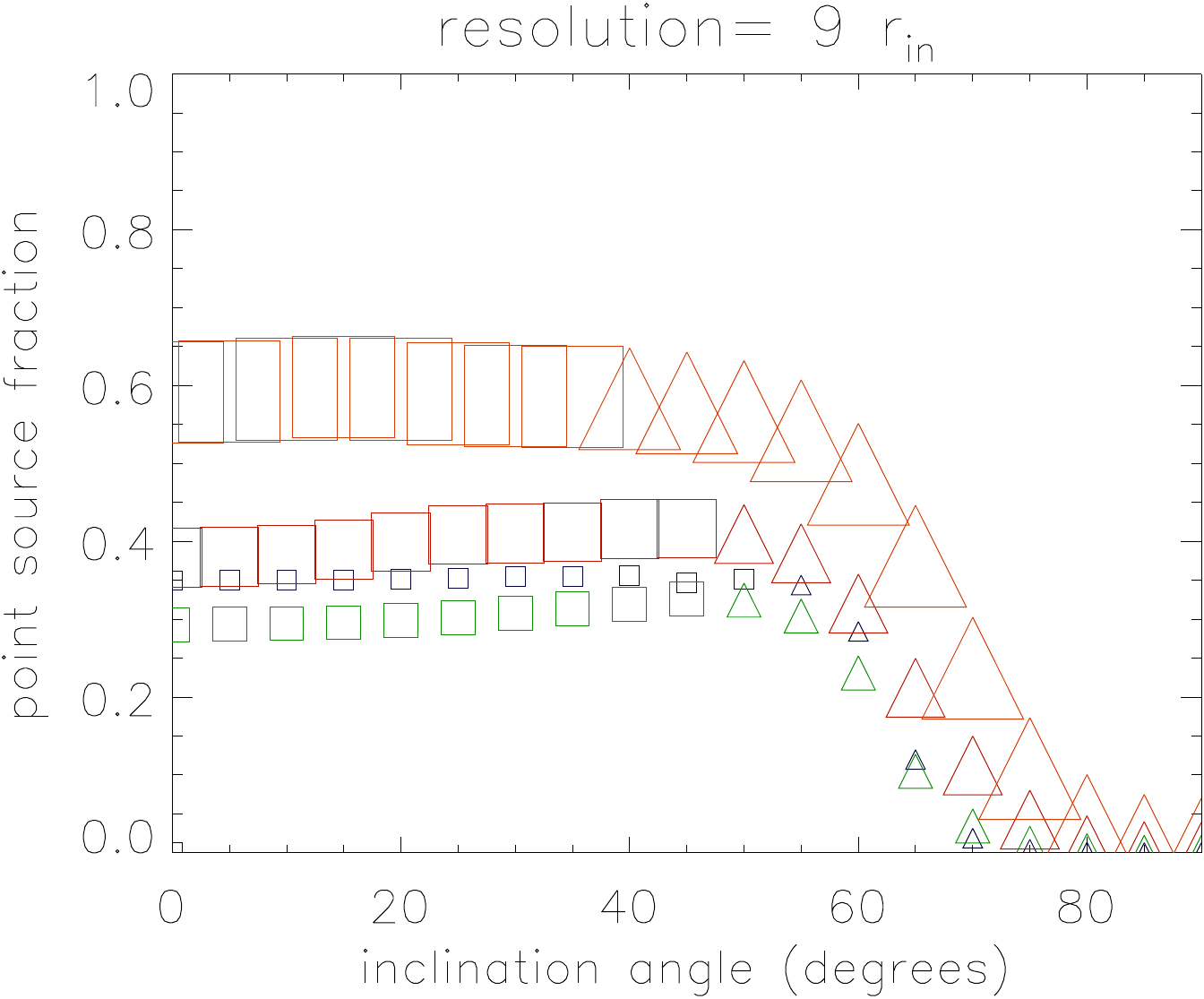}\includegraphics{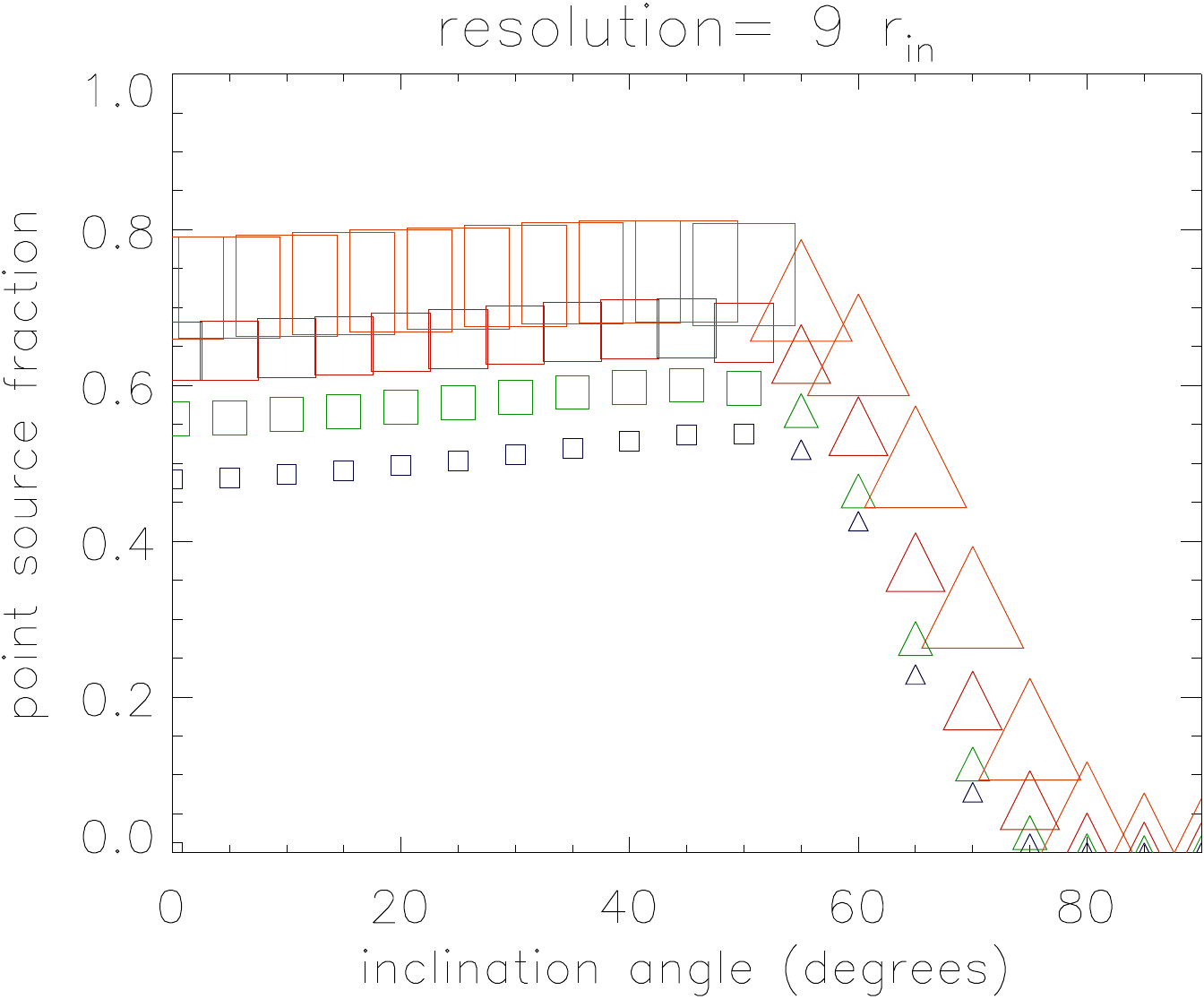}}
  \caption{Point source fraction ($9 r_{\rm in}$) for the Circinus model (upper row) and NGC~1068 model (lower row). Left column: $1/z$ wind (standard model); middle column: $1/z^2$ wind;
    right column: no wind. Boxes: type~1 objects; triangles: type~2 objects. The size of the symbols is proportional to the bolometric luminosity.
  \label{fig:psf16rin}}
\end{figure*}

\section{Model interferometry \label{sec:modelinterferometry}}

\begin{figure*}
  \centering
  \resizebox{\hsize}{!}{\includegraphics{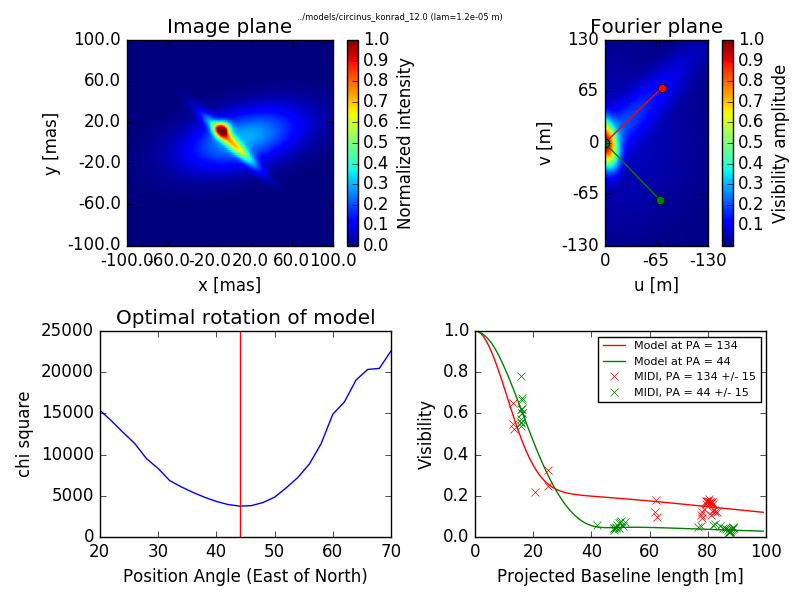}\includegraphics{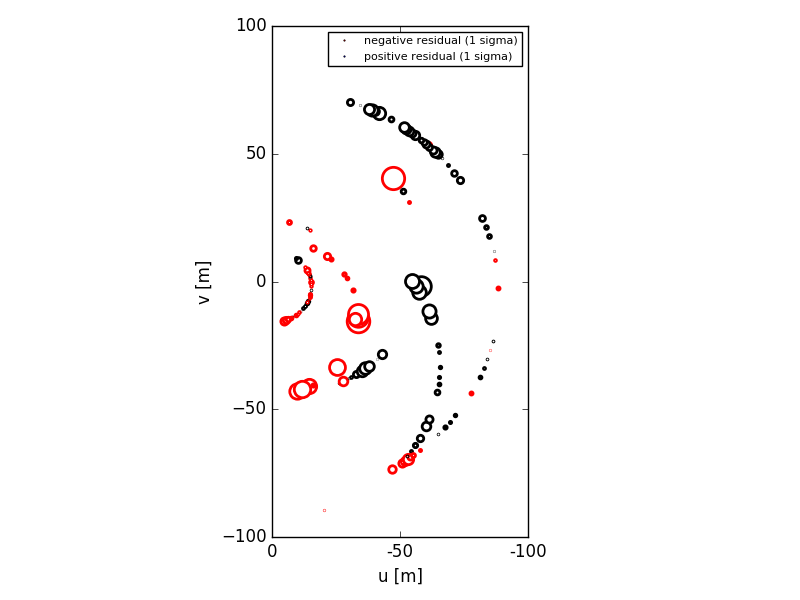}}
  \caption{Tristram et al. (2014) Circinus model. Upper left panel: image at $12~\mu$m; upper middle panel: visibility amplitudes; lower
    left panel: $\chi^2$ as a function of position angle; lower middle panel: visibility amplitudes as a function of projected 
    baseline length (solid lines: model, crosses: observations); right panel: difference between the model and observed 
    visibility amplitudes (red: (model-observations) $> 0$, black: (model-observations) $< 0$). The size of the rings
    is proportional to the value of the residual.
  \label{fig:circinus_konrad_11.0_1.00}}
\end{figure*}

\begin{figure*}
  \centering
  \resizebox{\hsize}{!}{\includegraphics{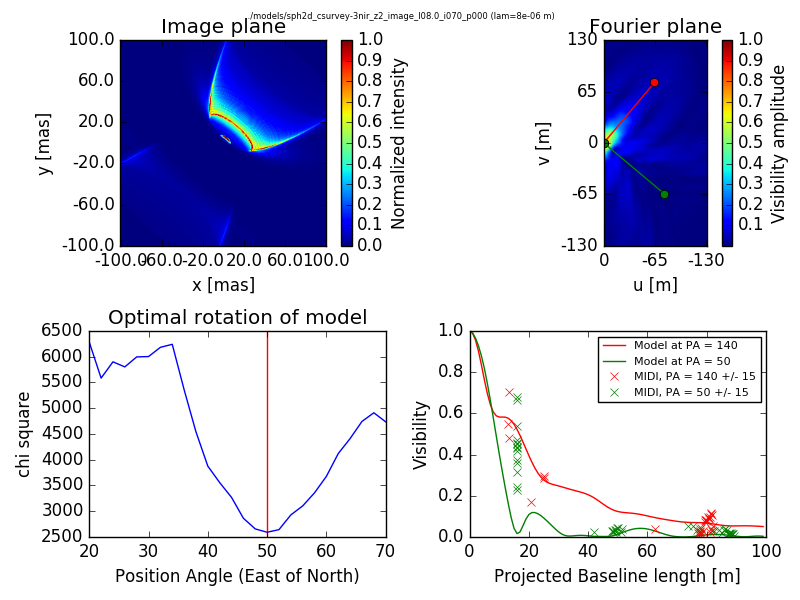}\includegraphics{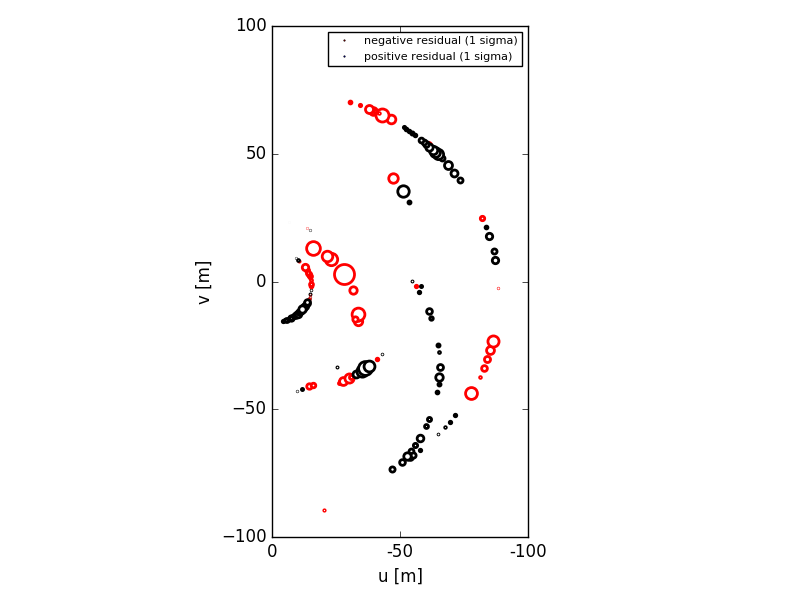}}
  \caption{Same as Fig.~\ref{fig:models_MARC_circinus_image_l12_i065_p000_1.00} for the Circnus radiative transfer model at $8~\mu$m.
  \label{fig:models_MARC_circinus_image_l08_i065_p000_1.00}}
\end{figure*}

\begin{figure*}
  \centering
  \resizebox{\hsize}{!}{\includegraphics{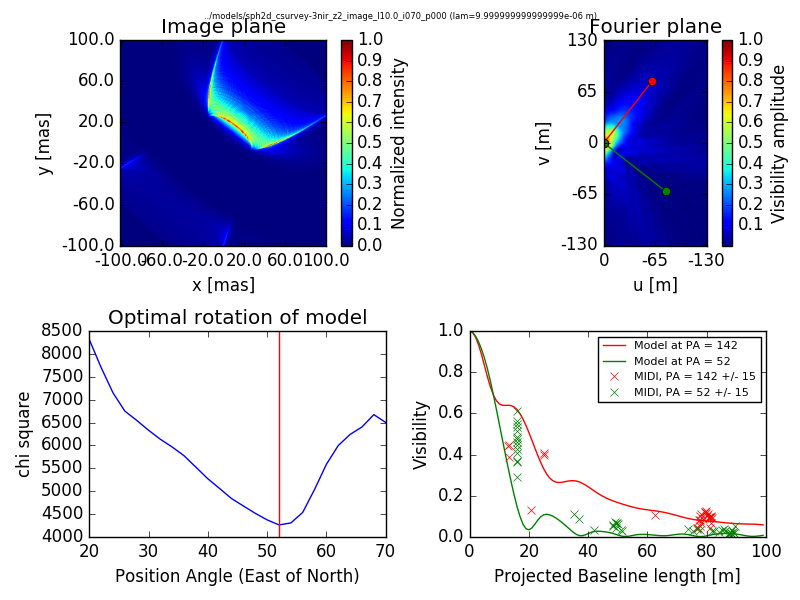}\includegraphics{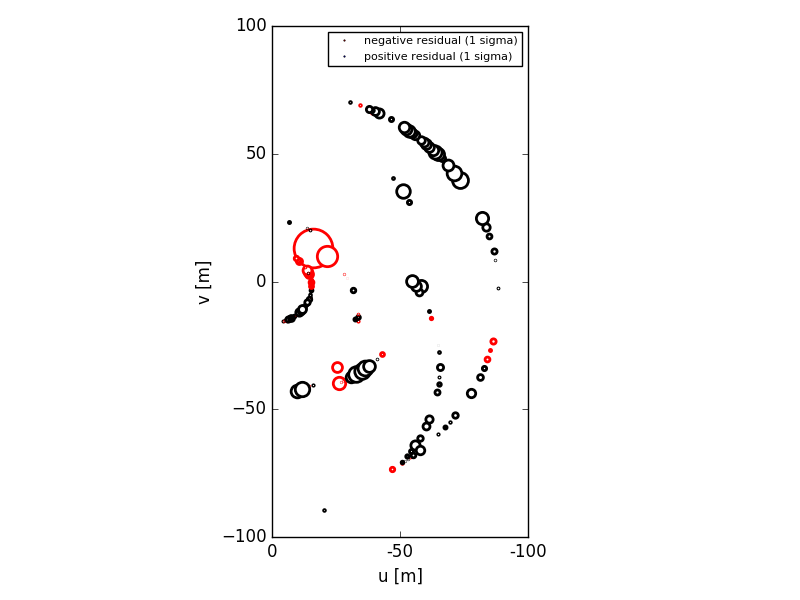}}
  \caption{Same as Fig.~\ref{fig:models_MARC_circinus_image_l12_i065_p000_1.00} for the Circnus radiative transfer model at $10~\mu$m.
  \label{fig:models_MARC_circinus_image_l10_i065_p000_1.00}}
\end{figure*}

\begin{figure*}
  \centering
  \resizebox{\hsize}{!}{\includegraphics{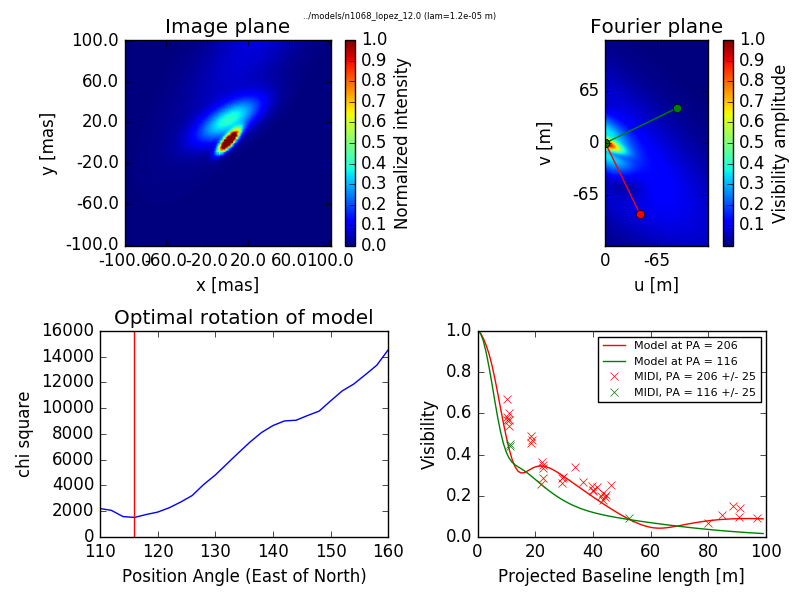}\includegraphics{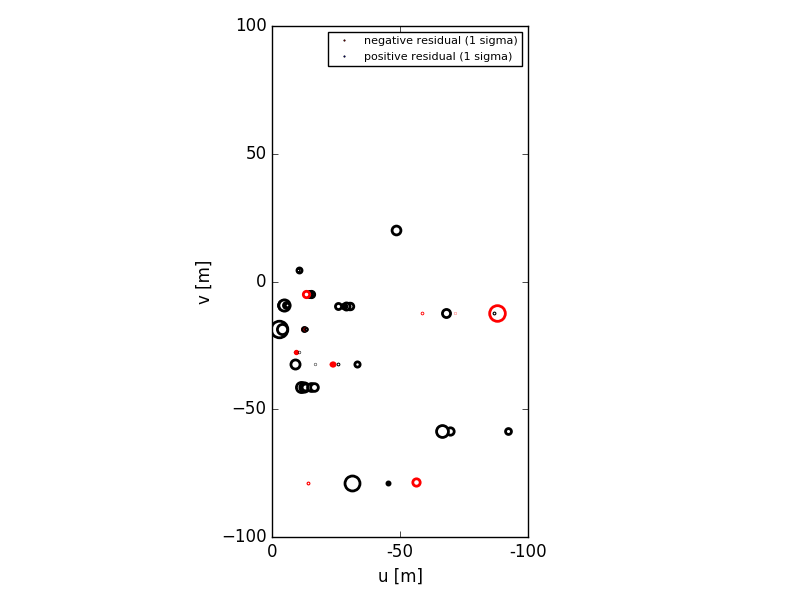}}
  \caption{NGC~1068. Same as Fig.~\ref{fig:circinus_konrad_11.0_1.00} for the Lopez-Gonzaga et al. (2014) NGC~1068 model. All three components are
    included. The third, extended component is barely visible in the left panel, but contributes to the visibilities at short baselines.
  \label{fig:n1068_lopez_12.0_1.00}}
\end{figure*}

\begin{figure*}
  \centering
  \resizebox{\hsize}{!}{\includegraphics{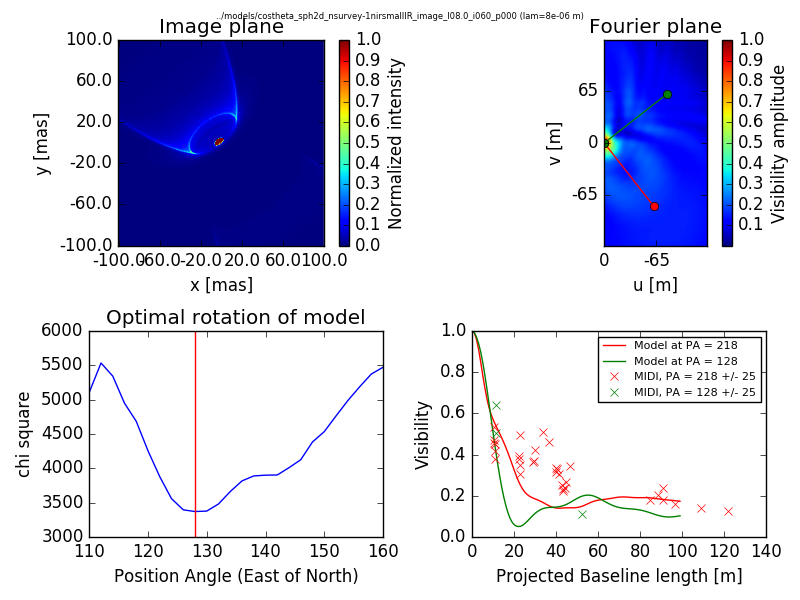}\includegraphics{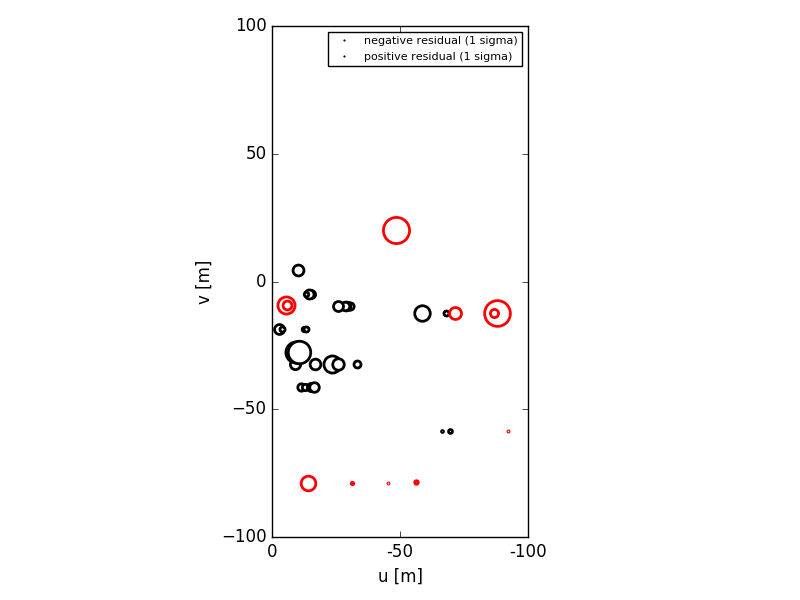}}
  \caption{Same as Fig.~\ref{fig:costheta_sph2d_nsurvey-1nirsmallIR_image_l12.0_i060_p000_1.25} for the NGC~1068 radiative transfer model at $8~\mu$m. 
    The puff-up of the inner thin disk is located at $r=3 \times r_{\rm sub}$.
  \label{fig:costheta_sph2d_nsurvey-1nirsmallIR_image_l08.0_i060_p000_1.25}}
\end{figure*}

\begin{figure*}
  \centering
  \resizebox{\hsize}{!}{\includegraphics{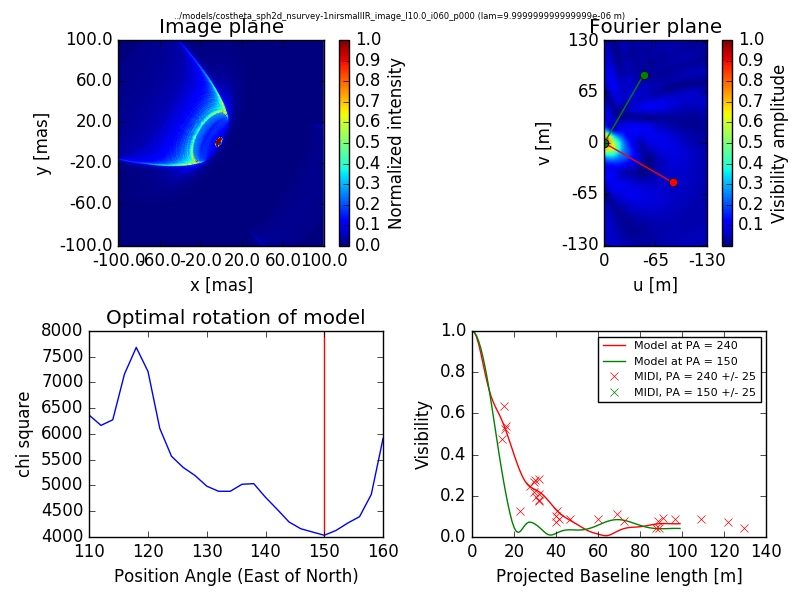}\includegraphics{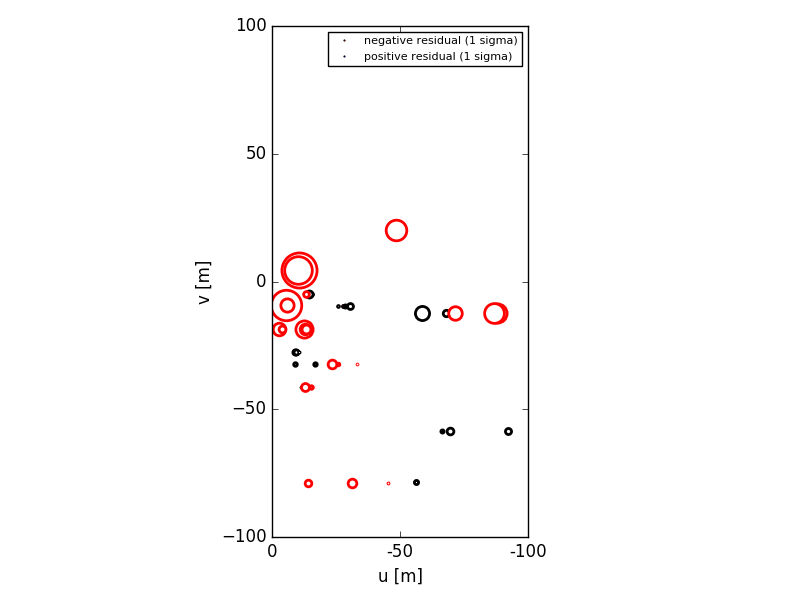}}
  \caption{Same as Fig.~\ref{fig:costheta_sph2d_nsurvey-1nirsmallIR_image_l12.0_i060_p000_1.25} for the NGC~1068 radiative transfer model at $10~\mu$m. 
    The puff-up of the inner thin disk is located at $r=3 \times r_{\rm sub}$. There are no observed MIDI visibilities for a position angle of $150^{\circ}$.
  \label{fig:costheta_sph2d_nsurvey-1nirsmallIR_image_l10.0_i060_p000_1.25}}
\end{figure*}

\end{appendix}


\begin{thebibliography}{}
 

\bibitem[Alexander et al.(2000)]{2000MNRAS.313..815A} Alexander, D.~M., Heisler, C.~A., Young, S., et al.\ 2000, MNRAS, 313, 815 

\bibitem[Antonucci \& Miller(1985)]{1985ApJ...297..621A} Antonucci, R.~R.~J., \& Miller, J.~S.\ 1985, ApJ, 297, 621

\bibitem[Antonucci(1993)]{1993ARA&A..31..473A} Antonucci, R.\ 1993, ARA\&A, 31, 473 

\bibitem[Asmus et al.(2015)]{2015MNRAS.454..766A} Asmus, D., Gandhi, P., H{\"o}nig, S.~F., Smette, A., \& Duschl, W.~J.\ 2015, MNRAS, 454, 766 
\bibitem[Asmus et al.(2016)]{2016ApJ...822..109A} Asmus, D., H{\"o}nig, S.~F., \& Gandhi, P.\ 2016, ApJ, 822, 109 

\bibitem[Barvainis(1987)]{1987ApJ...320..537B} Barvainis, R.\ 1987, ApJ, 320, 537 

\bibitem[Beck(2015)]{2016A&ARv..24....4B} Beck, R.\ 2015, A\&ARv, 24, 4 

\bibitem[Bjorkman \& Wood(2001)]{2001ApJ...554..615B} Bjorkman, J.~E., \& Wood, K.\ 2001, ApJ, 554, 615 

\bibitem[Blandford \& Payne(1982)]{1982MNRAS.199..883B} Blandford, R.~D., \& Payne, D.~G.\ 1982, MNRAS, 199, 883 

\bibitem[Blandford et al.(1990)]{1990agn..conf.....B} Blandford, R.~D., Netzer, H., Woltjer, L., Courvoisier, T.~J.-L., \& Mayor, M.\ 1990, Active Galactic Nuclei, 97

\bibitem[Burtscher et al.(2013)]{2013A&A...558A.149B} Burtscher, L., Meisenheimer, K., Tristram, K.~R.~W., et al.\ 2013, A\&A, 558, A149 

\bibitem[Burtscher et al.(2015)]{2015A&A...578A..47B} Burtscher, L., Orban de Xivry, G., Davies, R.~I., et al.\ 2015, A\&A, 578, A47 

\bibitem[Burtscher et al.(2016)]{2016arXiv160704533B} Burtscher, L., H{\"o}nig, S., Jaffe, W., et al.\ 2016, arXiv:1607.04533 

\bibitem[Chan \& Krolik(2017)]{2017ApJ...843...58C} Chan, C.-H., \& Krolik, J.~H.\ 2017, ApJ, 843, 58 

\bibitem[Chan \& Krolik(2016)]{2016ApJ...825...67C} Chan, C.-H., \& Krolik, J.~H.\ 2016, ApJ, 825, 67 

\bibitem[Code et al.(1993)]{1993ApJ...403L..63C} Code, A.~D., Meade, M.~R., Anderson, C.~M., et al.\ 1993, ApJL, 403, L63 

\bibitem[Das et al.(2006)]{2006AJ....132..620D} Das, V., Crenshaw, D.~M., Kraemer, S.~B., \& Deo, R.~P.\ 2006, AJ, 132, 620 

\bibitem[Davies et al.(2007)]{2007ApJ...671.1388D} Davies, R.~I., M{\"u}ller S{\'a}nchez, F., Genzel, R., et al.\ 2007, ApJ, 671, 1388 

\bibitem{a16a} Dobbs, C.~L., \& Pringle, J.~E.\ 2013, MNRAS, 1187 

\bibitem[Dorodnitsyn et al.(2016)]{2016ApJ...819..115D} Dorodnitsyn, A., Kallman, T., \& Proga, D.\ 2016, ApJ, 819, 115 

\bibitem[Draine \& Lee(1984)]{1984ApJ...285...89D} Draine, B.~T., \& Lee, H.~M.\ 1984, ApJ, 285, 89 

\bibitem[Draine(2003)]{2003ApJ...598.1026D} Draine, B.~T.\ 2003, ApJ, 598, 1026 

\bibitem[Draine et al.(2007)]{2007ApJ...663..866D} Draine, B.~T., Dale, D.~A., Bendo, G., et al.\ 2007, ApJ, 663, 866 

\bibitem[Dullemond et al.(2001)]{2001ApJ...560..957D} Dullemond, C.~P., Dominik, C., \& Natta, A.\ 2001, ApJ, 560, 957 

\bibitem[Dullemond(2012)]{2012ascl.soft02015D} Dullemond, C.~P.\ 2012, Astrophysics Source Code Library, ascl:1202.015 

\bibitem[Elitzur(2006)]{2006NewAR..50..728E} Elitzur, M.\ 2006, NewAR, 50, 728 

\bibitem[Elitzur(2012)]{2012ApJ...747L..33E} Elitzur, M.\ 2012, ApJL, 747, L33 

\bibitem[Elvis(2000)]{2000ApJ...545...63E} Elvis, M.\ 2000, ApJ, 545, 63 

\bibitem[Everett(2005)]{2005ApJ...631..689E} Everett, J.~E.\ 2005, ApJ, 631, 689 

\bibitem[Feruglio et al.(2010)]{2010A&A...518L.155F} Feruglio, C., Maiolino, R., Piconcelli, E., et al.\ 2010, A\&A, 518, L155 

\bibitem[Fischer et al.(2013)]{2013ApJS..209....1F} Fischer, T.~C., Crenshaw, D.~M., Kraemer, S.~B., \& Schmitt, H.~R.\ 2013, ApJS, 209, 1 

\bibitem[Fleck \& Canfield(1984)]{1984JCoPh..54..508F} Fleck, J.~A., Jr., \& Canfield, E.~H.\ 1984, Journal of Computational Physics, 54, 508 

\bibitem[Gallimore et al.(2016)]{2016ApJ...829L...7G} Gallimore, J.~F., Elitzur, M., Maiolino, R., et al.\ 2016, ApJL, 829, L7 

\bibitem[Gandhi et al.(2009)]{2009A&A...502..457G} Gandhi, P., Horst, H., Smette, A., et al.\ 2009, A\&A, 502, 457 

\bibitem[Garc{\'{\i}}a-Burillo et al.(2016)]{2016ApJ...823L..12G} Garc{\'{\i}}a-Burillo, S., Combes, F., Ramos Almeida, C., et al.\ 2016, ApJL, 823, L12 

\bibitem[Gaskell(2009)]{2009NewAR..53..140G} Gaskell, C.~M.\ 2009, NewAR, 53, 140 

\bibitem[Goosmann \& Gaskell(2007)]{2007A&A...465..129G} Goosmann, R.~W., \& Gaskell, C.~M.\ 2007, A\&A, 465, 129 

\bibitem[Greenhill et al.(1995)]{1995A&A...304...21G} Greenhill, L.~J., Henkel, C., Becker, R., Wilson, T.~L., \& Wouterloot, J.~G.~A.\ 1995, A\&A, 304, 21 

\bibitem[Greenhill et al.(1996)]{1996ApJ...472L..21G} Greenhill, L.~J., Gwinn, C.~R., Antonucci, R., \& Barvainis, R.\ 1996, ApJL, 472, L21 

\bibitem[Greenhill(1998)]{1998ASPC..144..221G} Greenhill, L.~J.\ 1998, IAU Colloq.~164: Radio Emission from Galactic and Extragalactic Compact Sources, 144, 221 

\bibitem[Greenhill et al.(2003)]{2003ApJ...590..162G} Greenhill, L.~J., Booth, R.~S., Ellingsen, S.~P., et al.\ 2003, ApJ, 590, 162 

\bibitem{a27} G\"{u}sten R., Genzel R., Wright M.C.H. et al., 1987, ApJ 318, 124

\bibitem[Hicks et al.(2009)]{2009ApJ...696..448H} Hicks, E.~K.~S., Davies, R.~I., Malkan, M.~A., et al.\ 2009, ApJ, 696, 448 

\bibitem[H{\"o}nig et al.(2008)]{2008A&A...485...33H} H{\"o}nig, S.~F., Prieto, M.~A., \& Beckert, T.\ 2008, A\&A, 485, 33 

\bibitem[H{\"o}nig \& Kishimoto(2010)]{2010A&A...523A..27H} H{\"o}nig, S.~F., \& Kishimoto, M.\ 2010, A\&A, 523, A27 

\bibitem[H{\"o}nig et al.(2011)]{2011ApJ...736...26H} H{\"o}nig, S.~F., Leipski, C., Antonucci, R., \& Haas, M.\ 2011, ApJ, 736, 26 

\bibitem[H{\"o}nig et al.(2012)]{2012ApJ...755..149H} H{\"o}nig, S.~F., Kishimoto, M., Antonucci, R., et al.\ 2012, ApJ, 755, 149 

\bibitem[H{\"o}nig et al.(2013)]{2013ApJ...771...87H} H{\"o}nig, S.~F., Kishimoto, M., Tristram, K.~R.~W., et al.\ 2013, ApJ, 771, 87 

\bibitem[H{\"o}nig \& Kishimoto(2017)]{2017ApJ...838L..20H} H{\"o}nig, S.~F., \& Kishimoto, M.\ 2017, ApJL, 838, L20 

\bibitem[Jud et al.(2017)]{2017MNRAS.465..248J} Jud, H., Schartmann, M., Mould, J., Burtscher, L., \& Tristram, K.~R.~W.\ 2017, MNRAS, 465, 248

\bibitem[Keating et al.(2012)]{2012ApJ...749...32K} Keating, S.~K., Everett, J.~E., Gallagher, S.~C., \& Deo, R.~P.\ 2012, ApJ, 749, 32 

\bibitem[Kelly et al.(2017)]{2017A&A...597A..11K} Kelly, G., Viti, S., Garc{\'{\i}}a-Burillo, S., et al.\ 2017, A\&A, 597, A11 

\bibitem[Kishimoto et al.(2007)]{2007A&A...476..713K} Kishimoto, M., H{\"o}nig, S.~F., Beckert, T., \& Weigelt, G.\ 2007, A\&A, 476, 713 

\bibitem[Kishimoto et al.(2011)]{2011A&A...536A..78K} Kishimoto, M., H{\"o}nig, S.~F., Antonucci, R., et al.\ 2011, A\&A, 536, A78 

\bibitem[K\"{o}nigl \& Pudritz(2000)]{2000prpl.conf..759K} Konigl, A., \& Pudritz, R.~E.\ 2000, Protostars and Planets IV, 759 


\bibitem{a36} Krolik J.~H., Begelman M.~C., 1988, ApJ, 329, 702 

\bibitem[Lamy \& Hutsem{\'e}kers(2004)]{2004A&A...427..107L} Lamy, H., \& Hutsem{\'e}kers, D.\ 2004, A\&A, 427, 107 

\bibitem[Laor \& Draine(1993)]{1993ApJ...402..441L} Laor, A., \& Draine, B.~T.\ 1993, ApJ, 402, 441 

\bibitem[Lin et al.(2016)]{2016MNRAS.458.1375L} Lin, M.-Y., Davies, R.~I., Burtscher, L., et al.\ 2016, MNRAS, 458, 1375 

\bibitem[Liu et al.(2014)]{2014ApJ...783..106L} Liu, T., Wang, J.-X., Yang, H., Zhu, F.-F., \& Zhou, Y.-Y.\ 2014, ApJ, 783, 106 

\bibitem[Lodato \& Bertin(2003)]{2003A&A...398..517L} Lodato, G., \& Bertin, G.\ 2003, A\&A, 398, 517 

\bibitem[L{\'o}pez-Gonzaga et al.(2014)]{2014A&A...565A..71L} L{\'o}pez-Gonzaga, N., Jaffe, W., Burtscher, L., Tristram, K.~R.~W., \& Meisenheimer, K.\ 2014, A\&A, 565, A71 

\bibitem[L{\'o}pez-Gonzaga et al.(2016)]{2016A&A...591A..47L} L{\'o}pez-Gonzaga, N., Burtscher, L., Tristram, K.~R.~W., Meisenheimer, K., \& Schartmann, M.\ 2016, A\&A, 591, A47
 
\bibitem[Lopez-Rodriguez et al.(2015)]{2015MNRAS.452.1902L} Lopez-Rodriguez, E., Packham, C., Jones, T.~J., et al.\ 2015, MNRAS, 452, 1902 

\bibitem[Lucy(1999)]{1999A&A...344..282L} Lucy, L.~B.\ 1999, A\&A, 344, 282 

\bibitem[Maiolino \& Rieke(1995)]{1995ApJ...454...95M} Maiolino, R., \& Rieke, G.~H.\ 1995, ApJ, 454, 95 

\bibitem[Marconi et al.(2004)]{2004MNRAS.351..169M} Marconi, A., Risaliti, G., Gilli, R., et al.\ 2004, MNRAS, 351, 169 

\bibitem[Marin et al.(2012)]{2012A&A...548A.121M} Marin, F., Goosmann, R.~W., Gaskell, C.~M., Porquet, D., \& Dov{\v c}iak, M.\ 2012, A\&A, 548, A121 

\bibitem[Marin \& Goosmann(2013)]{2013MNRAS.436.2522M} Marin, F., \& Goosmann, R.~W.\ 2013, MNRAS, 436, 2522

\bibitem[Marin(2014)]{2014MNRAS.441..551M} Marin, F.\ 2014, MNRAS, 441, 551 

\bibitem[Marin et al.(2015)]{2015A&A...577A..66M} Marin, F., Goosmann, R.~W., \& Gaskell, C.~M.\ 2015, A\&A, 577, A66 

\bibitem[Mateos et al.(2017)]{2017ApJ...841L..18M} Mateos, S., Carrera, F.~J., Barcons, X., et al.\ 2017, ApJL, 841, L18 

\bibitem[Mathis et al.(1977)]{1977ApJ...217..425M} Mathis, J.~S., Rumpl, W., \& Nordsieck, K.~H.\ 1977, ApJ, 217, 425 

\bibitem[Matt(2000)]{2000A&A...355L..31M} Matt, G.\ 2000, A\&A, 355, L31 

\bibitem[McKee et al.(2010)]{2010ApJ...720.1612M} McKee, C.~F., Li, P.~S., \& Klein, R.~I.\ 2010, ApJ, 720, 1612 

\bibitem[Meijerink \& Spaans(2005)]{2005A&A...436..397M} Meijerink, R., \& Spaans, M.\ 2005, A\&A, 436, 397 

\bibitem[Michel(1969)]{1969ApJ...158..727M} Michel, F.~C.\ 1969, ApJ, 158, 727 

\bibitem[Miller \& Antonucci(1983)]{1983ApJ...271L...7M} Miller, J.~S., \& Antonucci, R.~R.~J.\ 1983, ApJL, 271, L7 

\bibitem[Monnier et al.(2006)]{2006ApJ...647..444M} Monnier, J.~D., Berger, J.-P., Millan-Gabet, R., et al.\ 2006, ApJ, 647, 444

\bibitem[Moorwood et al.(1996)]{1996A&A...315L.109M} Moorwood, A.~F.~M., Lutz, D., Oliva, E., et al.\ 1996, A\&A, 315, L109 

\bibitem[M{\"u}ller-S{\'a}nchez et al.(2011)]{2011ApJ...739...69M} M{\"u}ller-S{\'a}nchez, F., Prieto, M.~A., Hicks, E.~K.~S., et al.\ 2011, ApJ, 739, 69 

\bibitem[M{\"u}ller S{\'a}nchez et al.(2009)]{2009ApJ...691..749M} M{\"u}ller S{\'a}nchez, F., Davies, R.~I., Genzel, R., et al.\ 2009, ApJ, 691, 749 

\bibitem[Natta et al.(2001)]{2001A&A...371..186N} Natta, A., Prusti, T., Neri, R., et al.\ 2001, A\&A, 371, 186 

\bibitem[Netzer(1987)]{1987MNRAS.225...55N} Netzer, H.\ 1987, MNRAS, 225, 55 

\bibitem[Netzer(2015)]{2015ARA&A..53..365N} Netzer, H.\ 2015, ARA\&A, 53, 365 

\bibitem[Netzer et al.(2016)]{2016ApJ...819..123N} Netzer, H., Lani, C., Nordon, R., et al.\ 2016, ApJ, 819, 123 

\bibitem[Oliva et al.(1998)]{1998A&A...329L..21O} Oliva, E., Marconi, A., Cimatti, A., \& di Serego Alighieri, S.\ 1998, A\&A, 329, L21 

\bibitem[Osterbrock(1991)]{1991RPPh...54..579O} Osterbrock, D.~E.\ 1991, Reports on Progress in Physics, 54, 579 

\bibitem[Pier et al.(1994)]{1994ApJ...428..124P} Pier, E.~A., Antonucci, R., Hurt, T., Kriss, G., \& Krolik, J.\ 1994, ApJ, 428, 124 

\bibitem[Prieto et al.(2010)]{2010MNRAS.402..724P} Prieto, M.~A., Reunanen, J., Tristram, K.~R.~W., et al.\ 2010, MNRAS, 402, 724 

\bibitem[Prieto et al.(2014)]{2014MNRAS.442.2145P} Prieto, M.~A., Mezcua, M., Fern{\'a}ndez-Ontiveros, J.~A., \& Schartmann, M.\ 2014, MNRAS, 442, 2145 

\bibitem{a45} Pringle J.~E., 1981, ARA\&A, 19, 137

\bibitem[Pudritz \& Norman(1983)]{1983ApJ...274..677P} Pudritz, R.~E., \& Norman, C.~A.\ 1983, ApJ, 274, 677 

\bibitem[Ramos Almeida et al.(2016)]{2016MNRAS.461.1387R} Ramos Almeida, C., Mart{\'{\i}}nez Gonz{\'a}lez, M.~J., Asensio Ramos, A., et al.\ 2016, MNRAS, 461, 1387 

\bibitem[Robitaille(2010)]{2010A&A...520A..70R} Robitaille, T.~P.\ 2010, A\&A, 520, A70 

\bibitem[Roth et al.(2012)]{2012ApJ...759...36R} Roth, N., Kasen, D., Hopkins, P.~F., \& Quataert, E.\ 2012, ApJ, 759, 36 

\bibitem[Ruiz et al.(2000)]{2000MNRAS.316...49R} Ruiz, M., Alexander, D.~M., Young, S., et al.\ 2000, MNRAS, 316, 49 

\bibitem[Sani et al.(2012)]{2012MNRAS.424.1963S} Sani, E., Davies, R.~I., Sternberg, A., et al.\ 2012, MNRAS, 424, 1963 

\bibitem[Sazonov et al.(2015)]{2015MNRAS.454.1202S} Sazonov, S., Churazov, E., \& Krivonos, R.\ 2015, MNRAS, 454, 1202 

\bibitem[Schartmann et al.(2005)]{2005A&A...437..861S} Schartmann, M., Meisenheimer, K., Camenzind, M., Wolf, S., \& Henning, T.\ 2005, A\&A, 437, 861

\bibitem[Schartmann et al.(2014)]{2014MNRAS.445.3878S} Schartmann, M., Wada, K., Prieto, M.~A., Burkert, A., \& Tristram, K.~R.~W.\ 2014, MNRAS, 445, 3878 

\bibitem[Smith et al.(2002)]{2002MNRAS.335..773S} Smith, J.~E., Young, S., Robinson, A., et al.\ 2002, MNRAS, 335, 773 

\bibitem[Stalevski et al.(2017)]{2017MNRAS.472.3854S} Stalevski, M., Asmus, D., \& Tristram, K.~R.~W.\ 2017, MNRAS, 472, 3854 

\bibitem[Stern \& Laor(2012)]{2012MNRAS.426.2703S} Stern, J., \& Laor, A.\ 2012, MNRAS, 426, 2703 

\bibitem[Suganuma et al.(2006)]{2006ApJ...639...46S} Suganuma, M., Yoshii, Y., Kobayashi, Y., et al.\ 2006, ApJ, 639, 46 

\bibitem[Tristram et al.(2014)]{2014A&A...563A..82T} Tristram, K.~R.~W., Burtscher, L., Jaffe, W., et al.\ 2014, A\&A, 563, A82 

\bibitem[Veilleux \& Bland-Hawthorn(1997)]{1997ApJ...479L.105V} Veilleux, S., \& Bland-Hawthorn, J.\ 1997, ApJL, 479, L105 

\bibitem[Vollmer \& Duschl(2001)]{2001A&A...367...72V} Vollmer, B., \& Duschl, W.~J.\ 2001, A\&A, 367, 72 

\bibitem[\protect\citeauthoryear{Vollmer \& Beckert}{2002}]{VollmerBeckert} Vollmer B., Beckert T. 2002, A\&A, 382, 872

\bibitem[Vollmer et al.(2004)]{2004A&A...413..949V} Vollmer, B., Beckert, T., \& Duschl, W.~J.\ 2004, A\&A, 413, 949 

\bibitem[Vollmer et al.(2008)]{2008A&A...491..441V} Vollmer, B., Beckert, T., \& Davies, R.~I.\ 2008, A\&A, 491, 441 

\bibitem[Vollmer \& Davies(2013)]{2013A&A...556A..31V} Vollmer, B., \& Davies, R.~I.\ 2013, A\&A, 556, A31 

\bibitem[Wada et al.(2002)]{2002ApJ...577..197W} Wada, K., Meurer, G., \& Norman, C.~A.\ 2002, ApJ, 577, 197 

\bibitem[Wada et al.(2009)]{2009ApJ...702...63W} Wada, K., Papadopoulos, P.~P., \& Spaans, M.\ 2009, ApJ, 702, 63 

\bibitem[Wada(2012)]{2012ApJ...758...66W} Wada, K.\ 2012, ApJ, 758, 66 

\bibitem[Wada et al.(2016)]{2016ApJ...828L..19W} Wada, K., Schartmann, M., \& Meijerink, R.\ 2016, ApJL, 828, L19 

\bibitem[Wardle \& Koenigl(1993)]{1993ApJ...410..218W} Wardle, M., \& Koenigl, A.\ 1993, ApJ, 410, 218 

\bibitem[Weingartner \& Draine(2001)]{2001ApJ...548..296W} Weingartner, J.~C., \& Draine, B.~T.\ 2001, ApJ, 548, 296 

\bibitem[Wilson et al.(2000)]{2000AJ....120.1325W} Wilson, A.~S., Shopbell, P.~L., Simpson, C., et al.\ 2000, AJ, 120, 1325 

\bibitem[Yang et al.(2015)]{2015ApJ...799...91Y} Yang, H., Wang, J., \& Liu, T.\ 2015, ApJ, 799, 91 

\end{thebibliography}
\end{document}